\documentclass[times]{cnmauth}

\usepackage{lineno,hyperref}
\usepackage{arydshln}
\modulolinenumbers[5]

\newtheorem{algorithm}{Algorithm}[section]

\usepackage{exscale}
\usepackage{graphicx,amsmath,bm}
\usepackage{amssymb,amscd,verbatim}
\usepackage{color}
\usepackage{float}
\usepackage{algpseudocode,algorithm,algorithmicx}

\algrenewcommand\algorithmicrequire{\textbf{Precondition:}}
\algrenewcommand\algorithmicensure{\textbf{Postcondition:}}

\newcommand{\ds}{\,\mathrm{d}s}

\newcommand{\p}{\partial}

\newcommand{\gradd}{\nabla}

\DeclareMathAlphabet{\mathsfsl}{OT1}{cmss}{m}{sl}

\renewcommand{\vec}[1]{\mbox{\boldmath$#1$}}
\newcommand{\oo}{\ensuremath{\Omega}}
\newcommand{\G}{\Gamma}

\newcommand{\mj}{\,\mathcal{J}}

\newcommand{\mf}{\,\mathcal{F}}

\newcommand{\vv}{\vec{V}}

\newcommand{\vvv}{\vec{v}}

\newcommand{\vu}{{\vec{u}}}

\newcommand{\vf}{\vec{f}}
\newcommand{\vg}{\vec{g}}

\newcommand{\vF}{\vec{F}}

\newcommand{\vS}{\vec{S}}
\newcommand{\vI}{\vec{I}}
\newcommand{\vE}{\vec{E}}

\newcommand{\vsigma}{\vec{\sigma}}
\newcommand{\vPi}{\vec{\Pi}}

\newcommand{\vvarepsilon}{\vec{\varepsilon}}
\newcommand{\vn}{\vec{n}}
\newcommand{\vx}{\vec{x}}
\newcommand{\vX}{\vec{X}}

\newcommand{\vd}{\vec{d}}

\begin{document}

\title{Simulation of unsteady blood  flows  in a patient-specific compliant pulmonary artery with a highly parallel monolithically coupled fluid-structure interaction algorithm}

\runningheads{Fande Kong, Vitaly Kheyfets, and Xiao-Chuan Cai}{A monolithic coupling fluid-structure interaction simulation}

\author{Fande Kong\footnotemark[2], Vitaly Kheyfets\footnotemark[3], Ender Finol\footnotemark[4] and  Xiao-Chuan Cai\footnotemark[5]\corrauth}

\address{\center \footnotemark[2]{Modeling and Simulation, Idaho National Laboratory,  P.O. Box 1625, Idaho Falls, ID 83415-3840, USA }
~\\
\center \footnotemark[3]{School of Medicine, University of Colorado Denver, Aurora, CO 80045-7109, USA }
~\\
\center \footnotemark[4]{Department of Mechanical Engineering, University of Texas at San Antonio, San Antonio, TX 78249, USA }\\
~\\
\footnotemark[5]{Department of Computer Science, University of Colorado Boulder, Boulder, CO 80309-0430, USA}} 

\corraddr{Department of Computer Science, University of Colorado Boulder, Boulder, CO 80309-0430, USA}

\begin{abstract}
 Computational fluid dynamics (CFD) is increasingly used to study  blood  flows in patient-specific arteries for  understanding  certain cardiovascular diseases.  The techniques work quite well for relatively simple problems, but need improvements when the problems become harder in the case when (1) the geometry becomes complex (from a few branches to a full pulmonary artery), (2) the model becomes more complex (from fluid-only calculation to coupled fluid-structure interaction  calculation), (3) both the fluid and wall models become highly nonlinear, and (4) the computer on which we run the simulation is a supercomputer with tens of thousands of processor cores. To push the limit of CFD in all four fronts, in this paper, we develop and study a highly parallel algorithm for solving a monolithically coupled fluid-structure  system for the modeling of the interaction of the blood flow and the arterial wall. As a case study, we consider a patient-specific, full size pulmonary artery obtained from  CT (Computed Tomography) images, with an artificially added layer of wall with a fixed thickness. The fluid is modeled with a system of incompressible Navier-Stokes equations and the wall is modeled by a geometrically nonlinear  elasticity equation. As far as we know this is the first time the unsteady blood flow in a full pulmonary artery is simulated without assuming a rigid wall. The proposed numerical algorithm and software scale well beyond 10,000 processor cores on a supercomputer for solving the fluid-structure interaction problem discretized with a stabilized finite element method in space and an implicit scheme in time  involving hundreds of millions of unknowns.
\end{abstract}


\maketitle

\section{Introduction}
Millions of people die every year from cardiovascular diseases, representing more than $30\%$ of all global  deaths \cite{wang2016global}. Computational fluid dynamics (CFD) is useful for understanding certain cardiovascular diseases, for example, in  \cite{kheyfets2015patient}, it was shown that the wall shear stress obtained through a steady blood flow simulation   is highly correlated to the resistive pulmonary arterial impedance. In most numerical simulations, the wall of the artery is assumed to be rigid, but   it is widely believed that a simulation with an elastic wall would produce  more valuable results than a fluid-only simulation \cite{crosetto2011fluid},  even though, such calculations are computationally very expensive,  and often take days or months  of computing time on small computers without lots of processor cores.  With the recent development of supercomputers, high-resolution fluid-structure interaction (FSI) computation  becomes feasible and there have been several successful applications in, for example, respiratory mechanics \cite{verdugo2017efficient}, aeroelasticity  \cite{farhat1998load}, and hemodynamics \cite{kong2016scalability, kong2017scalable, bazilevs2006isogeometric}.  However, simulating the unsteady blood flow in a deformable pulmonary artery is challenging because the coupled FSI system is highly nonlinear  and the geometry of the  computational domain is complex.  The nonlinearities come from the nonlinear fluid and solid equations, and also from  the dependency of the blood velocity and pressure on the domain movement governed by a harmonic extension equation.  Moreover, the  complexity of the computational domain makes  the generation of a matching fluid and solid mesh rather  difficult.  In this paper, we  extend and investigate   a monolithically coupled  Newton-Krylov-Schwarz (NKS) algorithm for the  FSI simulation of the three-dimensional  pulmonary vasculature  on a supercomputer with more than 10,000 processor cores.

Generally speaking, there are two approaches for coupling a fluid problem   with a solid problem, namely, loose coupling  (partitioned) and full coupling   (monolithic). In the partitioned method,  a fluid problem or a solid equation  is calculated first, and  then its solution is provided as the boundary condition of  the other domain. The partitioned algorithm is essentially a Gauss-Seidel  iteration, and it has been successfully applied to  a few engineering areas, for example,  aeroelasticity \cite{farhat1998load, farhat2006provably}.  But there are two major drawbacks in the partitioned method. First,  it is not suitable for  parallel computing  because the Gauss-Seidel  iteration is sequential and has a low concurrency \cite{kongfully}. Second, there is an ``added-mass effect", in other words,   significant numerical instabilities are introduced  when the densities of the fluid and the solid  are close to each other \cite{causin2005added}.  In this paper, a monolithic approach is employed since it  does not suffer from such difficulties and often offers a stable and scalable parallel algorithm. The monolithic  approach has been  used in several computational  hemodynamics applications \cite{crosetto2011fluid, kong2016scalability, kong2017scalable, bazilevs2006isogeometric}.  In our current work,   a nonlinear elasticity  equation is used to model the wall of the artery,   an incompressible Navier-Stokes system  is employed  to model the blood flow,  and a third equation is used to describe the moving fluid domain.   All three partial differential  equations are monolithically coupled together based on an  Arbitrary Lagrangian Eulerian (ALE) framework \cite{souli2000ale}. To  discretize  the coupled FSI system, a $P_1$ finite element is utilized for both the solid and domain movement equations  and a stabilized $P_1-P_1$ finite element  pair is employed  for the incompressible Navier-Stokes equations.  The resulting semi-discretized  FSI system of equations is further discretized in time using an backward Euler scheme.   After the spatial and temporal discretization, a large and highly nonlinear system of algebraic equations is produced   and its solution requires  an efficient parallel algorithm since we are aiming for a supercomputer  with a large number of processor cores, without which the calculation may take days or months of computing time.  To tackle the discretized FSI system, we develop  an inexact Newton method  for solving the nonlinear system, during each Newton iteration a Krylov subspace method together with a Schwarz preconditioner is carefully chosen for the solution of the Jacobian system.   

We next briefly review some recent progresses of numerical simulation of blood flows in the pulmonary artery.  For the fluid-only calculation,  in \cite{spilker2007morphometry}, the blood flow is described by  a  one-dimensional model discretized  with a finite element method, and the corresponding  system of nonlinear  equations is solved via a quasi Newton  method.  In \cite{tang2011three}, a  three-dimensional  simulation of unsteady blood flows of a   patient-specific pulmonary artery is obtained using a stabilized finite element method for  the incompressible Navier-Stokes equations, and the sequential simulation takes a couple of days of computing time for a problem with a million elements.   In \cite{qureshi2014numerical}, a multiscale  model of  a one-dimensional   pulmonary network  is presented and used to analyze  the  arterial and venous pressure, and the flow.  
For the FSI simulation,  in \cite{su2012impact, hunter2006simulations},  the CFD-ACE multiphysics package \cite{CFD_ACE} is used for a two-dimensional  unsteady blood flow simulation, where  a finite volume method is used for the fluid equations and a finite element method is used for the arterial wall.  The resulting system of equations is solved with an algebraic multigrid method for the fluid equations and a direct method for the solid equations.   A three-dimensional pulmonary arterial bifurcation with simple geometry is simulated by  solving   a steady state FSI  problem in \cite{yang2007fluid}, where an  in-house FEM code is used to calculate the nonlinear deformation of the thin-walled structure and a commercial CFD solver, ANSYS \cite{Ansys2017Fluent}, is used to resolve the fluid equations.

Most of these published works  focus   on either  the fluid only simulation or FSI simulations  with simple geometry; i.e., a small number of branches.   To best of our knowledge,  unsteady 3D FSI simulation with the full patient-specific  pulmonary artery is still not available in the existing literature because the required scalable parallel algorithm and software are difficult   to develop.  Existing commercial software such as  ANSYS \cite{Ansys2017Fluent} scales only to a few hundred processor cores, and that's not enough to carry out these large calculations in a reasonable  amount of time. The monolithically  coupled NKS  was previously  applied to a  FSI simulation  with a small portion of a three-dimensional  artery  \cite{kong2016scalability, kong2017scalable, wu2014fully,  kong2016parallel}, but  the approach is not easy to  be used for the full artery case  since  some of the algorithmic   parameters  are  geometry-dependent and problem-dependent; as the complexity of the geometry increases the convergence becomes problematic.   With the right choices of parameters, including  the  subdomain overlap, the fill-in level of incomplete LU factorization,  the reordering schemes of the subdomain matrices, the lag of the Jacobian computation and the inexactness of   the Newton iterations,  we show experimentally that the proposed  version of NKS  method is  scalable with up to 10,240 processor cores for the FSI simulation of a full  pulmonary network.   We also want to mention that  NKS has also been successfully used in  different applications in our previous work;   unsteady blood flow simulation \cite{kong2017efficient},  nonlinear elasticity equations \cite{kong2016highly},  and transient multigroup neutron diffusion equations \cite{kongfully}.

The remainder of this paper is organized as follows. In Section 2,  we   present  the physics models used in  the FSI coupling and their spatial and temporal discretization.    A fully implicit,  monolithically coupled parallel  Newton-Krylov-Schwarz method is described  in Section 3. In Section 4, some numerical experiments and observations  
are presented, and we focus mainly on the parallel performance of the proposed approach.  Lastly,
some concluding remarks are given in Section 5. 

\section{Mathematical models of the fluid and the wall}
\begin{figure}
   \centering
   \includegraphics[width=3.5in]{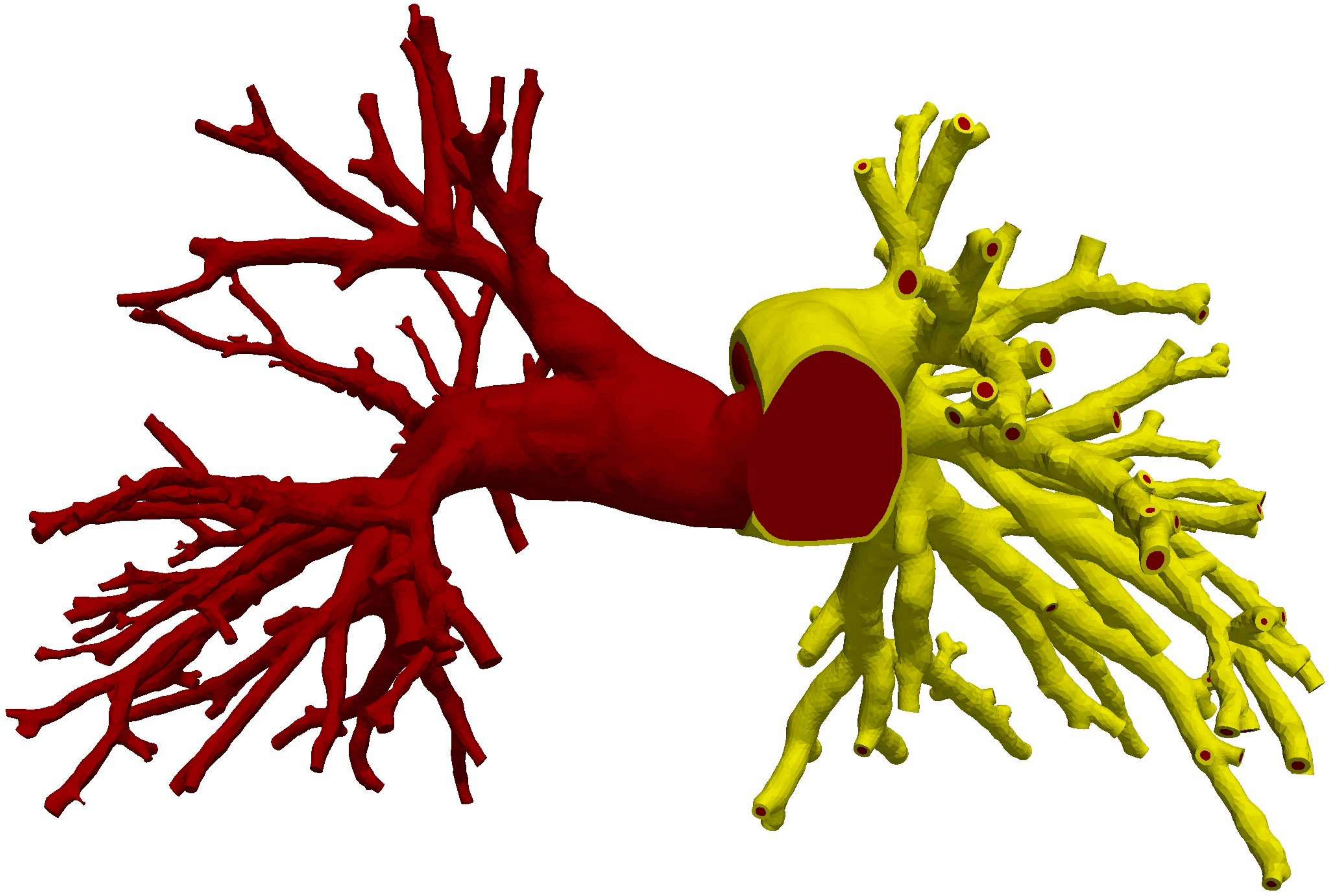} 
   \caption{A patient-specific pulmonary artery. The red part is the blood flow domain and the yellow is the arterial wall.} \label{computational_domain}
 \end{figure}

The  pulmonary circulation  carries deoxygenated blood from the right ventricle  to the lungs.  A patient-specific  complete pulmonary tree, shown in Fig.~\ref{computational_domain},  is considered in this work. 
In Fig.~\ref{computational_domain},  the red part is the blood flow domain, and the yellow part is the arterial wall.  The cut out  is for the visualization, and
the entire arterial  wall is included in the actual FSI simulation. In this section we describe the models for the blood flow and the arterial wall, as well as the spatial and temporal discretization of the equations.

\subsection{Mathematical models for blood flow, arterial wall, and moving fluid domain}
\begin{figure}
   \centering
   \includegraphics[width=4.5in]{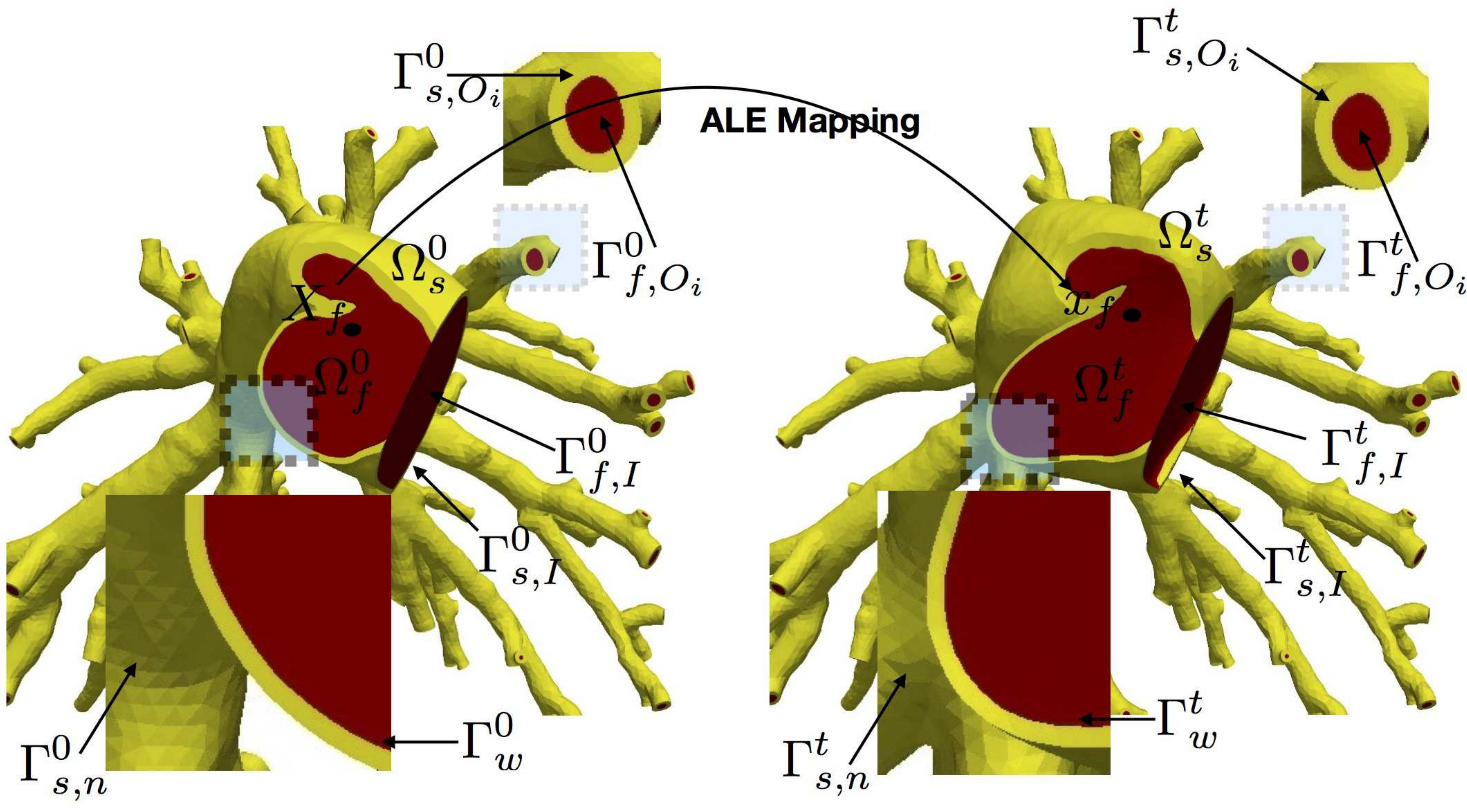} 
   \caption{The figure shows the  ALE mapping $A_t$ from the initial configuration (left) to the configuration at time $t$ (right). In the right figure one can also see the deformation  of the arterial wall and the blood flow domain.} \label{fsiconfiguration}
\end{figure}

We begin by introducing  some notations.  At time $t \in [0, T]$,   let $\oo_f^t \in R^3$ be the fluid  domain,  
$\oo_s^t \in R^3$   the arterial wall,  $ \G^t_{f,I}$ the  fluid inlet,  $\G^t_{f,o_i}, i = 1, 2, ....$  the fluid  outlets,  $ \G^t_{s,I}$  the wall inlet,  
 $\G^t_{s,o_i}$  the wall outlets, $\G^t_{s,n}$ the wall outer surface   and $\G^t_{w}$  the  wet interface between the fluid and the arterial  wall. Note  that $t=0$ corresponds to the initial configuration.   The FSI configuration is shown in Fig.~\ref{fsiconfiguration}, where an ALE mapping is defined for tracking the movement of the fluid domain: 
 $$
A_t: \vx_f = A_t(\vX_f)  \equiv \vX_f+\vd_m,  \vx_f \in \oo_f^t, \vX_f \in \oo_f^0.
$$
We denote by $\vd_m \in R^3 $ as the fluid domain displacement  at time $t$, and  assume it satisfies  the following harmonic  extension equation
\begin{equation}\label{domainmotion}
\left \{
\begin{array}{lllll}
-\Delta \vd_m  =  \boldsymbol{0} &\text{~in~} \oo_f^0,\\
\vd_m = \boldsymbol{0} &\text{~on~} \G_f^0  \equiv \G^0_{f,I} \cup \G^0_{f,O_i}, i =1,2, ...\\
\vd_m = \vd_s & \text{~on~} \G_w^0.
\end{array}
\right .
\end{equation}
Note that this equation doesn't have any particular physical meaning, and it is used to describe the motion of the fluid domain. We denote by  $\vd_s \in R^3$ as the arterial  wall displacement  governed by an unsteady, geometrically nonlinear elasticity equation \cite{howell2009applied} as follows:
\begin{equation}\label{solidequation}
\left \{
\begin{array}{llll}
\rho_s \displaystyle \frac{\p^2 \vd_s}{\p t^2} +\eta_s \frac{\p \vd_s}{\p t}- \gradd \cdot \vPi_s= \rho_s \vf_s &\text{~in~} \oo_s^0,\\
\vPi_s \vn_s = \vg_s & \text{~on~} \G_{s,n}^0,\\
\vd_s  = \boldsymbol{0} & \text{~on~} \G_{s,I}^0 \cup \G_{s,O_i}^0,  i =1,2, ...\\
\vsigma_s {\hat{\vn}}_s = - \vsigma_f \vn_f &\text{~on~} \G_w^t.
\end{array}
\right .
\end{equation}
In (\ref{solidequation}), $\rho_s$ is the  wall density, $\vf_s$ is a volumetric force per unit of mass,  $\eta_s \p \vd_s /\p t$ is a damping term used to mimic the impact of  the surrounding tissues,   $\eta_s$ is a damping parameter,  $\vn_s$ and ${\hat{\vn}}_s$ are the unit outward normal  vectors (they are related by  the Nanson formula \cite{howell2009applied}) under the initial configuration and the deformed domain, respectively, $\vsigma_s$ is the Cauchy stress tensor of the  wall,   and  $\vsigma_f$  is the Cauchy stress tensor for the fluid domain to be defined shortly.  Here $ \vPi_s$  is  the  nonsymmetric first Piola-Kirchhoff stress tensor for the Saint Venant-Kirchhoff material,
$$
\begin{array}{llll}
 \vF = (\vI+\nabla\vd_s), \\
 \displaystyle \vE = \frac{1}{2} ({\vF}^T \vF -\vI), \\
  \vS = \lambda_s \text{trace}(\vE)\vI + 2\mu_s \vE,  \\
  \vPi_s = \vF \vS, 
 \end{array}
$$
where $\vI$ is a $3\times3$ identity matrix, $\vF$ is the deformation gradient tensor, $\vE$ is the Green-Lagrangian strain tensor, $\vS$ is the second Piola-Kirchhoff stress tensor,
and $\mu_s$ and $\lambda_s$  are the material Lam$\acute{e}$ constants expressed as functions of Young's modulus, $E_s$, and Poisson's ratio, $\nu_s$, by
$$
\mu_s =\frac{E_s}{2\left(1+\nu_s \right)}  \text{ and }\lambda_s=\frac{E_s\nu_s}{\left(1+\nu_s\right)\left(1-2
\nu_s\right)}.
$$ 
The nonsymmetric first Piola-Kirchhoff stress tensor $\vPi_s$ and the  Cauchy stress tensor $\vsigma_s$ are related by the formula:
$$
\vsigma_s=\frac{\vPi_s \vF^T} {det(\vF)}. 
$$
Here $det(\vF)$ is the determinant of the tensor $\vF$. 
For the blood flows, let $\vu_f$ and $p_f$ denote the  velocity and the pressure, respectively, and the incompressible Navier-Stokes equations on the moving domain are presented  as follows
\begin{equation}\label{fluidequation}
%\footnotesize
\centering
\left \{
\begin{array}{llllll}
\displaystyle \left . \rho_f \frac{\p \vu_f}{\p t} \right |_{\vX_f} +\rho_f \left[\left(\vu_f-\frac{\p \vd_m}{\p t}\right)\cdot \gradd\right]\vu_f  \displaystyle-\gradd \cdot \vsigma_f = \rho_f \vf_f &\text{~in~} \oo^t_{f}, \\
\gradd \cdot \vu_f = 0 & \text{~in~} \oo^t_{f},\\
\displaystyle \vu_f     = \vvv_f^d & \text{~on~} \G^t_{f,I}, \\
\displaystyle \vsigma_f \vn_f = \vg_f & \text{~on~} \G^t_{f,O_i},\\
\displaystyle  \vu_f  = \frac{\p \vd_s(A_t^{-1})}{ \p t} & \text{~on~} \G^t_{w},
\end{array}
\right .
\end{equation}
where $\rho_f$ is the fluid density, $|_{\vX_f} $ indicates that the time derivative is taken under the ALE configuration, $\p \vd_m/ \p t$ is the velocity of the mesh movement, $\vf_f$ is a  volumetric force per unit of mass,    $\vg_f$ is a traction applied to the outlets, $\vvv_f^d$ is a velocity profile prescribed at the inlet,     $A_t^{-1}$ is a pull-back operator that maps the current coordinates to the original configuration, $\vn_f$ is the unit outward normal  vector for  the fluid domain and $\vsigma_f$ is the Cauchy stress tensor for the fluid defined as 
$$
\vsigma_f = -p_f \vI + 2 \nu_f \vvarepsilon_f,  ~ \vvarepsilon_f = \frac{1}{2} \left(\gradd \vu_f + \gradd \vu_f^T\right),
$$
where   $\vvarepsilon_f$ is the strain rate tensor and $\nu_f$ is the viscosity coefficient.  On the wet interface,  three coupling conditions are imposed to couple the solid and  fluid equations. The first  condition is the continuity of the velocity: $\vu_f = \p \vd_s / \p t$. The second  condition is the continuity of the displacement: $\vd_m = \vd_s$. Lastly, the traction forces from the fluid and the solid are the same: $\vsigma_s {\hat{\vn}}_s = - \vsigma_f \vn_f$. These conditions are included in (\ref{domainmotion}), (\ref{solidequation}) and (\ref{fluidequation}), through which the coupled FSI system is formed. 

\subsection{Seamless coupling discretization}
To  discretize (\ref{domainmotion}), (\ref{solidequation}) and (\ref{fluidequation}), we  consider a $P_1-P_1$ stabilized finite element pair  \cite{kong2017efficient, taylor1998finite, whiting2001stabilized} for the incompressible Navier-Stokes equations and a $P_1$ finite element method for both the solid equation and the fluid domain moving equation. Interested readers are referred to 
 \cite{wu2014fully, kong2016parallel} for more  details.

There are two approaches  to implement the continuity conditions (the velocity continuity and the displacement continuity) on the wet interface for the displacement and the velocity, that is, they can be formed either weakly or strongly.  Let $\vu_f^h, \vu_s^h, \vd_m^h$ and $\vd_s^h$ be their counterparts in the finite element spaces. The  constraints can be enforced weakly through the following weak forms:
\begin{equation}\label{weak_continuity_velocity}
 \int_{\G_w^h}  (\vu_f^h - \vu_s^h) \vvv  \ds  = 0, ~\forall \vvv \in \vv^h
\end{equation}
and 
\begin{equation}\label{weak_continuity_displacement}
 \int_{\G_w^h}  (\vd_m^h - \vd_s^h) \vvv \ds  = 0, ~\forall \vvv \in \vv^h. 
\end{equation}
Here $\vv^h$ is a  $P_1$ finite element function space defined on the wet interface.  
In the strong form of the interface condition, the  constraints are enforced at every mesh points,  that is, $\vd_m^h |_{\G_w^h} = \vd_s^h |_{\G_w^h}$ and $\vu_f^h  |_{\G_w^h} = \vu_s^h |_{\G_w^h}$. Here we think of $\vu_f^h, \vu_s^h, \vd_m^h$ and $\vd_s^h$ as nodal values of their corresponding finite element functions without introducing any confusion.  More precisely, let  $\vd^h_{\G,m}, \vd^h_{\G,s}$ be the unknowns of the fluid and solid displacements on the wet interface, respectively, and $\vd^h_{I,m}$,  $\vd^h_{I,s}$ correspond to the unknowns  defined in the interior of the domain.  To show the matrix structure of the coupling condition on the interface, we take $\vd^h_{\G_w,m} = \vd^h_{\G_w,s}$ as an example, and   the structure of the coupled FSI matrix for the mesh movement  and the solid deformation looks like 
\begin{equation}\label{coupled_mesh_solid}
\left [
\begin{array}{llllll}
J^m_{I,I} &  J^m_{I,\G_w}  \\
0 &  I&  -I& \\ 
0 & B & 0 & \\
0 &  0 & J^s_{I,\G_w}  & J^s_{I, I}  \\ 
\end{array}
\right ]
\left [
\begin{array}{llllll}
\vd^h_{I,m}  \\
\vd^h_{\G_w,m}   \\
 \vd^h_{\G_w,s}\\
\vd^h_{I,s}\\
\end{array}
\right ]
=
\left [
\begin{array}{llllll}
0  \\
0 \\
0 \\
\vf_s \\
\end{array}
\right ].
\end{equation}
(\ref{coupled_mesh_solid}) is reduced to (\ref{reduced_mesh_solid}) if we replace the wall displacement unknowns   $\vd_{\G_w,s}^h$ by  the fluid displacement unknowns  $\vd_{\G_w,m}^h$ as follows: 
\begin{equation}\label{reduced_mesh_solid}
\left [
\begin{array}{llllll}
J^m_{I,I} &  J^m_{I,\G_w}  \\
0 & B & 0 & \\
 0 & J^s_{I,\G_w}  & J^s_{I, I} \\ 
\end{array}
\right ]
\left [
\begin{array}{llllll}
\vd^h_{I,m}  \\
\vd^h_{\G_w,m}   \\
\vd^h_{I,s}\\
\end{array}
\right ]
=
\left [
\begin{array}{llllll}
0 \\
0 \\
\vf_s \\
\end{array}
\right ].
\end{equation}
A similar  procedure  can be applied to the continuity condition on the velocity   as well. This implementation  of the continuity  conditions  makes the postprocessing  more convenient  because there are no duplicate unknowns on the interface mesh.  For the detailed structure of the  coupled FSI system,  we refer to our  previous work \cite{barker2010scalable}.

 After the discretization in space, the corresponding  semi-discretized system is  a time-dependent nonlinear  system
\begin{equation}\label{fsialgebra}
\frac{\p y(t)}{ \p t} + N( y(t))  = F, 
\end{equation}
where $F$ is the right-hand side and   $N(\cdot)$ is a  nonlinear function, $y(\cdot)$ is the time-dependent vector of nodal values of the fluid velocity and pressure,  the solid velocity  and displacement, and the displacement of the moving fluid domain.  Using  an implicit first-order  backward Euler scheme,  (\ref{fsialgebra}) is further discretized  in time
as:   
\begin{equation}\label{fsialgebrabdf}
M_n y_{n} + \delta t N(y_{n})  =  \delta t F+ M_n y_{n-1},
\end{equation}  
where  $\delta t $ is the time step size,  $y_n$ is the approximation of $y$ at the $n$th time step,  $M_n$ is the mass matrix dependent of $y_{n}$ since the computational fluid domain is moving. 
The ALE velocity $\p \vd_m/ \p t$ is approximated by  the first-order  backward Euler scheme in (\ref{fsialgebrabdf}) as well. 
For convenience, we rewrite (\ref{fsialgebrabdf}) at the $n$th  time step  as a nonlinear algebraic system:
\begin{equation}\label{nonlinearequation}
\mf(y) = 0,
\end{equation}
where $\mf(\cdot)$ is the combination of four terms in (\ref{fsialgebrabdf}),  and  $y$  (we drop the subscript $n$ here for simplicity) is the vector of nodal values  at the $n$th time step.

Because we want a high resolution simulation of the FSI system,  (\ref{nonlinearequation}) is usually very large, and is quite difficult to solve since all the high nonlinearities are  coupled in this single system. In the next section, we discuss a highly parallel, domain decomposition based solver for  (\ref{nonlinearequation}).

\section{Monolithic  coupling   parallel algorithm}
The parallel algorithm consists of a Newton method \cite{dembo1982inexact} for the coupled nonlinear system, a Krylov subspace method \cite{saad2003iterative} for the Jocabian system and an overlapping domain decomposition  preconditioner   \cite{smith2004domain} for the acceleration  of the linear solver.  With a given initial guess, $y^{(0)}$, inexact Newton obtains a new approximation as follows:
\begin{equation}\label{newton_method}
y^{(n+1)} = y^{(n)} + \alpha^{(n)} \delta y^{(n)},
\end{equation}
where $y^{(n)}$ is the approximate solution at the $n$th Newton step, $ \alpha^{(n)}$ is a Newton step size computed using a line search scheme such as backtracking \cite{dennis1996numerical}, and $\delta y^{(n)}$ is the Newton direction obtained by solving the  Jacobian system:
\begin{equation}\label{Jacobian_sytem}
\mj(y^{(\tilde{n})}) \delta y^{(n)} = -\mf(y^{(n)}). 
\end{equation}
Here $\mj(y^{(\tilde{n})})$ is the Jacobian matrix evaluated at $y^{(\tilde{n})}$, and $\mf(y^{(n)})$ is the nonlinear function residual evaluated at $y^{(n)}$. $\tilde{n}$ is smaller than or equal to $n$. The Jacobian matrix  from a previous step  is reused if $\tilde{n}$ is strictly smaller than $n$.  

The convergence of (\ref{newton_method}) depends critically on how the Jacobian matrix $\mj$ is constructed and how the Jacobian system (\ref{Jacobian_sytem}) is solved. The construction and the solve of the Jacobian are both expensive  and therefore carefully designed algorithms are extremely important. To have good nonlinear convergence, the Jacobian matrix $\mf$ is analytically derived and hand-coded  in the C++ code.  It is  a challenging task to form all derivatives exactly instead of using a finite difference method,  but it worth doing so because Newton equipped with an  exact Jacobian  often has a more robust convergence.  To save the compute time, the Jacobian from the previous Newton step may be reused as the evaluation of the Jacobian matrix is expensive.  As stated earlier,  the system~(\ref{nonlinearequation}) is highly nonlinear, and so the resulting Jacobian system is ill-conditioned. To overcome this difficulty, we propose a Krylov method, GMRES \cite{saad2003iterative} together  with a preconditioner based on an overlapping domain decomposition method.  The Krylov subspace method is generally understood in the  existing  literatures, but for it to work well for a specific application, a preconditioner has to be constructed carefully,  especially these problem dependent parameters need to be  considered.  More precisely, instead of (\ref{Jacobian_sytem}),  the following preconditioned linear system is  solved 
\begin{equation}\label{pre_Jacobian_sytem}
B^{-1}\mj  \delta y = - B^{-1}\mf. 
\end{equation}
For simplicity, the arguments of $\mj(\cdot)$ and $\mf(\cdot)$ are dropped here.  $B^{-1}$ is a parallel preconditioner to be constructed based on an overlapping domain decomposition method  below. Note that the same preconditioner can be applied from the right  side as well.

The basic idea of domain decomposition methods \cite{smith2004domain, toselli2006domain} is to divide the   mesh $\oo_h = \oo_{h,s} \cup \oo_{h,f}$ into $np$ submeshes $\oo_{h,i}, i=0, 1, 2... $, and each  submesh $\oo_{h,i}$ is assigned to a processor core and all subproblems are solved simultaneously in parallel.  For the overlapping version of domain decomposition methods, the submeshes are  extended to overlap with their neighboring submeshes by $\delta$ layers of mesh points.  The overlapping submeshes are denoted  as $\oo_{h,i}^{\delta}$.  To partition the  mesh $\oo_h$ across different processor cores, we employ a hierarchical partitioning method because mesh  partitioning software  such as ParMETIS/ METIS \cite{karypis1997parmetis} doesn't work well for the full    pulmonary tree   due to the complexity of geometry.  The idea of a  hierarchical partition  is quite simple but very effective. The  mesh $\oo_h$ is first partitioned into $np_1$ submeshes ($np_1$ is the number of compute nodes),  then each submesh is further partitioned into $np_2$ smaller submeshes ($np_2$ is the number of processor cores per compute node), and finally we have $np=np_1 \times np_2$ submeshes in total.  The advantage of the hierarchical partitioning  method is to take the architecture of modern computers into consideration, and the communication among different compute nodes are minimized.  More details of  the hierarchical partitioning  method are provided in  \cite{kong2016scalability, kong2017scalable, kong2016highly}. 

To describe the preconditioning technique, we denote the vector and submatrix   associated with submesh $\oo_{h,i},$ as $y_{h,i}$ and $\mj_{h,i}$, and that for the overlapping submesh  as  $\oo_{h,i}^{\delta},$ as $y_{h,i}^{\delta}$ and $\mj_{h,i}^{\delta}$. Let us define a restriction operator $R_i^{\delta}$ as 
\begin{equation}\label{restrict_operator}
y_{h,i}^{\delta} = R_i^{\delta} y = (\vI ~ 0) (y^{\delta}_{h,i}  ~ y/y^{\delta}_{h,i})^T, 
\end{equation}
where $\vI$ is an identity matrix whose size is the same as $y_{h,i}^{\delta}$. The restriction operator $R_i^{\delta}$ is used to extract a subverter $y_{h,i}^{\delta}$ from the global vector $y$ by selecting the corresponding components.  Using $R_i^{\delta}$, the overlapping submesh matrix $\mj_{h,i}^{\delta}$ is written as 
\begin{equation}\label{submatrix_operator}
\mj_{h,i}^{\delta} = (R_i^{\delta})^T \mj R_i^{\delta}. 
\end{equation}
With these ingredients, a restricted Schwarz preconditioner (overlapping domain decomposition method) reads as 
\begin{equation}\label{submatrix_operator}
B^{-1} = \sum_{i=0}^{np-1} (R_i^{0})^T (\mj_{h,i}^{\delta})^{-1}  (R_i^{\delta})^T,
\end{equation}
where $R_i^{0}$ is a restriction  operator without any overlap, and $(\mj_{h,i}^{\delta})^{-1}$ represents a subdomain  solver that is an incomplete LU factorization in this work. 

\section{Numerical experiments and observations}
In this section, we discuss some results obtained by applying the algorithms developed in the previous sections to the full pulmonary artery. The geometry of the interior of the artery is obtained from the segmentation of  a contrast-enhanced CT image of a healthy 19 year-old subject, and the arterial wall is added  to the resulting artery manually. The wall thickness is assumed to be $1~mm$ everywhere. Note that because of the lack of imaging data for the arterial wall, the artificially added wall may not be physiologically correct, however, the algorithmic framework introduced in this study can be easily applied  when the correct wall geometry becomes available.  

The main emphasis of the section is the parallel performance of the algorithm which is one of the key factors for obtaining high resolution simulations of patient-specific pulmonary arteries in a reasonably amount of time. 
For convenience, we introduce   some  notations and default parameters used in the following study. ``NI" is used to represent the number of Newton iterations per time step, ``LI" denotes  the averaged  number of GMRES(fGMRES) iterations pert Newton step, ``T" is the total compute time in second for all 10 time steps, ``MEM" in megabytes (MB) is the estimated memory usage per processor core, and ``efficiency" is the parallel efficiency of the proposed algorithm.   ILU(1) is adopted as the subdomain solver, the subdomain overlapping size is 1, and the relative tolerances for Newton and GMRES are $10^{-6}$ and $10^{-3}$ respectively.  These parameters are used through the whole performance study, unless otherwise specified. The proposed algorithm is implemented based on PETSc \cite{petsc-user-ref}.

\subsection{Simulation results  and discussions}
\begin{figure}
   \centering
    \includegraphics[width=2.6in]{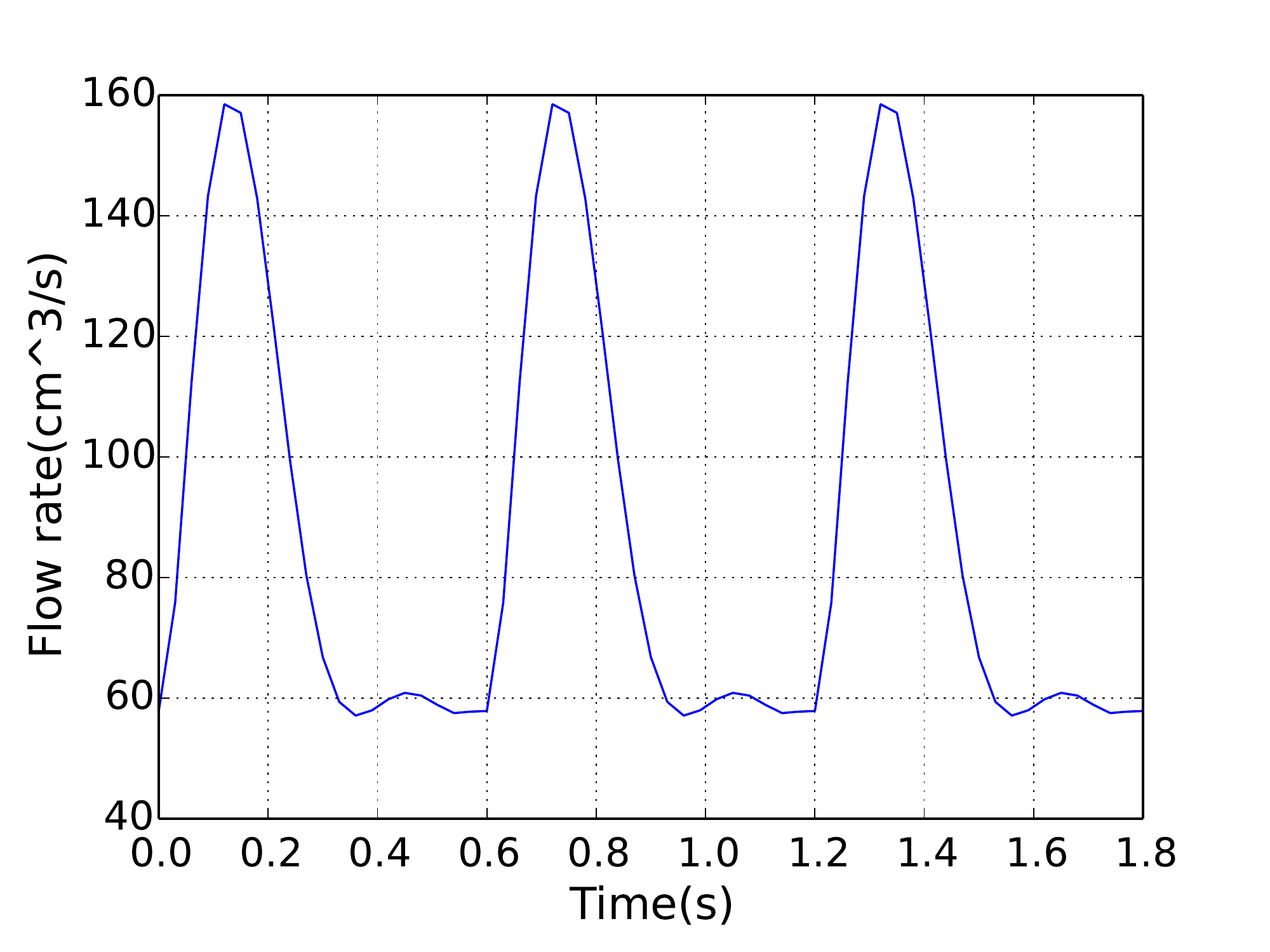}  
    \includegraphics[width=2.6in]{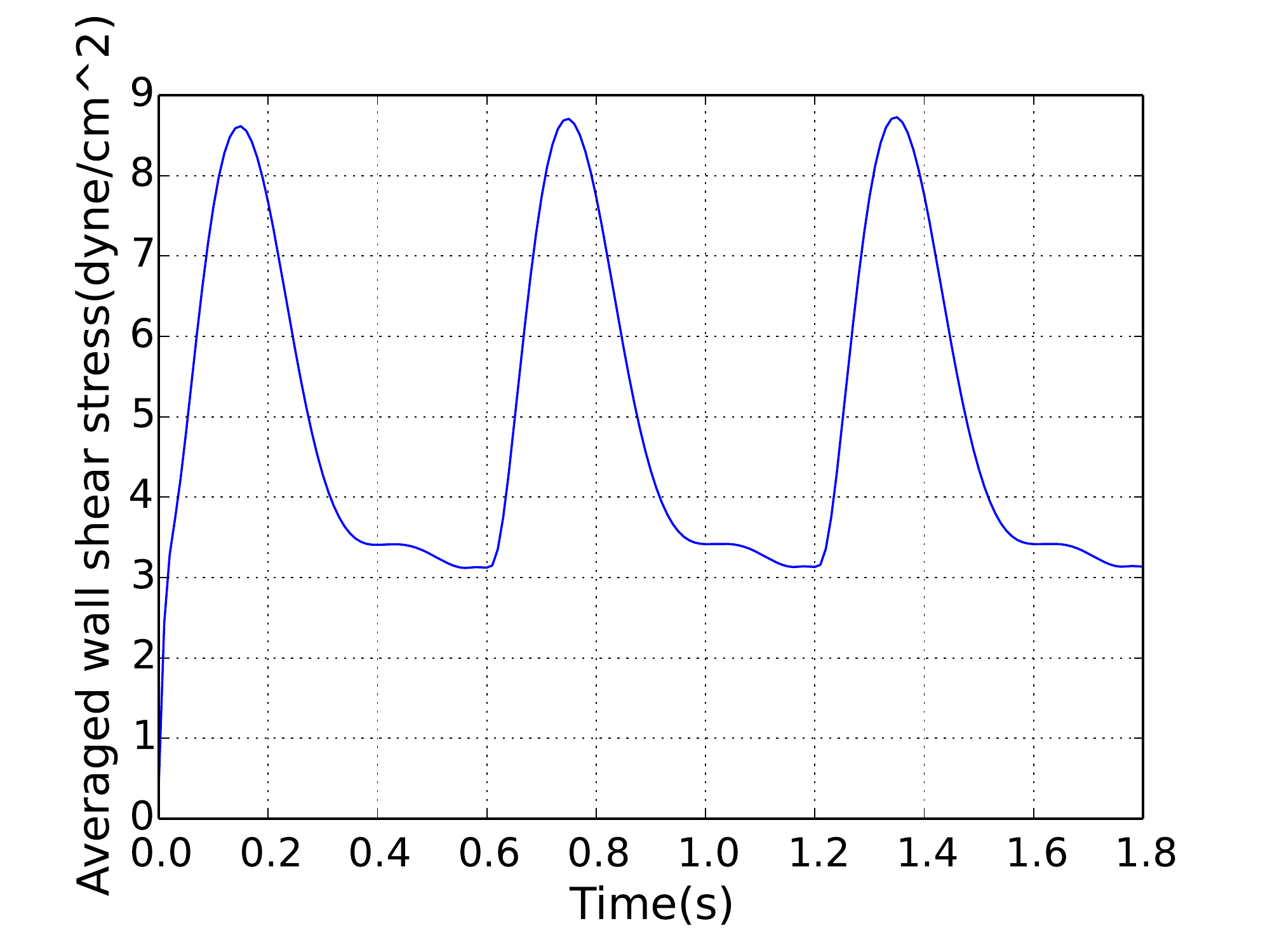}  
   \caption{Left: inflow rate for  three cardiac cycles, $[0,1.8]$, with 0.6 second per cycle.  Right:  spatially averaged  wall sheer stress. } \label{heartflow_rate}
\end{figure}

\begin{figure}
   \centering
    \includegraphics[width=2.6in]{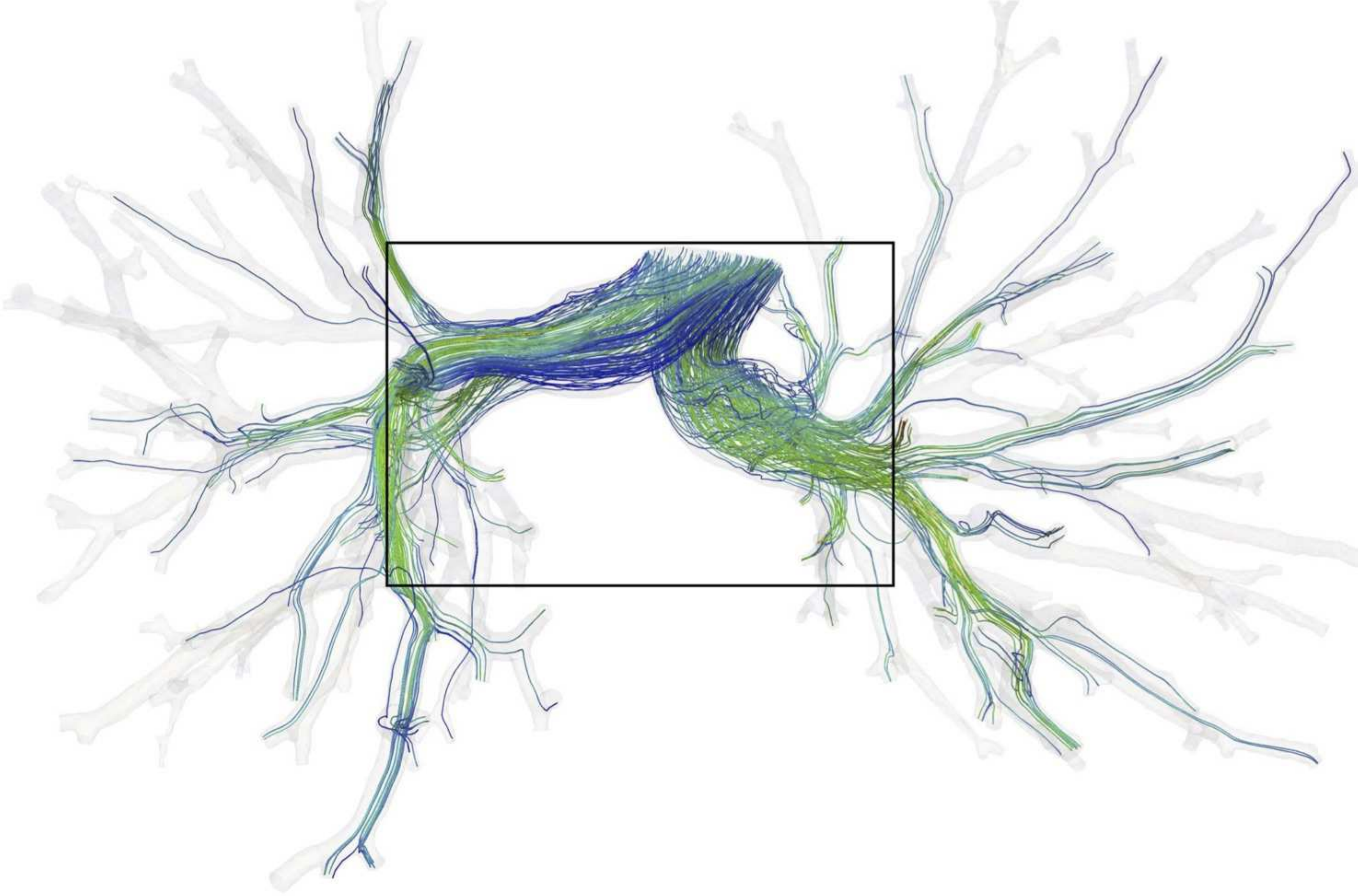}  
    \includegraphics[width=2.6in]{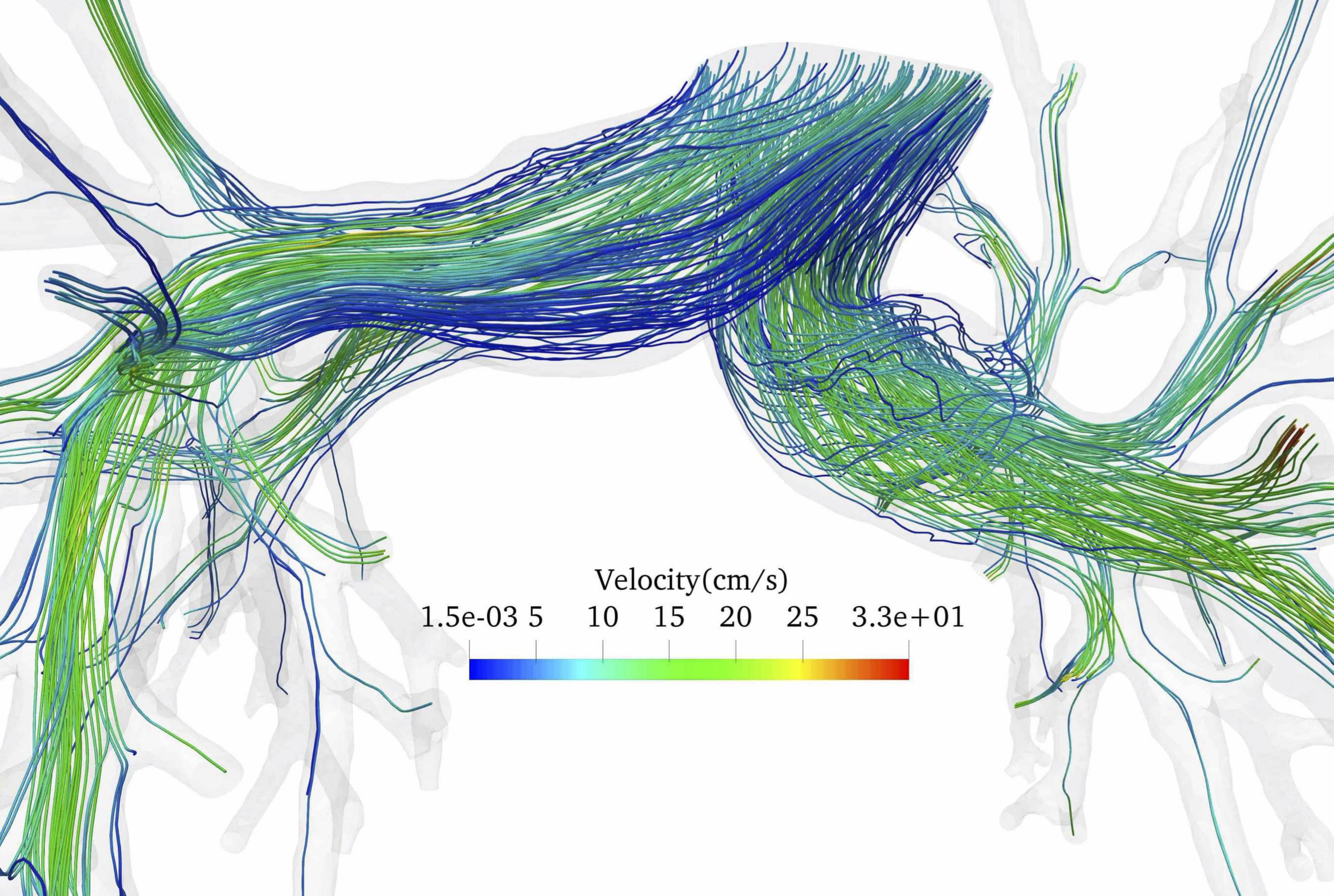} \\
    \includegraphics[width=2.6in]{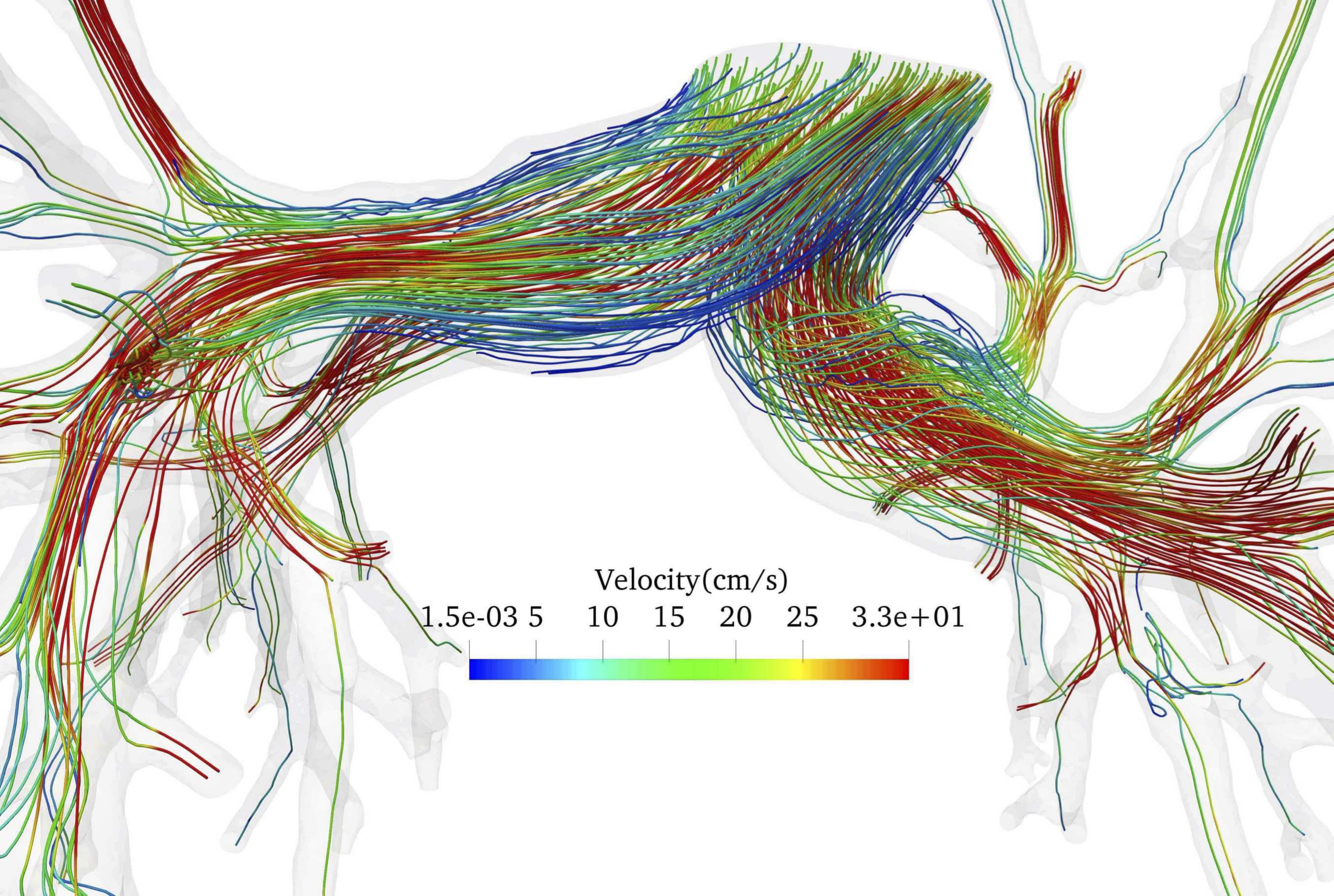}
    \includegraphics[width=2.6in]{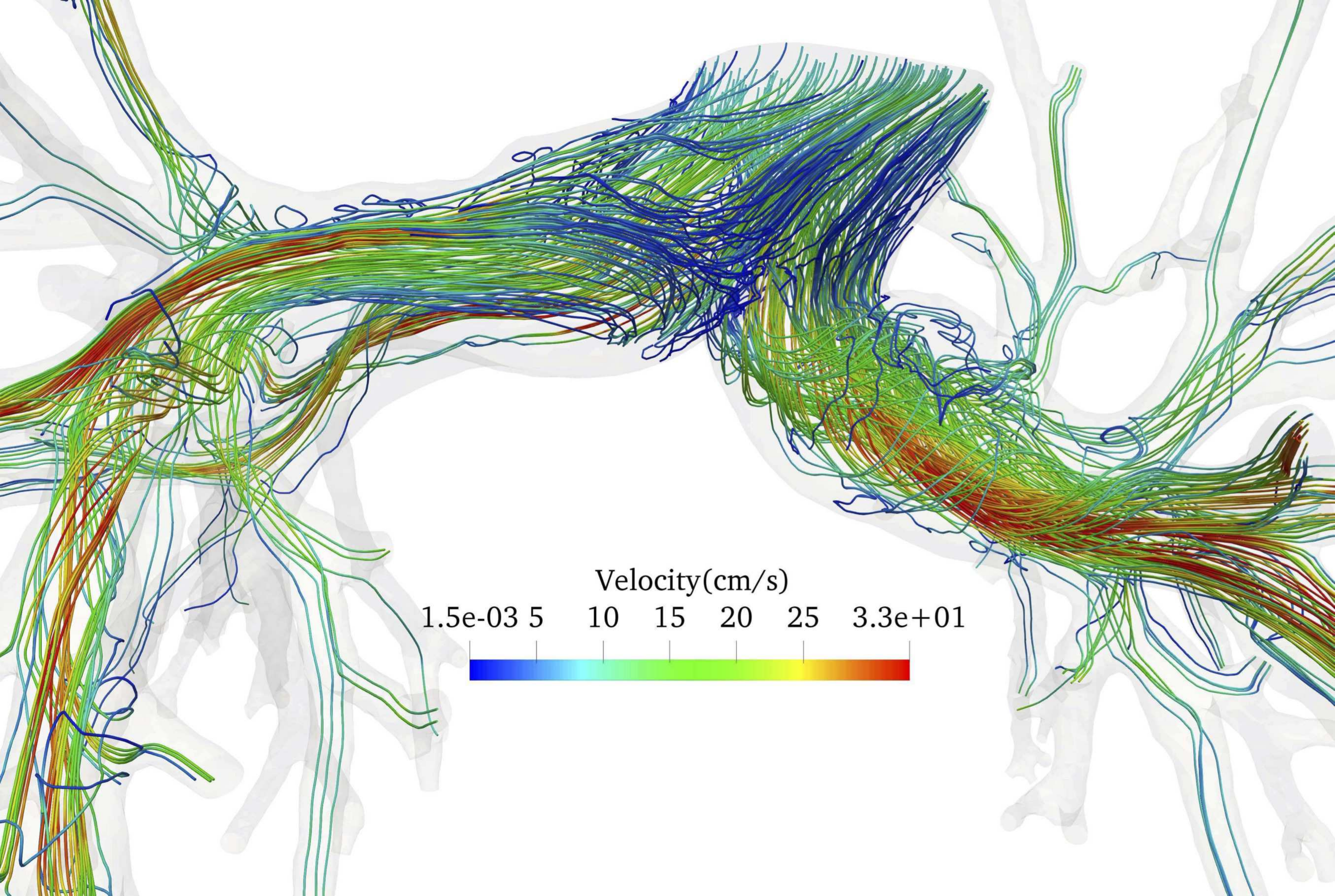}\\
    \includegraphics[width=2.6in]{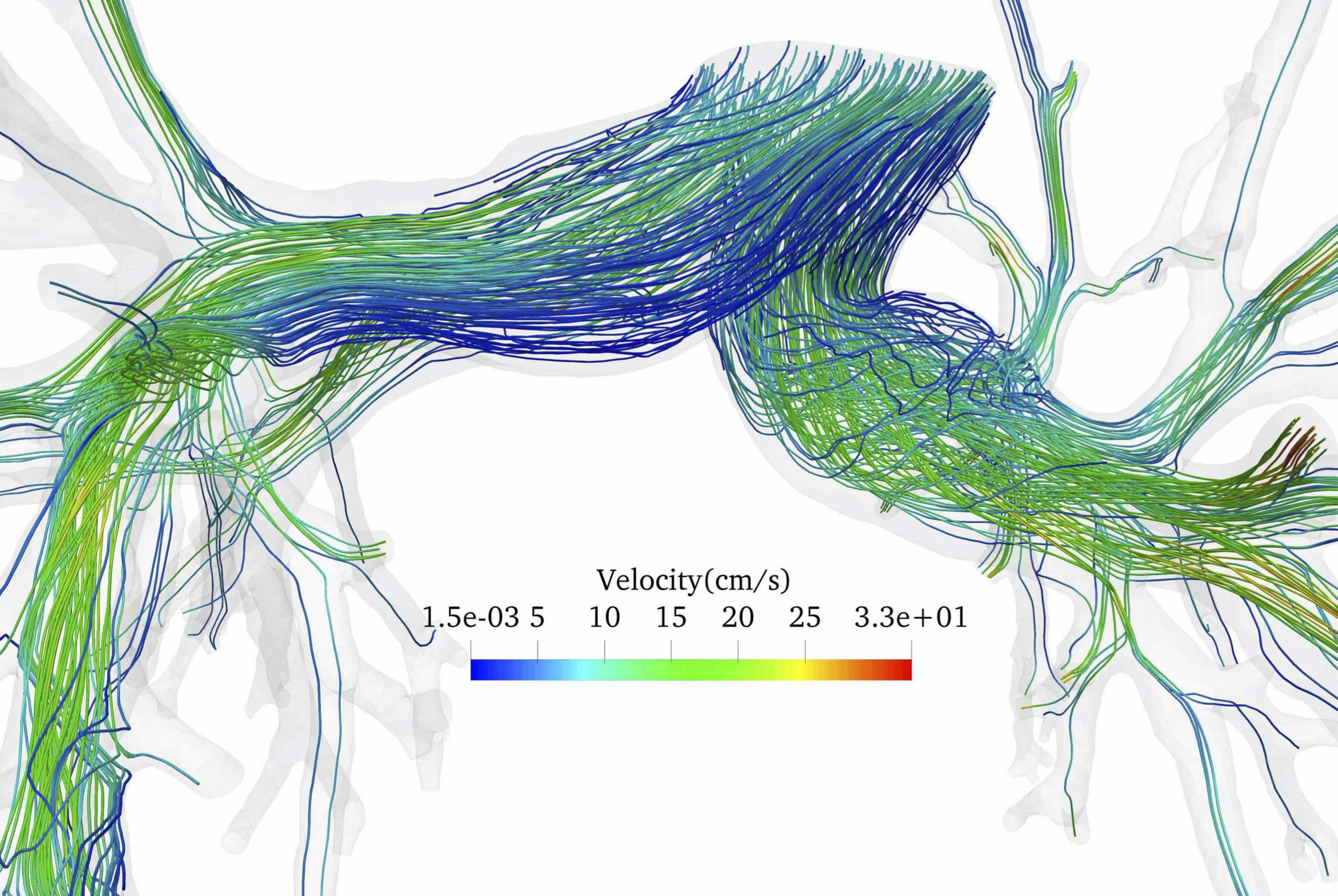}
    \includegraphics[width=2.6in]{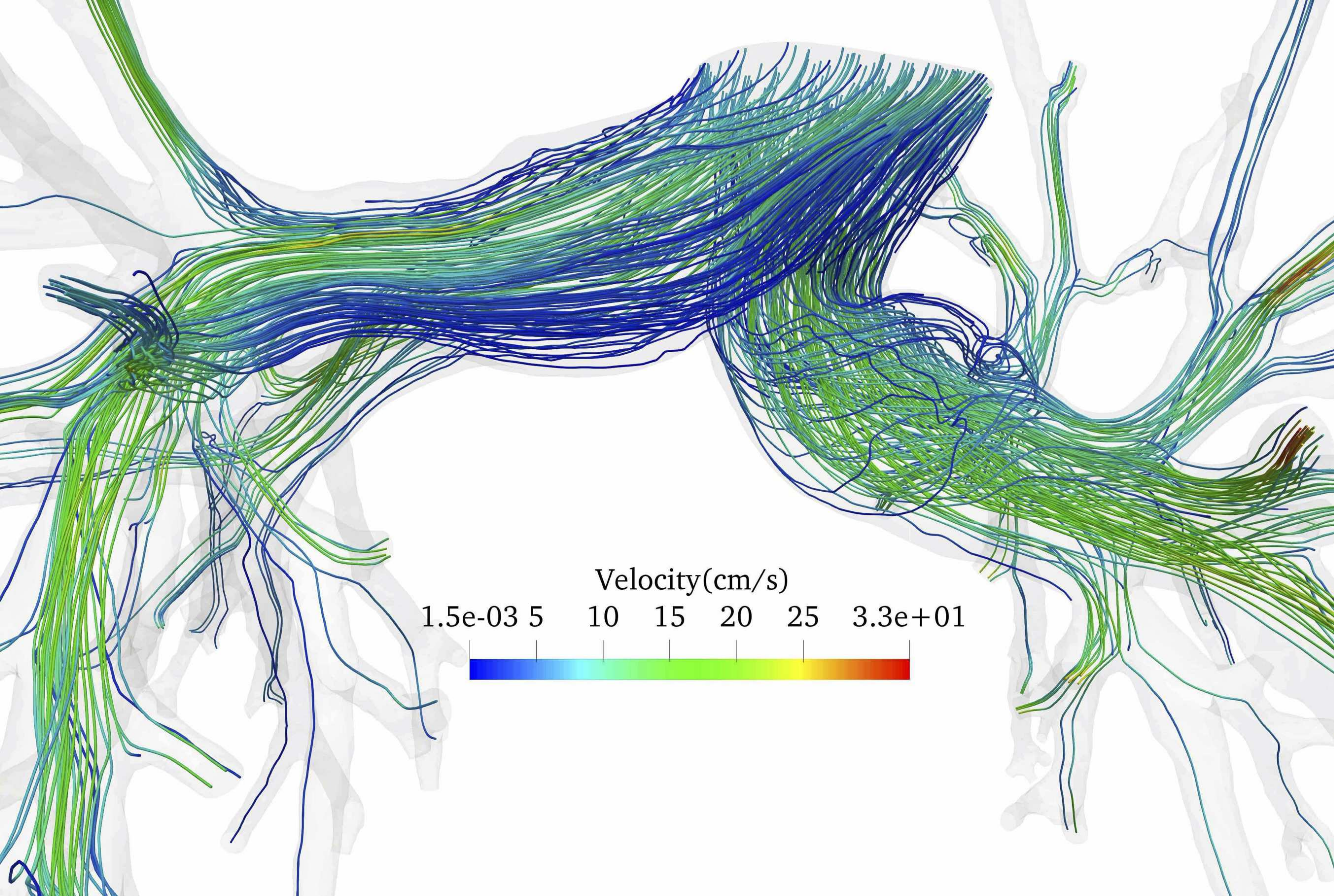}
   \caption{Streamlines of the flow field at $1.2s$ (top right), $1.35s$ (middle left), $1.5s$ (middle right), $1.65s$ (bottom left) and $1.8s$ (bottom right). Top left: the entire blood flow domain; all others: zoom-in pictures for streamlines } \label{streamline}
\end{figure}

\begin{figure}
   \centering
    \includegraphics[width=2.6in]{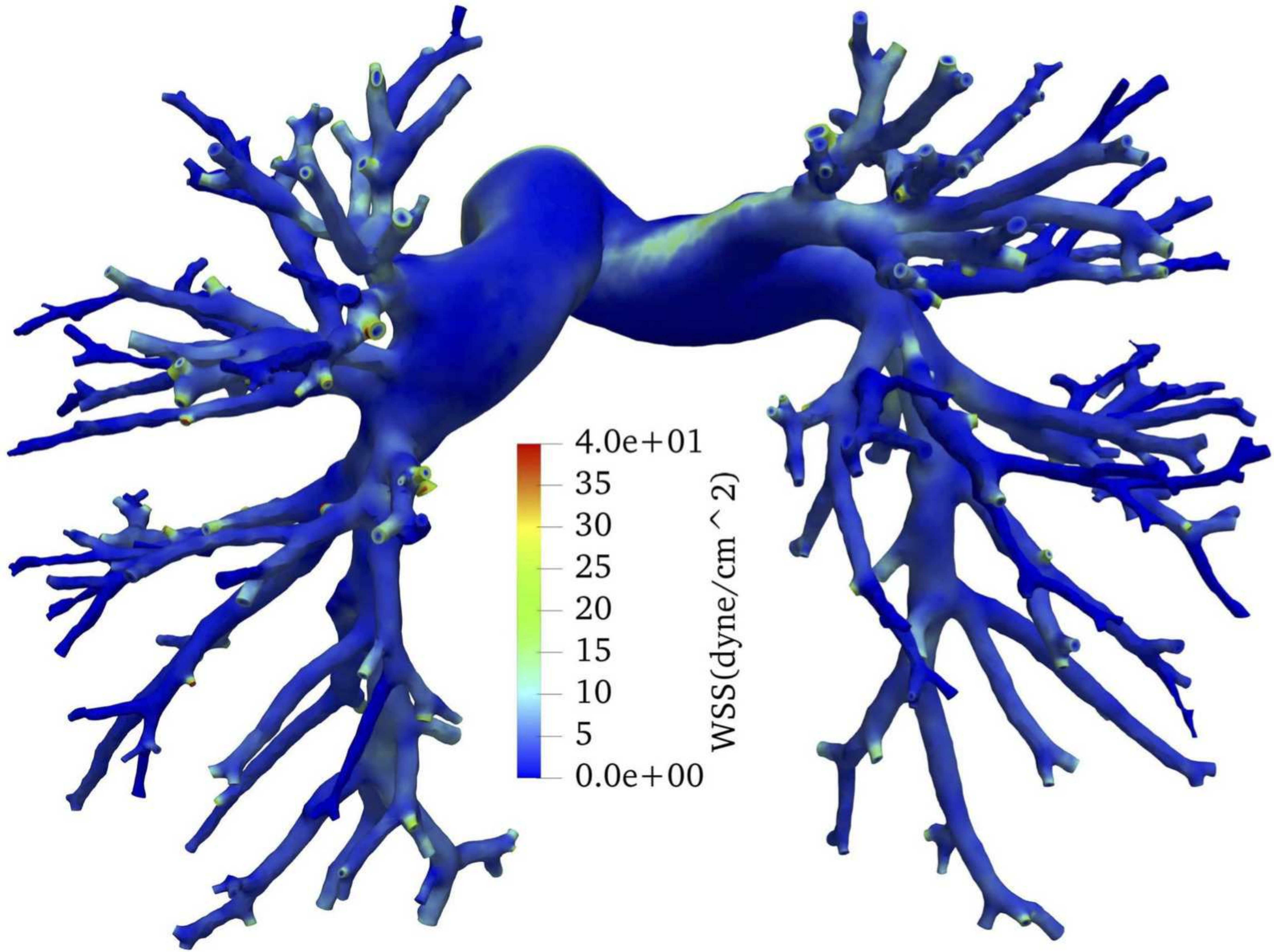}  
    \includegraphics[width=2.6in]{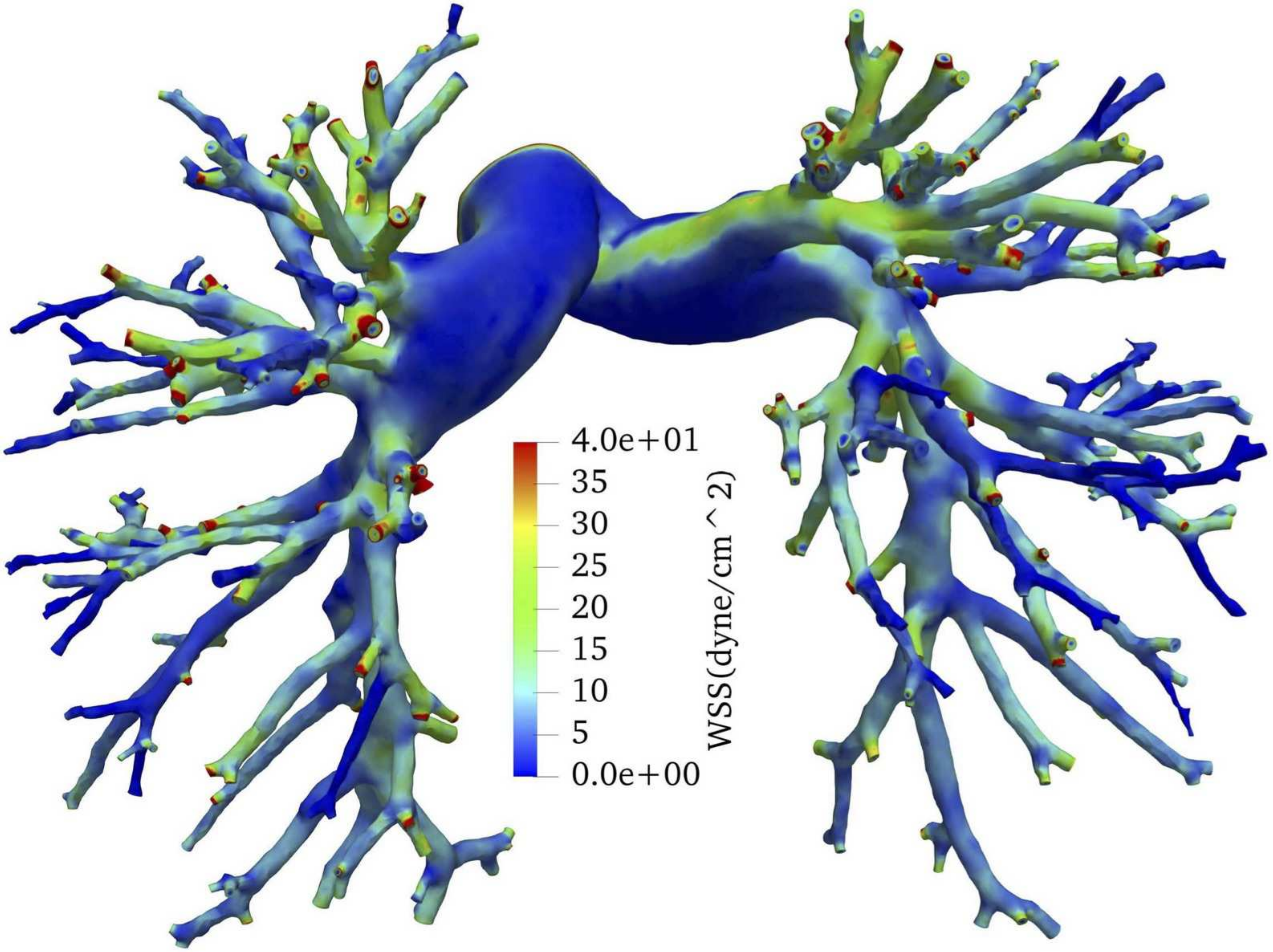} \\
    \includegraphics[width=2.6in]{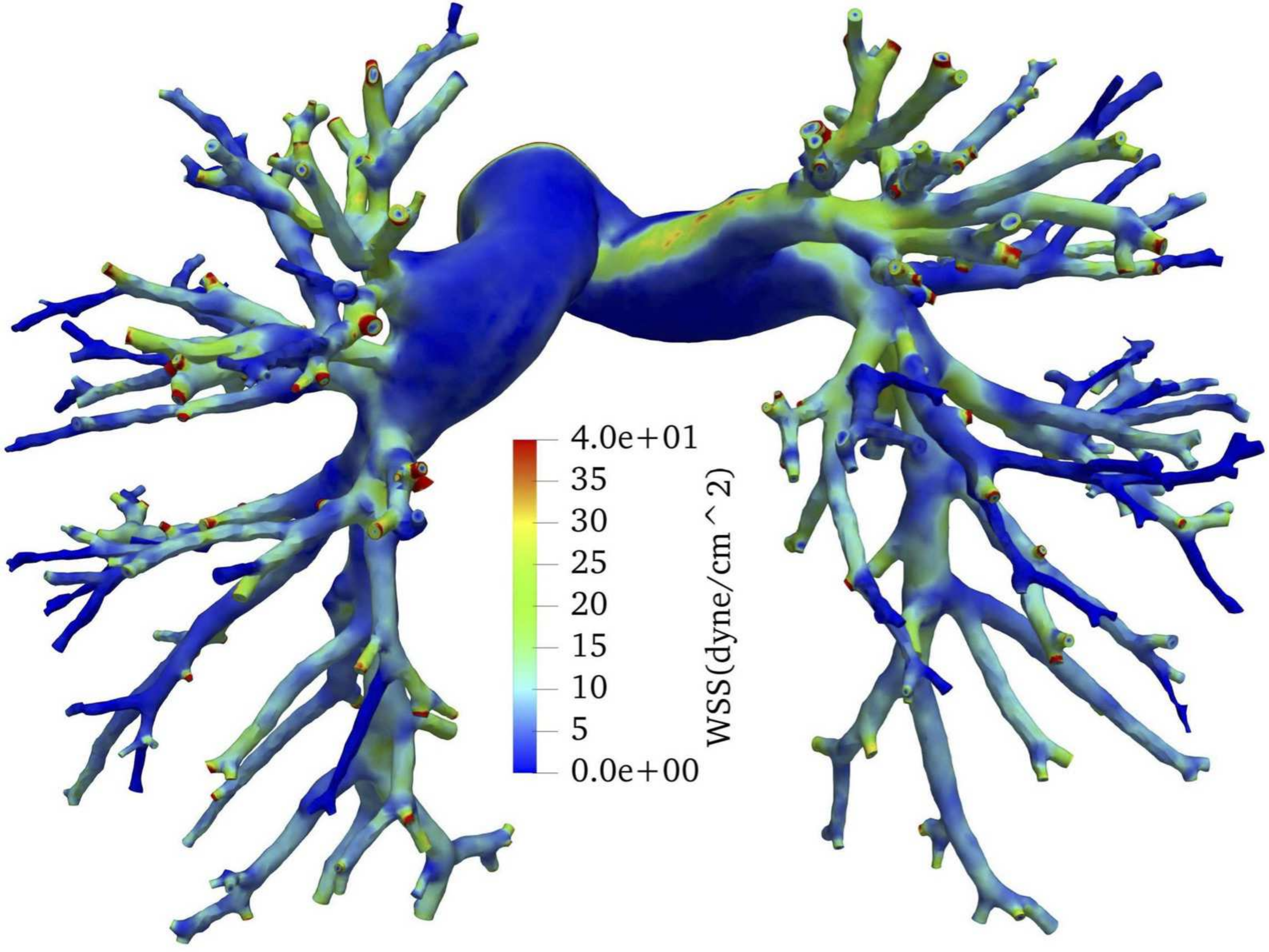}
    \includegraphics[width=2.6in]{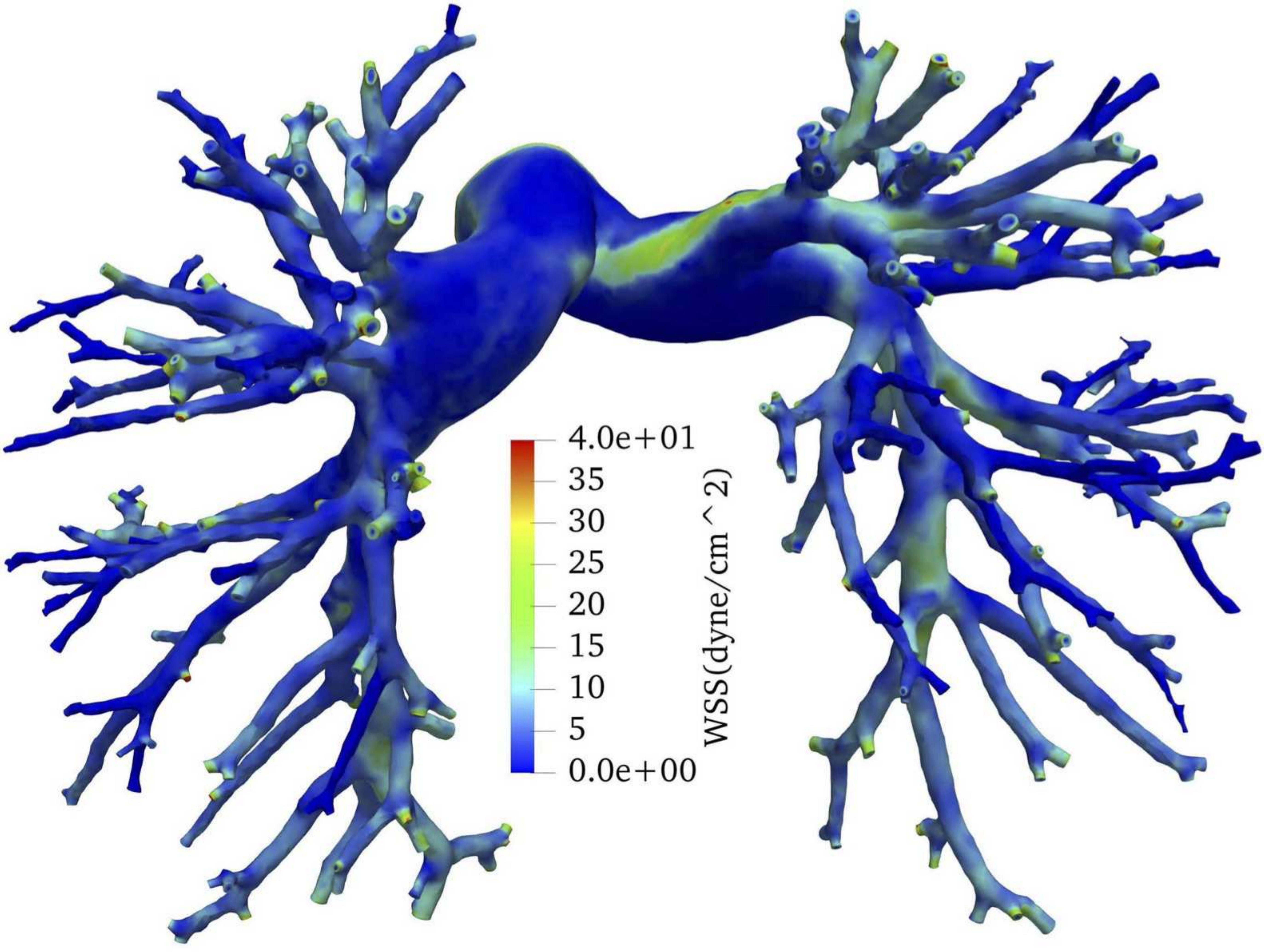}\\
    \includegraphics[width=2.6in]{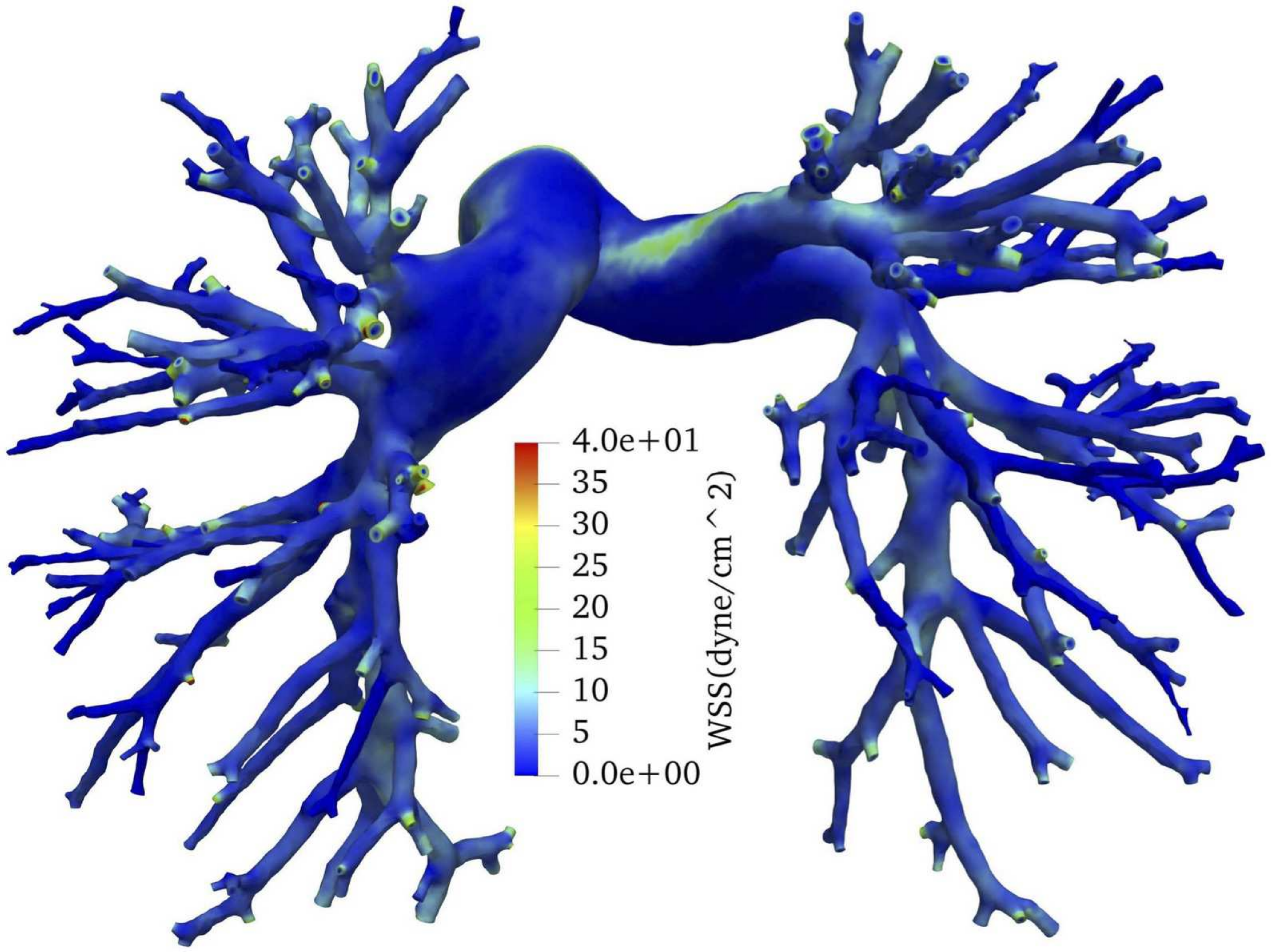}
    \includegraphics[width=2.6in]{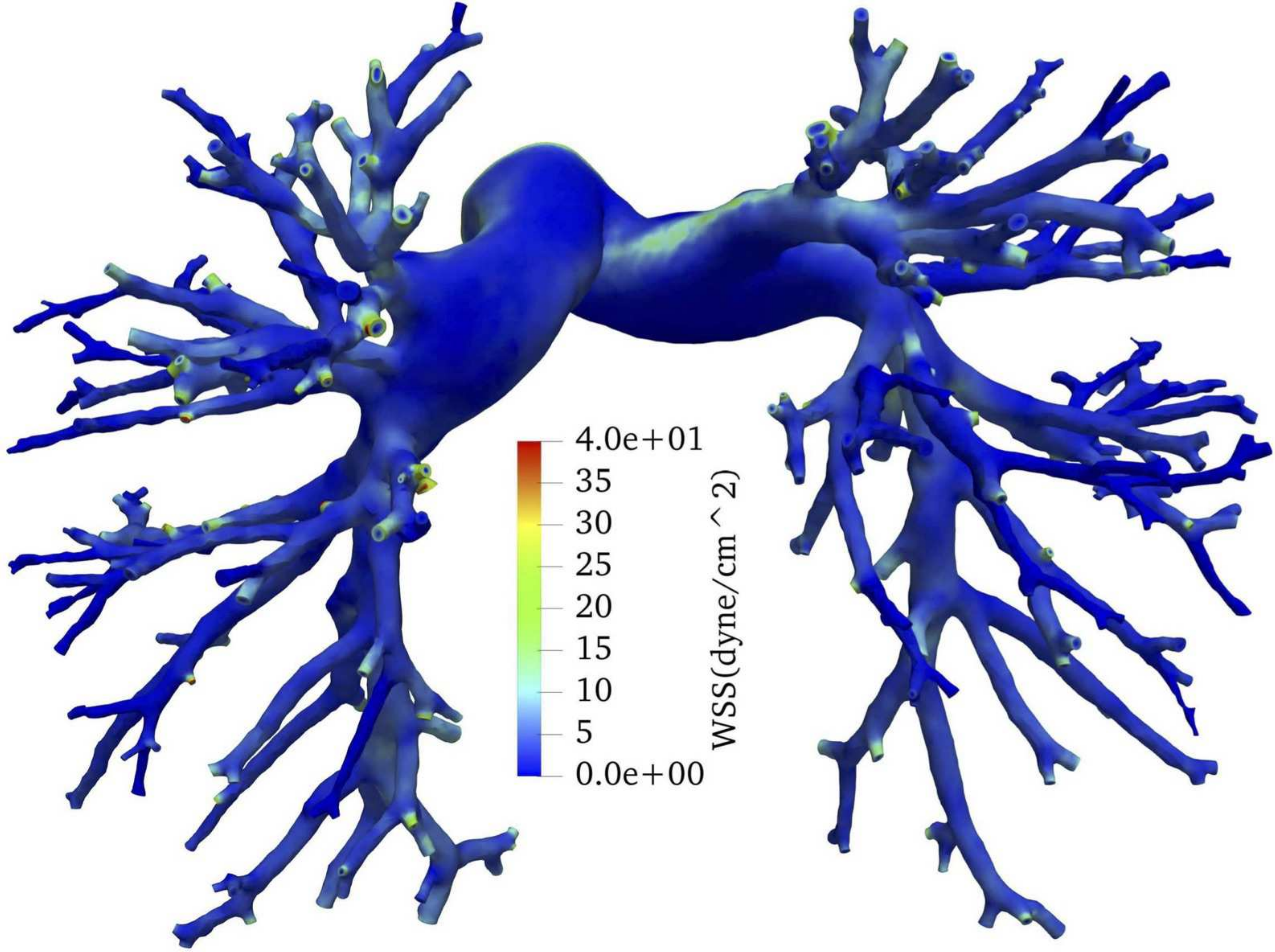}
   \caption{Wall shear stress at $1.2s$ (top left), $1.35s$ (top right), $1.4s$ (middle left), $1.5s$ (middle right), $1.65s$ (bottom left) and $1.8s$ (bottom right).  } \label{WSS}
\end{figure}

\begin{figure}
   \centering
    \includegraphics[width=2.6in]{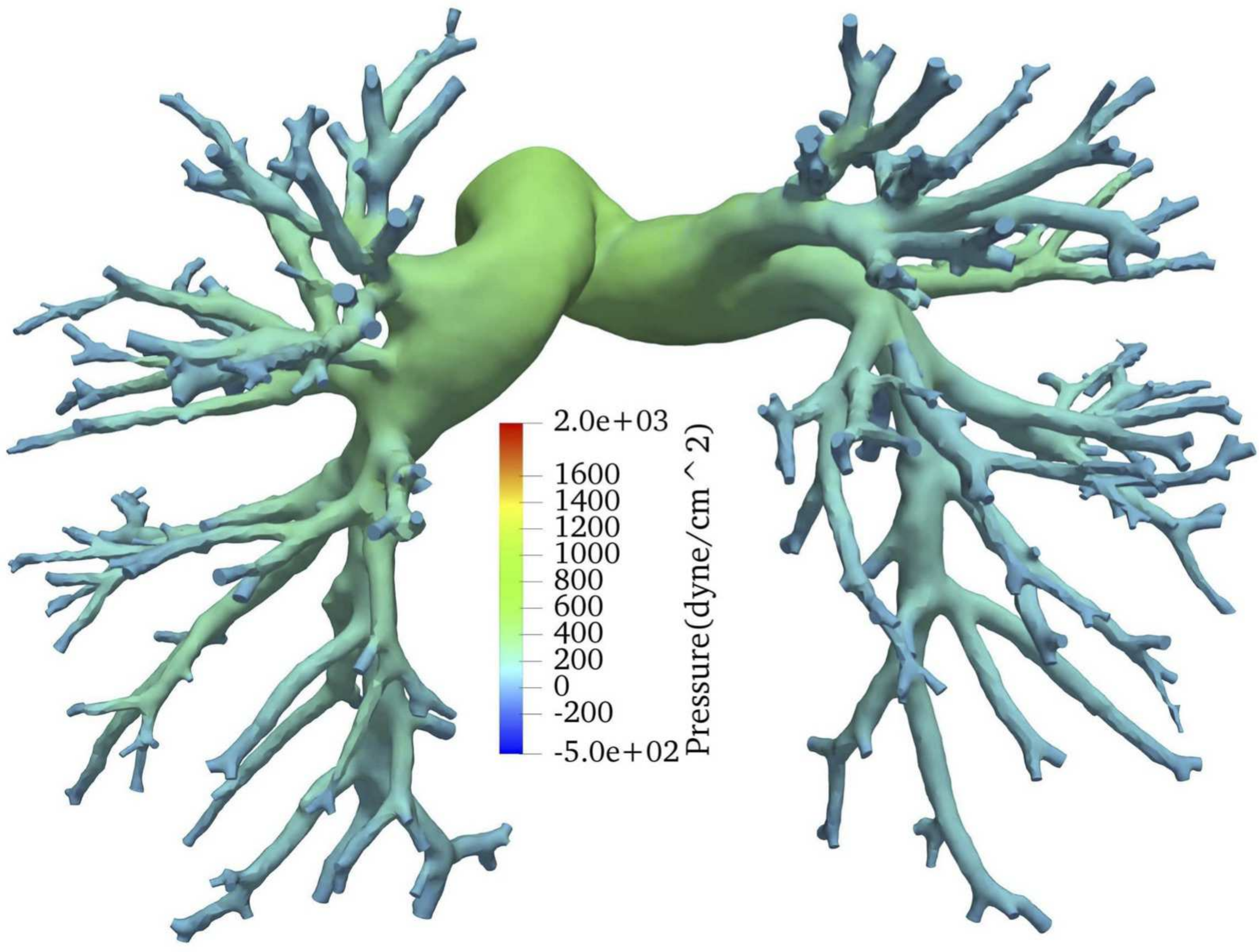}  
    \includegraphics[width=2.6in]{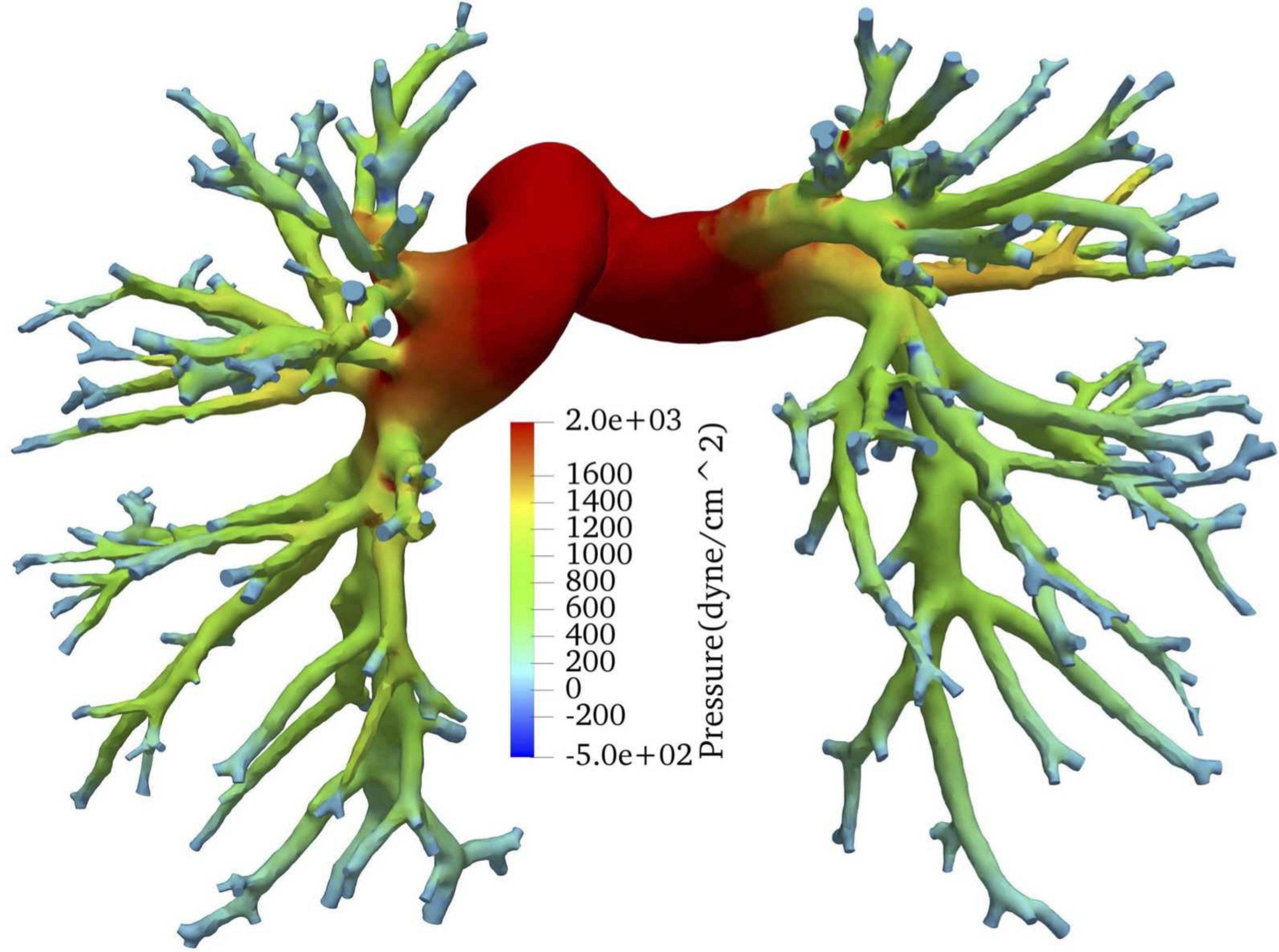} \\
    \includegraphics[width=2.6in]{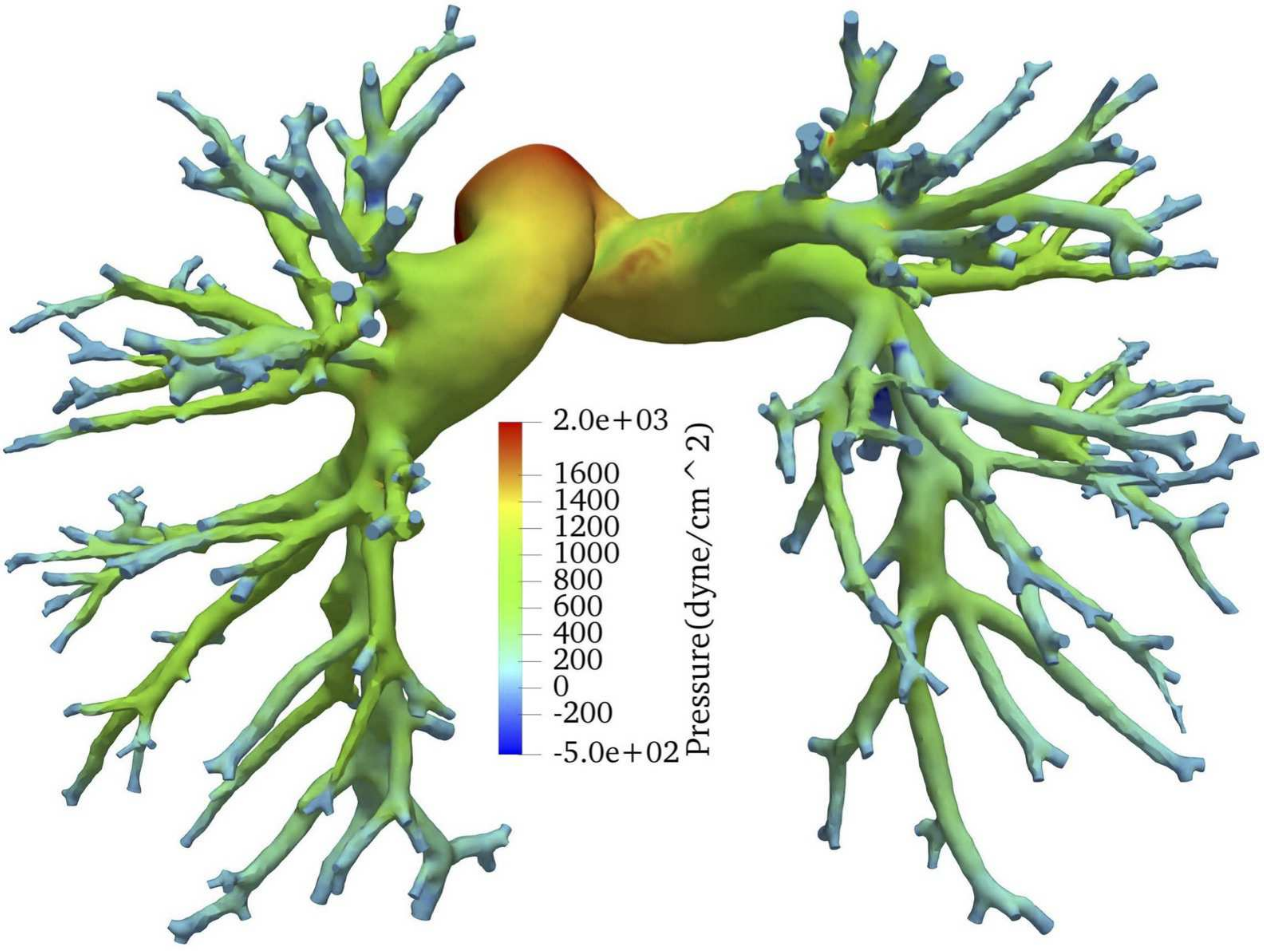}
    \includegraphics[width=2.6in]{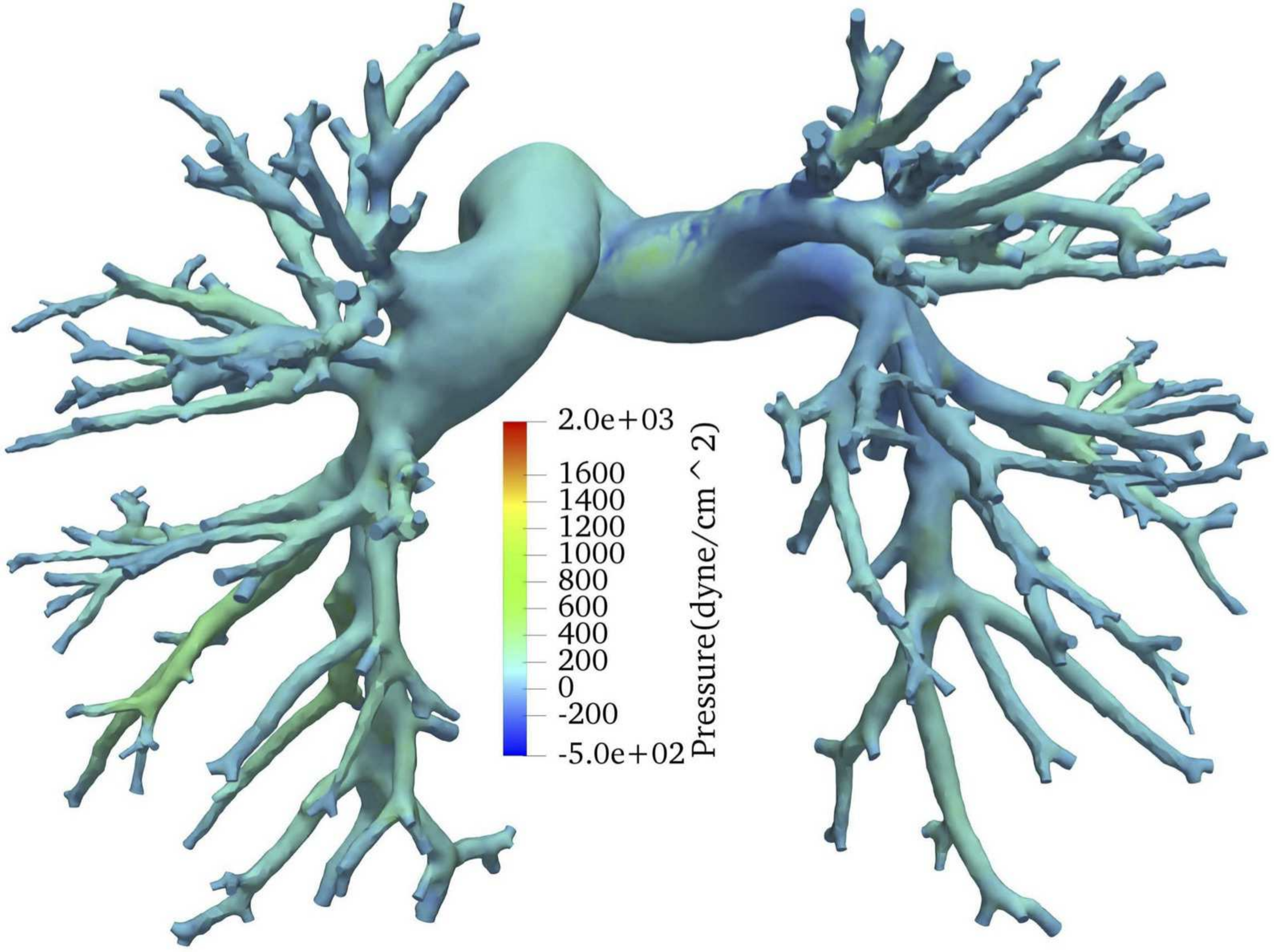}\\
    \includegraphics[width=2.6in]{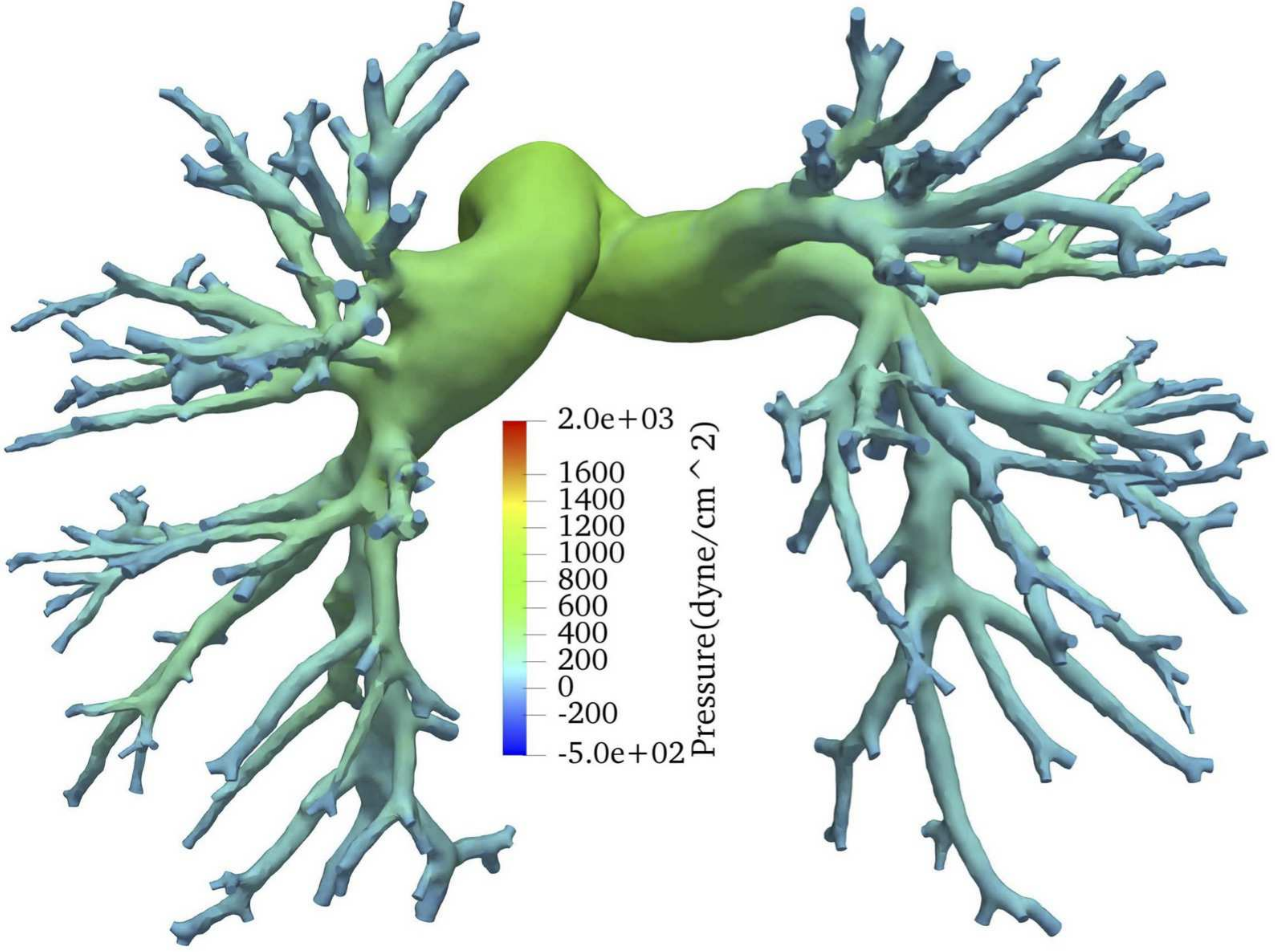}
    \includegraphics[width=2.6in]{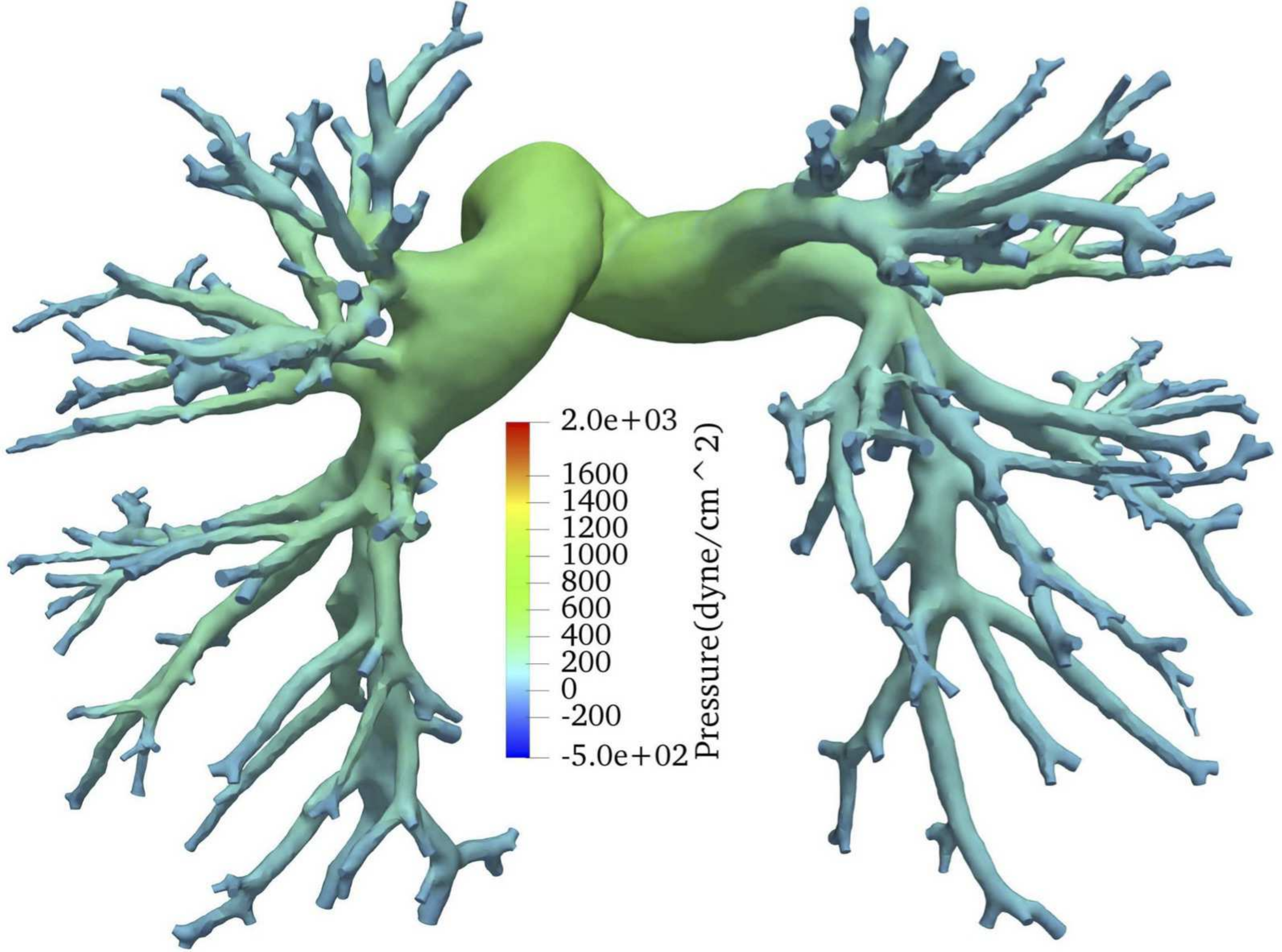}
   \caption{Blood flow pressure at $1.2s$ (top left), $1.35s$ (top right), $1.4s$ (middle left), $1.5s$ (middle right), $1.65s$ (bottom left) and $1.8s$ (bottom right).  } \label{bloodflow_pressure}
\end{figure}

\begin{figure}
   \centering
    \includegraphics[width=2.6in]{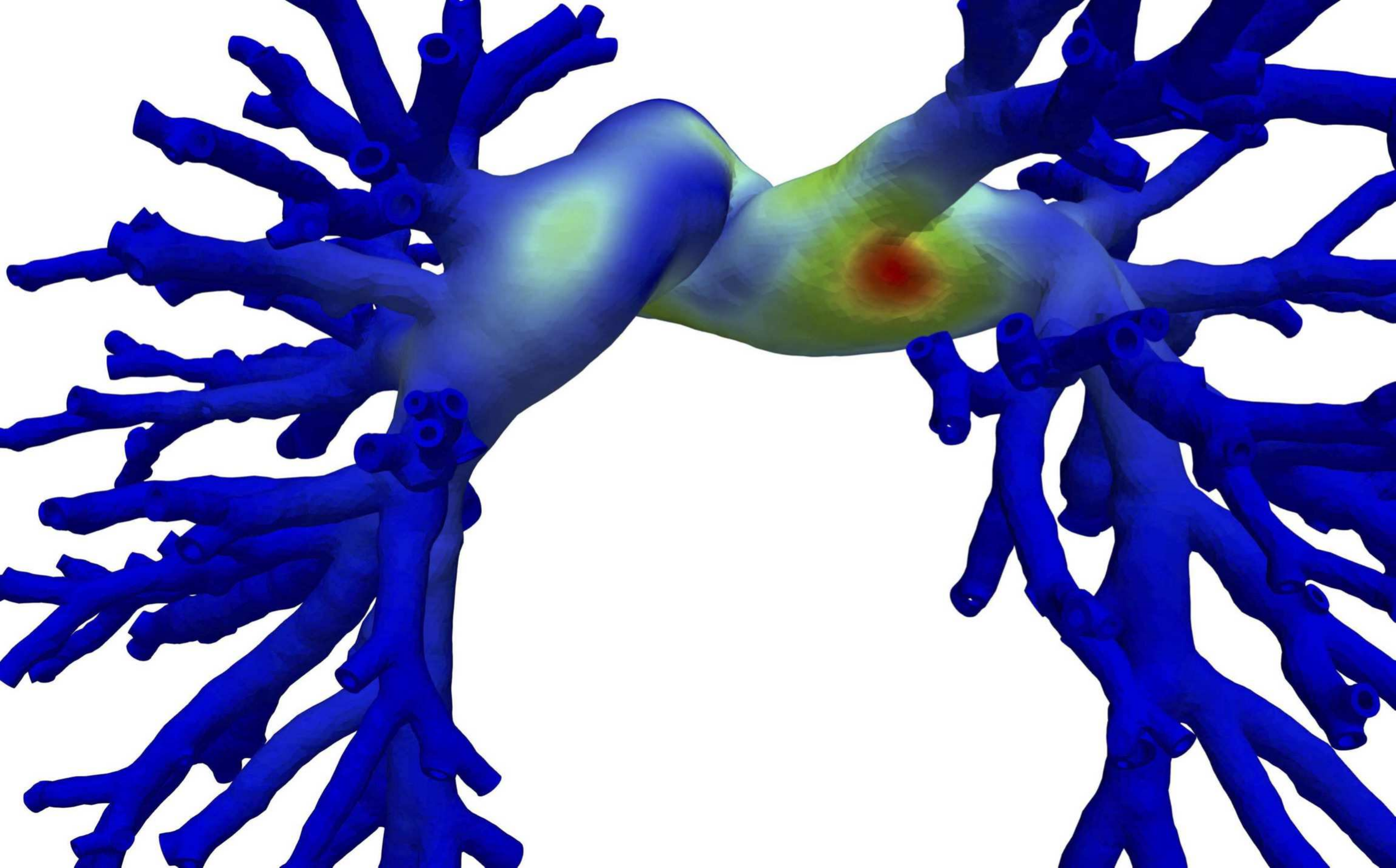}  
    \includegraphics[width=2.6in]{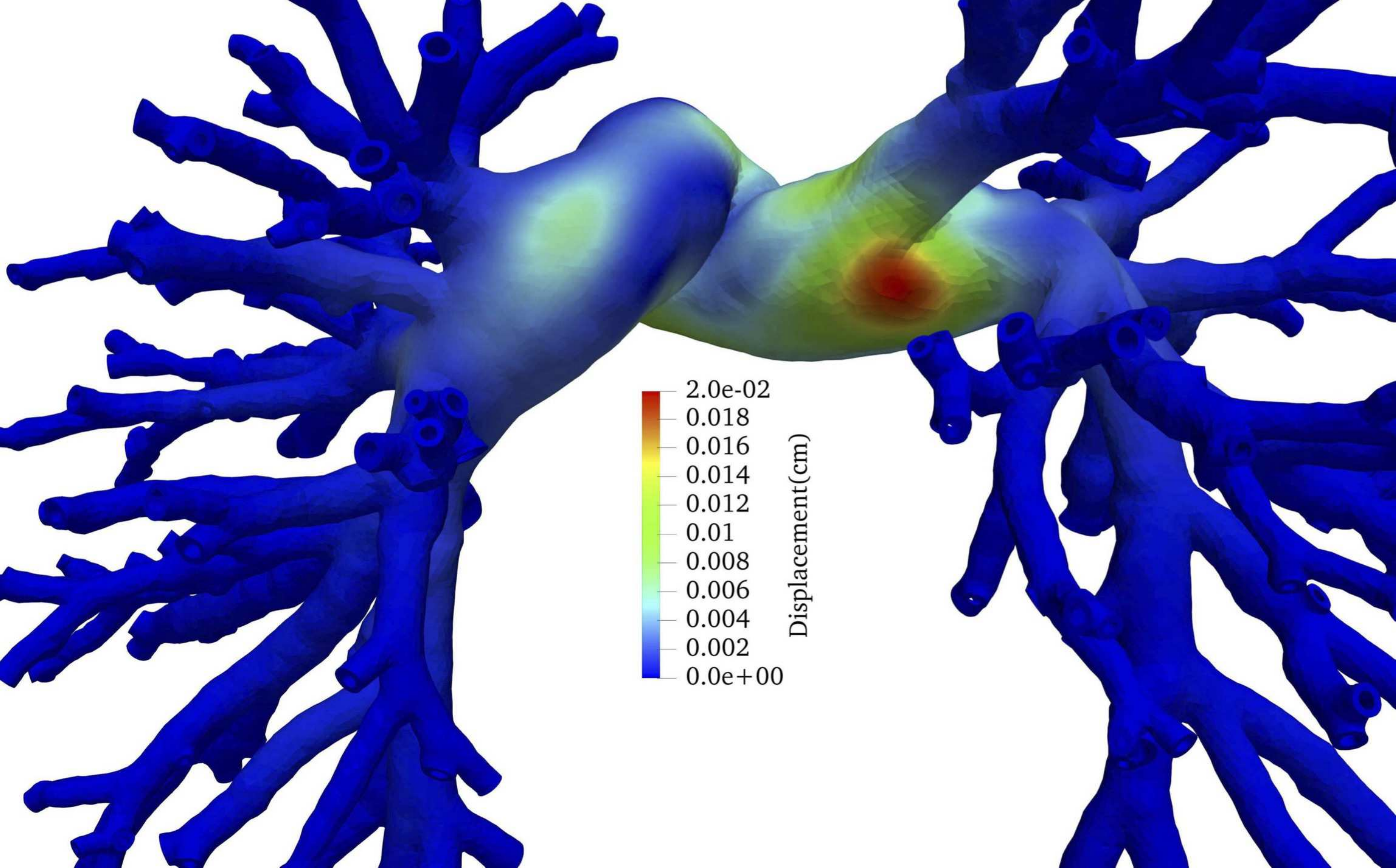} \\
    \includegraphics[width=2.6in]{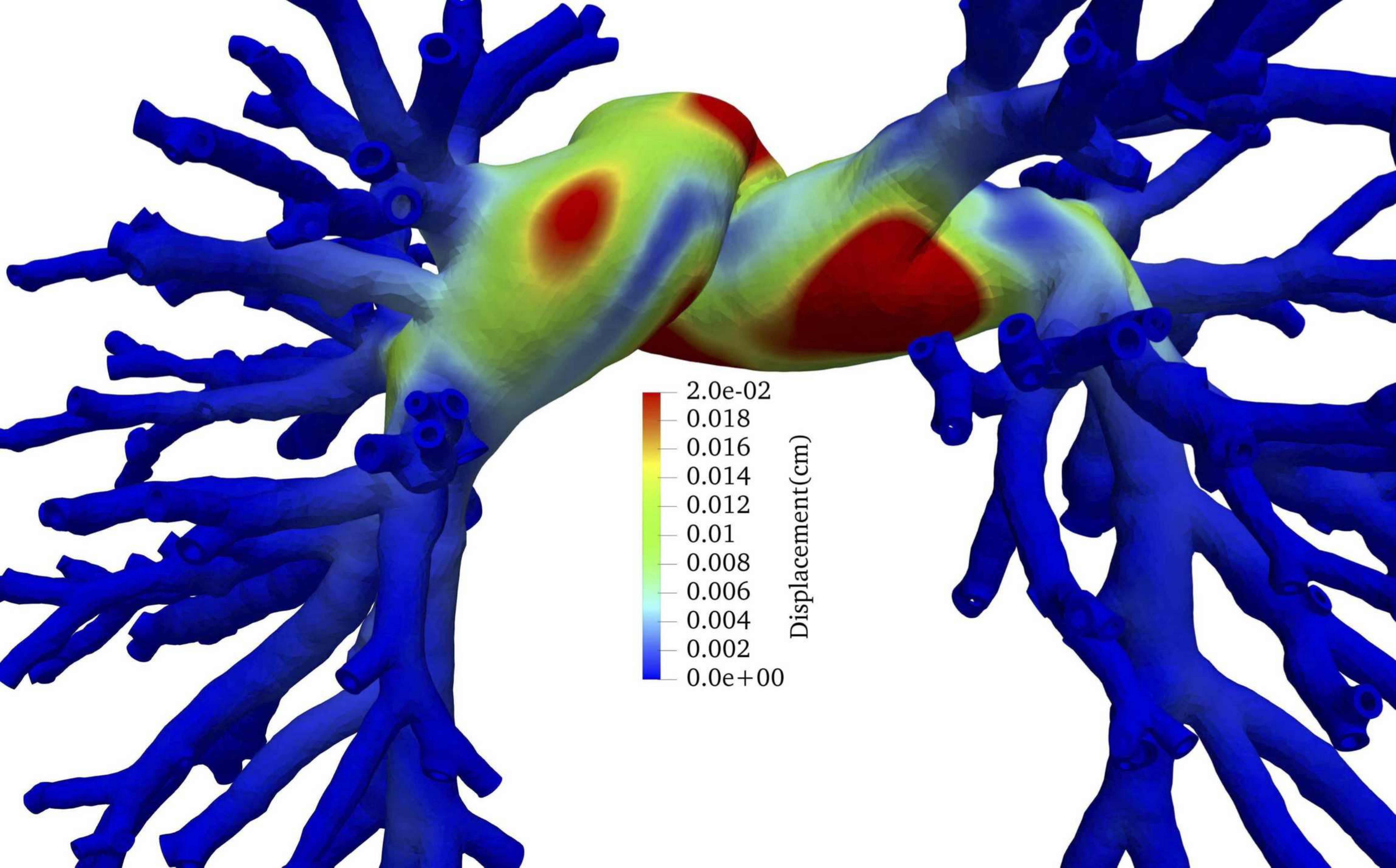}
    \includegraphics[width=2.6in]{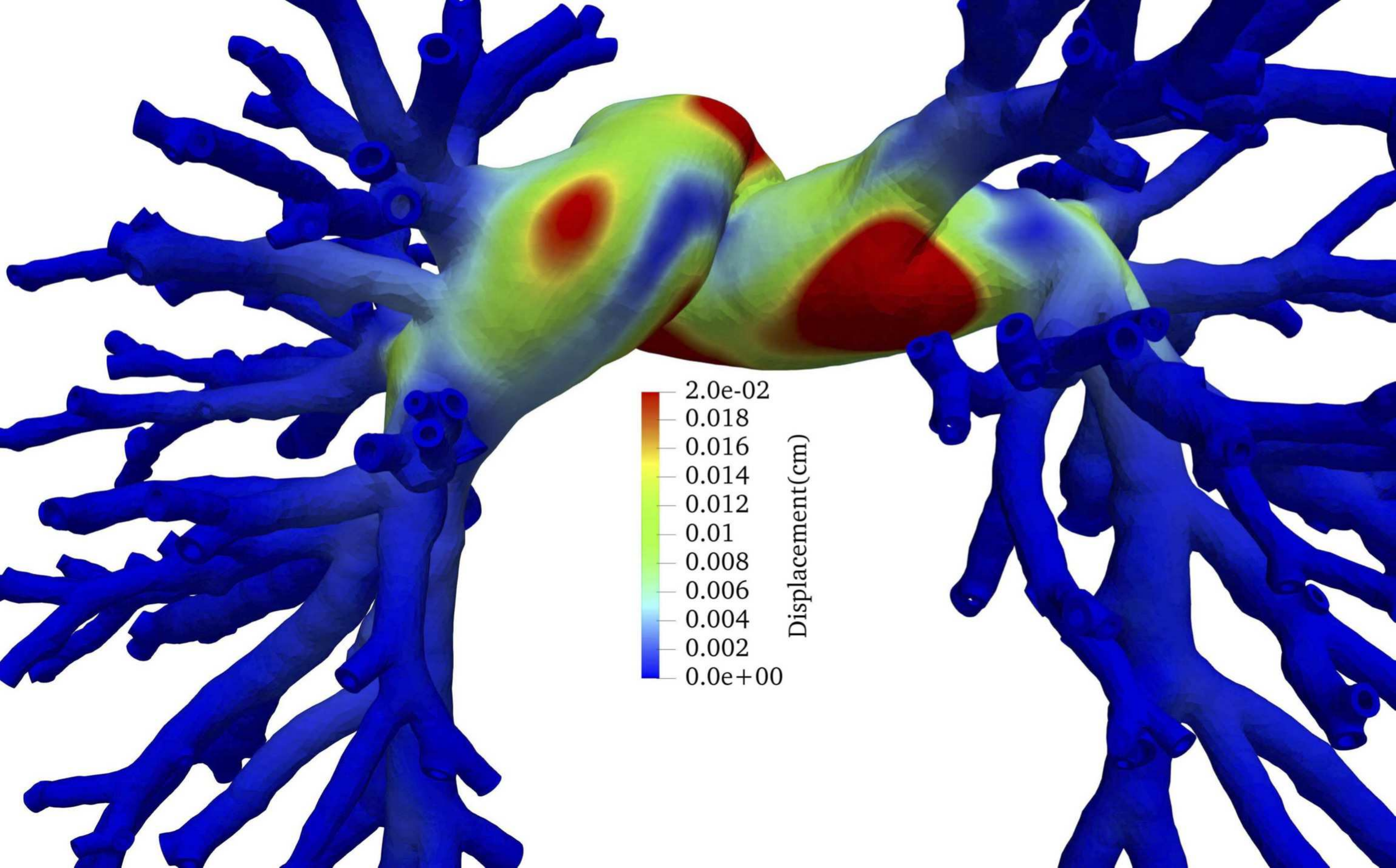}\\
    \includegraphics[width=2.6in]{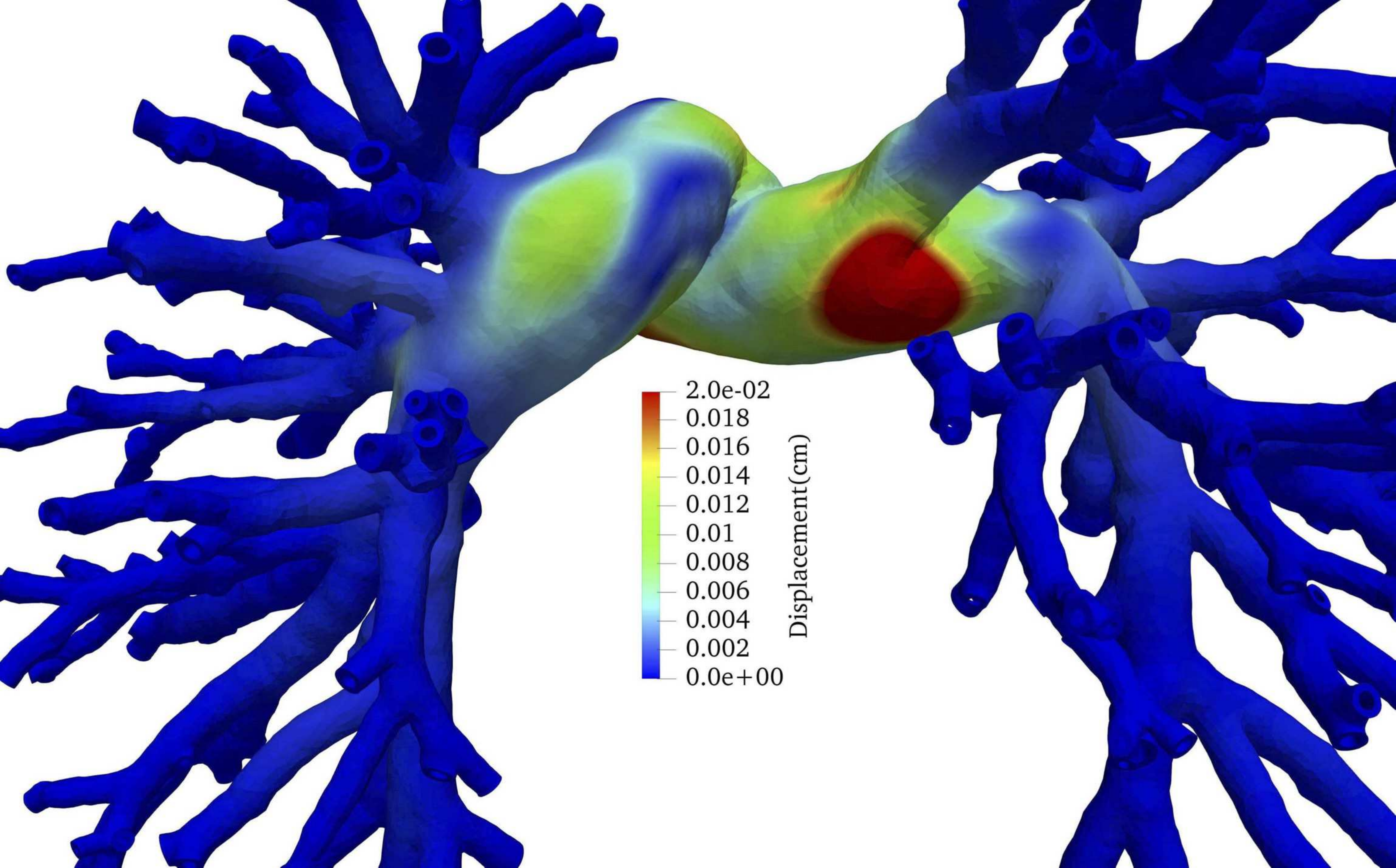}
    \includegraphics[width=2.6in]{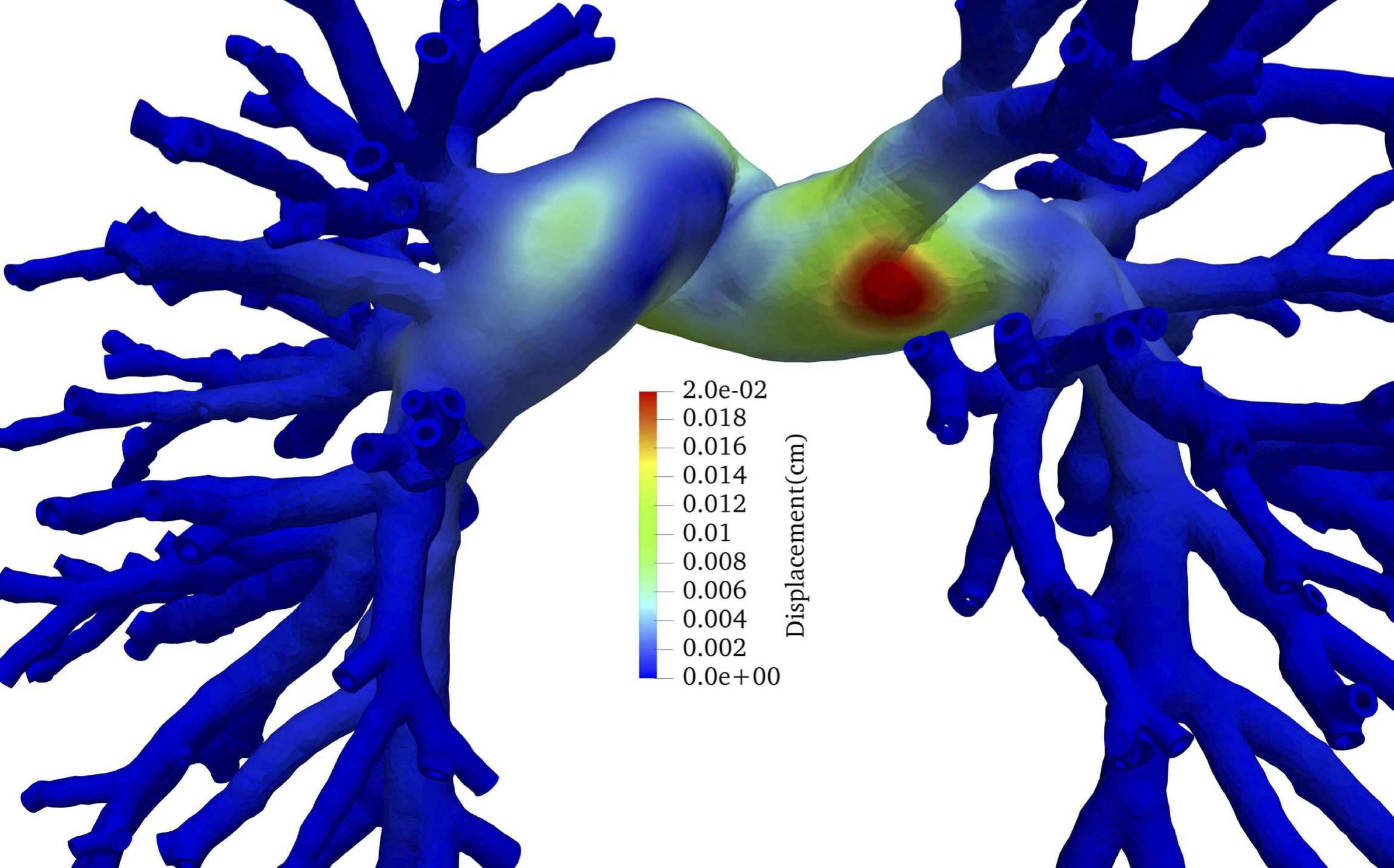}
   \caption{The displacement of the arterial wall at $1.2s$ (top right), $1.35s$ (middle left), $1.4s$ (middle right), $1.5s$ (bottom left) and $1.8s$ (bottom right). For visualization,  the displacement is  amplified by a factor of 20. The top left  picture represents the undeformed  domain.} \label{artery_displacement}
\end{figure}

\begin{figure}
   \centering
    \includegraphics[width=2.6in]{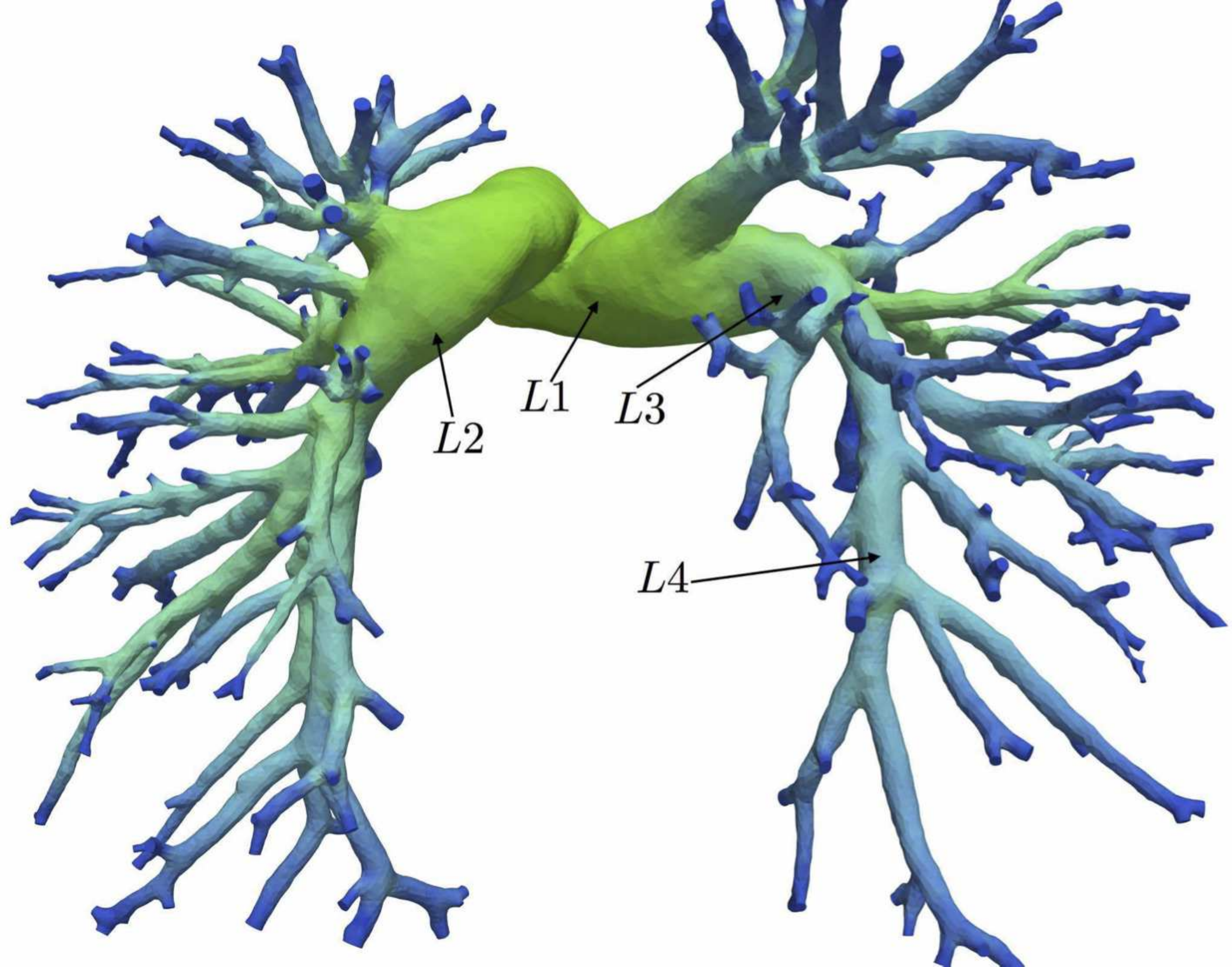}  \\
    \includegraphics[width=2.6in]{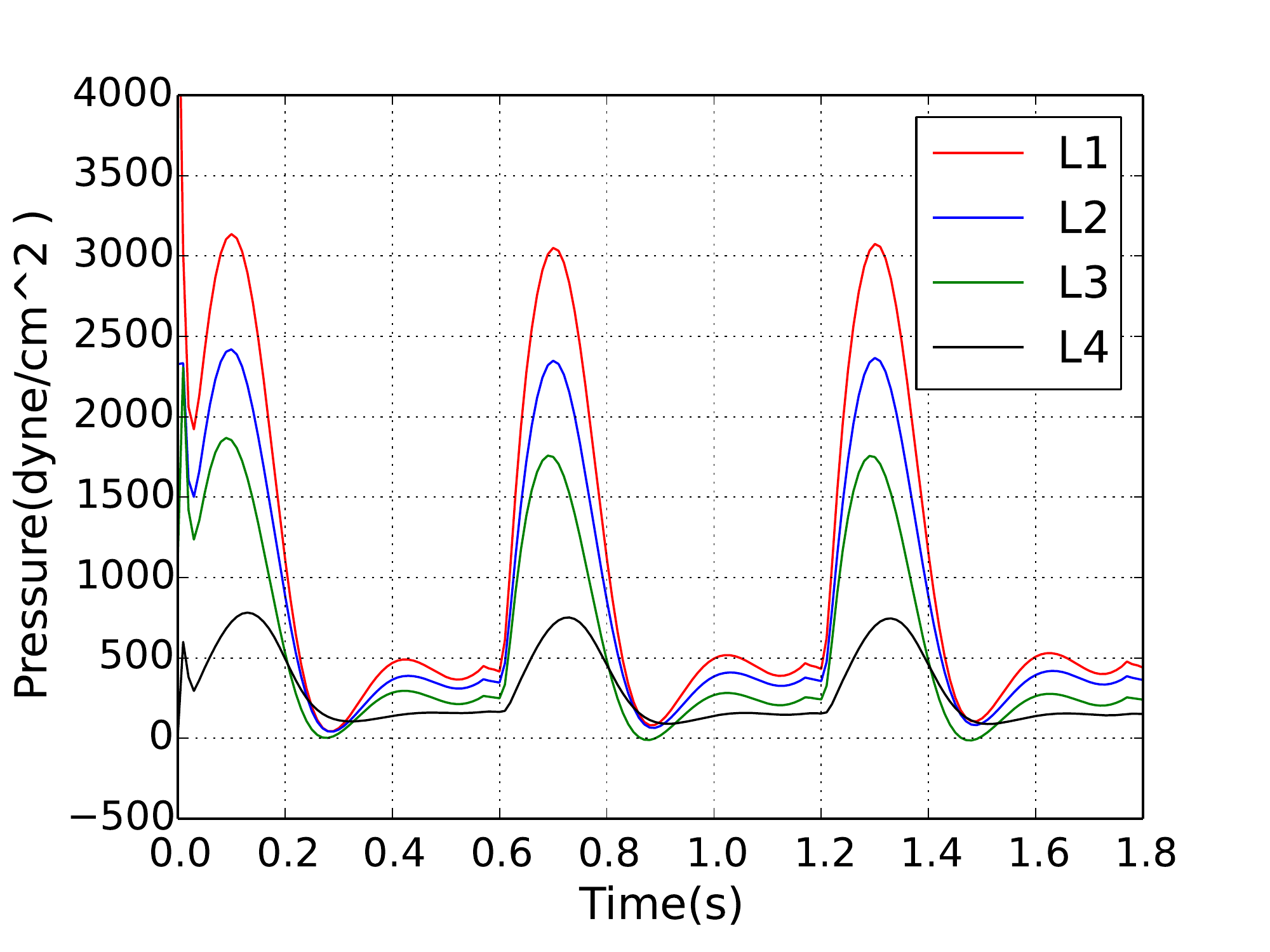} 
    \includegraphics[width=2.6in]{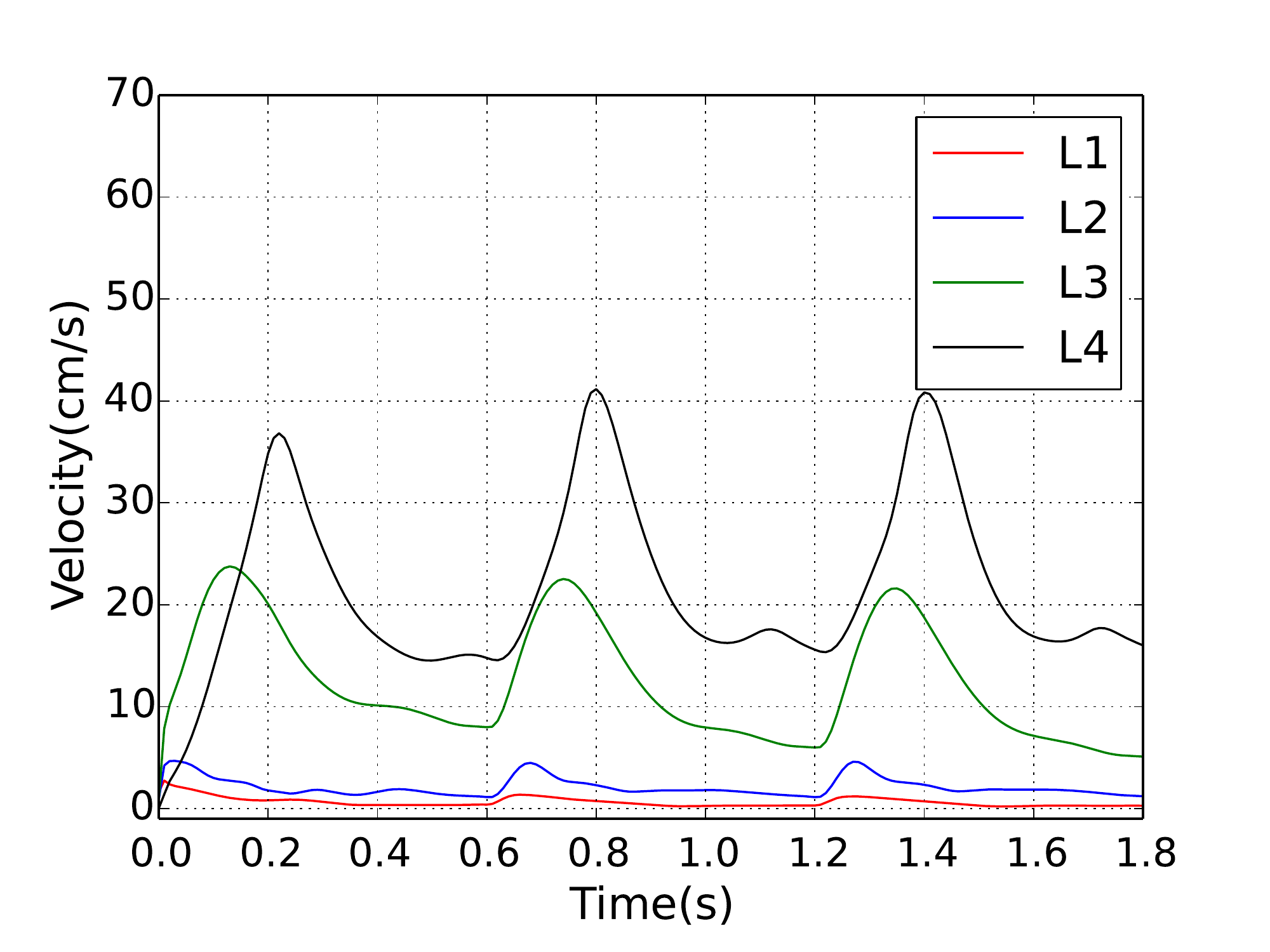}
   \caption{Pressure (left) and velocity (right) at different locations marked in the top picture.} \label{location_probe}
\end{figure}

In this section, we report some simulation results obtained by applying the proposed algorithms to the pulmonary artery shown in Fig~\ref{computational_domain}.   A velocity profile derived from a given inflow rate shown in Fig~\ref{heartflow_rate},  is applied to the inlet, while the flows at all outlets are supposed to be traction-free.  The fluid is characterized with viscosity $\mu_f = 0.03~g/(cm~s)$ and density $\rho_f =1~g/cm^3$.  The material parameters of the  arterial  wall are Young's modulus $E_s= 4 \times 10^6 ~g/(cm~s^2)$,  Poisson's  ratio $\mu_s = 0.42$, and density $\rho_s = 1.2~g/cm^3$.   The simulation is carried out with  the time step size $\delta t = 10^{-3}$ for three cardiac cycles, $[0, 1.8]$ seconds, and the solution is shown in Fig~\ref{streamline}, \ref{WSS}, \ref{bloodflow_pressure} and~\ref{artery_displacement}. 
The wall sheer stress (WSS) is an important metric, and it is calculated by the following formula
$$
WSS=\vsigma_f \vn_f -  (\vsigma_f \vn_f \cdot \vn_f) \vn_f,
$$
and the  spatially averaged WSS (SAWSS) is obtained by integrating the WSS on the entire  inner surface of  the artery and then normalized over the area, that is,
$$
SAWSS=\frac{1}{A} \int_{\p \oo_f} WSS dA,
$$
where $A$ is the total area of the inner surface  of the pulmonary artery. In Fig.~\ref{heartflow_rate}, we observe that  SAWSS is highly correlated with the input flow rate; SAWSS increases   when the input flow rate  increases  with time, and it decreases when the input flow rate decreases.    
In Fig.~\ref{streamline}, we see that the velocity is maximized at $1.35s$ when the right ventricle contracts  and the inflow rate reaches the maximum, and the flow pattern becomes more complicated at the end of   systole, $1.5s$. The flow is periodic in time, and it is restored to a similar pattern  at $1.8s$ that is the start of the next cycle.  
In Fig~\ref{WSS}, at $1.35s$ and $1.4s$, the inner artery wall has the largest wall shear stress in  magnitude, and it is correlated to the inflow pattern.  
In Fig.~\ref{bloodflow_pressure},  similarly, the pressure is also highly correlated to the inflow rate, that is, the pressure is high at  systole and low at diastole.  
From Fig.~\ref{artery_displacement},  it is easily observed that  the largest displacements are at  $1.35s$ and  $1.4s$, and the proximal blood vessels  have larger displacements while the ones at the distal do not deform much. In order to further understand how blood flows are different at the proximal and distal vessels, we probe the blood velocity  and the pressure  at different locations, and the results are shown in Fig.~\ref{location_probe}, which shows clearly that the pressure decreases from the proximal to the distal, and it is correlated to the inflow wave.  The third cycle of pressure wave  has the same pattern as  the second cycle, which indicates the simulation has reached a quasi steady state.  We also have a similar observation for the velocity, except that the velocity  magnitude  at $L4$ is larger than all other three locations.  

\subsection{Linear solver impact on the outer Newton iteration}
The accuracy of the linear Jacobian solver has a major impact on   the convergence of the outer Newton iterations.  More Newton iterations are usually  required  if the Jacobian system is not solved accurate enough, on the other hand the linear problems should not be over-solved because  beyond certain point it does not help the nonlinear solver any more and  simply wastes computing time.  The accuracy of the linear solver is controlled using a relative tolerance, $\text{L}_{rel}$, where a smaller value indicates a more accurate Newton direction.  We next  carry out a test to investigate the impact of the linear solver.  The mesh used in this test has 2,014,726 vertices and 9,464,723 elements, the problem has 12,793,688 unknowns, and  the simulation is carried out for 10 time steps.  The overlapping domain decomposition method with ILU(1) as the subdomain solver and $\delta=1$ is used as the preconditioner to accelerate the convergence of the Krylov subspace method GMRES.  The algorithm performance using 128 to 1,240 processor cores is summarized in Table~\ref{inexact_newton_method}. 
\begin{table}
\centering
\caption{Impact of different tolerances of linear solver. A nonlinear system of equations with 12,793,688 unknowns  is solved by inexact Newton-Krylov method together with a one-level Schwarz preconditioner.  ``$\ast $" indicates no convergence }\label{inexact_newton_method}
\begin{tabular}{c c c c c c  c c c c}
\cline{1-7}
$np$  &$\text{L}_{rel}$ & NI & LI &  T (second)  &MEM (MB) & EFF  \\
\cline{1-7}
128 &  $10^{-1}$& 6.8  &$13.2$ &1255.9 &600.4   & 100\%\\

128 & $10^{-2}$    & 4 &19.2 &771.8 &  608.3& 100\%\\

128 & $10^{-3}$ & 3.5 &23.6  &690.3&  616.3&  100\%\\

128 & $10^{-4}$ & 3 &30.4  &630.8&  616.3&  100\%\\

128 & $10^{-5}$ & 2.9 &39  &632&  624.2&  100\%\\
\cline{1-7}
256 &  $10^{-1}$& 7.2  &$13.9$ &719.8 &299.6   & 87\%\\

256 & $10^{-2}$  & 4.8 &19.5 & 509.8 &  303.6& 76\%\\

256 & $10^{-3}$ & 3.9 &23.7  &420.5&  307.6&  82\%\\

256 & $10^{-4}$ & 3.4 &30.8 &384.9&  311.6&  82\%\\

256 & $10^{-5}$ & 3 &40 &366.4&  311.6&  86\%\\

\cline{1-7}
512 &  $10^{-1}$& $\ast $ & $\ast $ & $\ast $ & $\ast $  & $\ast$\\

512 & $10^{-2}$     & 4.8 &20.4& 312.3 &  181& 62\%\\

512 & $10^{-3}$ & 3.9 &24.9  &261.1&  183&  66\%\\

512 & $10^{-4}$ & 3.2 &32.9 &224.9&  185.1&  70\%\\

512 & $10^{-5}$ & 3 &41.6  &221.2&  187.2&  71\%\\
\cline{1-7}
1,024 &  $10^{-1}$&7.2  & 15.1 &  244.3&  82.7 & 64\%\\

1,024 & $10^{-2}$ & 4.5 &20.3& 155.4 &  83.7& 62\%\\

1,024 & $10^{-3}$ & 3.9 &25.3 &138.6&  84.7&  62\%\\

1,024 & $10^{-4}$ & 3 &33.3 &112&  85.7&  70\%\\

1,024 & $10^{-5}$ & 2.9 &42.1  &113.2&  86.7&  70\%\\
\cline{1-7}
\end{tabular}
\end{table}
It is easily observed from  Table~\ref{inexact_newton_method} that the number of Newton iterations is decreased quickly at the beginning and then slowly when we increase the accuracy of the linear solver   by reducing  the relative tolerance. For the 128-core case, the averaged number of Newton iterations is decreased from $6.8$ to $4$ when $\text{L}_{rel}$ is reduced from $10^{-1}$  to $10^{-2}$, and when we continue to decrease $\text{L}_{rel}$  by another order of magnitude from $10^{-2}$ to $10^{-3}$, the number of Newton  iterations is decreased by only $0.5$.  The number of linear iterations per Newton using a loose tolerance is smaller than that using a tight tolerance since more effort is needed for achieving a more accurate Newton direction.  Sometimes, if the tolerance is too loose, the overall algorithm may not converge at all because a minimum accuracy is required for  Newton to   converge.  For instance, Newton does not work for the 512-core case when the tolerance is chosen as $\text{L}_{rel} = 10^{-1}$.   We keep decreasing  the tolerance from $10^{-3}$ to $10^{-4}$, and the number of Newton iterations is first reduced by half or one iteration,  and it does not change much with a  decrease in the tolerance  from $10^{-4}$ to $10^{-5}$. The compute time is reduced  significantly at the very beginning because  the number of Newton iterations is decreased a lot due to a tighter tolerance, and it is decreased slowly thereafter.   For example, in the 128-core case, the compute time is $1255.9$ seconds when $\text{L}_{rel} = 10^{-1}$, while it is reduced to $771.8$ seconds by almost half with $\text{L}_{rel} = 10^{-2}$.  The compute time does not reduce much and sometimes slightly increases  when the accuracy of the linear solver reaches a certain level.  Let us look at the 128-core case again: the compute time is $630.8$ with  $\text{L}_{rel} = 10^{-4}$, and it is increased by a few seconds to $632$ when we use $\text{L}_{rel} = 10^{-4}$. A tighter linear solver helps maintain  a great scalability.  The parallel efficiency is kept at about $70\%$ when the relative tolerance is chosen as $10^{-4}$ or $10^{-3}$. In all, the difference  of compute time using  the tolerance of $10^{-3}$ to $10^{-5}$ is small, and hence an arbitrary  relative tolerance ranging from $10^{-3}$ to $10^{-5}$ should work well for the problems at hand.   The scalability is good for all cases  as long as the tolerance is not too loose.

\subsection{Subdomain solver: incomplete LU factorization}
Incomplete LU (ILU) factorization is a popular subdomain solver for overlapping  domain decomposition methods. Like all other solvers, there are a few parameters, in ILU, that affect the overall convergence of the linear solver. Among these parameters, the most important one is the fill-in  level that represents how much  extra allocation is  allowed to store new values introduced by the factorization. Level 0, denoted as ILU(0), indicates that all new extra values are discarded, and the factorized matrix has the same sparsity as the original submatrix.  ILU($l$) represents that $l$ layers of extra entries are kept.  Larger $l$ usually leads to a better and more robust converge but meanwhile it may slow down the entire solver because more operations and more memory are needed. An optimal fill-in level is also problem-dependent. We perform a test for different fill-in levels, and the results are summarized in Table~\ref{incomplete_lu}.  The same configuration as in the previous test is used. 
\begin{table}
\centering
\caption{Different fill-in levels  for ILU.  A nonlinear system of equations  with 12,793,688 unknowns is calculated by an one-level Schwarz preconditioner with ILU($l$) as the subdomain solver. }\label{incomplete_lu}
\begin{tabular}{c c c c c c  c c c c}
\cline{1-7}
$np$  &subsolver& NI & LI &  T  &MEM & EFF  \\
\cline{1-7}
128 &  ILU(0)&4.1 &72.7  &  1015.1&  552 & 100\% \\

128 &  ILU(1)&3.5 &23.6  &  689.4&  616.3 &  100\%\\

128 &  ILU(2)&3.6&17.3  &  797.9&  800.3 &  100\%\\

\cline{1-7}

256 &  ILU(0)&3.9 &69.2  &  507.7&  274.7 & 100\%\\

256 &  ILU(1)&3.9 &23.7  &  426.6&  307.6 & 81\%\\

256 &  ILU(2)&3.9 &18.2  &  466.6&  397 &  86\%\\
\cline{1-7}

512 &  ILU(0)&3.9 &68.9  &  302.3&  157.9 &  84\% \\

512 &  ILU(1)&3.9 &24.9  &  265.6&  183 &  65\%\\

512 &  ILU(2)&3.9 &18.5  &  289&  242&  69\%\\

\cline{1-7}
1,024 & ILU(0)& $ \ast $ & $ \ast $  & $\ast$ & $\ast $ &  $\ast$  \\

1,024 &  ILU(1)&3.9 &25.3 &  138&  84.7 &  62\%\\

1,024 &  ILU(2)&3.9 &20.4 &  152.9&  109.2 &  65\%\\

\cline{1-7}
\end{tabular}
\end{table}
For  the128-core case, when the fill-in level is increased from $0$  to $1$ the number of GRMES iterations is reduced to one-third, which results in the reduction of the compute time by half. When we continue increasing the fill-in level from $1$ to $2$, the number of GMRES iterations is decreased by only $5$, and the compute time is increased due to the increased cost of LU factorization per iteration.  For all cases, the number of Newton iterations  is kept close to a constant, $4$, because the relative tolerance of GRMES is fixed as $10^{-3}$ regardless of the ILU fill-in level.  ILU($0$) sometimes is   unstable.  For example,  Newton does not converge for the case of 1,024 processor cores when ILU(0) is used, while it performs well using  ILU(1) and ILU(2).  The memory usage is increased slightly when ILU(1) is used instead of ILU(0), while it is increased a lot with the increase of fill-in level from $1$ to $2$.  In  the128-core case, the memory usage is increased by $70$ MB when we increase the fill-in level from $0$ to $1$, while it is increased by almost $200$ MB with the increase of fill-in level from $1$ to $2$. ILU(1) is the best subdomain solver in this test when we take both the compute time and the memory usage into consideration.  For all cases, the overall algorithm scales well when the number of processor cores is increased from $128$ to $1,024$. 

\subsection{Jacobian reevaluation}
By default,  the Jacobian matrix is updated at every Newton iteration.  Since  the evaluation of the Jacobian matrix is  expensive, sometimes the overall efficiency of the algorithm  is   improved by reusing  the Jacobian matrix for  several Newton iterations. In the rest of this subsection, we denote by "$lag$" as the number of times that the Jacobian matrix is reused. In this test, we use the same configuration as before to illustrate the  behavior of the algorithm when  using  different lags.  The results are shown in Table~\ref{newton_lag}, which shows clearly that
\begin{table}
\centering
\caption{Impact of Newton lags on the compute time.  A nonlinear system of equations  with 12,793,688 unknowns is solved using inexat Newton and each Jacobian matrix is used for $lag$ iterations. }\label{newton_lag}
\begin{tabular}{c c c c c c  c c c c}
\cline{1-7}
$np$  &$lag$ & NI & LI &  T  &MEM & EFF  \\
\cline{1-7}
128 &  1&3.5 &23.6  &  689.5&  616.3 &  100\%\\

128 &  2&3.9 &22.6  &  492.5&  616.3 &  100\%\\

128 &  3&4.1 &21.4  &  485.2&  616.3 &  100\%\\

128 &  4&4.1 &21.1  &  351.1&  616.3 &  100\%\\
\cline{1-7}

256 &  1&3.9 &23.7  &  427.8&  307.6 &  81\%\\

256 &  2&4.3 &23.3  &  304.2&  307.6 &  81\%\\

256 &  3&4.6 &22.4  &  284.1&  307.6 &  85\%\\

256 &  4&4.8 &21.6  &  250.3&  307.6 &  70\%\\
\cline{1-7}

512 &  1&3.9 &24.9  &  261.2&  183 &  66\%\\

512 &  2&4.1 &24.6 &  169.3&  183 &  73\%\\

512 &  3&4.8 &23.5  &  171.5&  183 &  71\%\\

512 &  4&4.9 &22.6  &  160.5&  183 &  55\%\\
\cline{1-7}

1024 &  1&3.9 &25.3  &  138.7&  84.7 &  62\%\\

1024 &  2&4.0 &25  &  86.5&  84.7 &  71\%\\

1024 &  3&4.4 &23.8  &  88.2&  84.7 &  69\%\\

1024 &  4&4.4 &23.1  &  70.8&  84.7 &  62\%\\
\cline{1-7}
\end{tabular}
\end{table}
 the compute time is significantly reduced when we reuse the Jacobian matrix for  two Newton steps instead of one, and then it is improved very little when we further increase   $lag$ for all processor counts except the case when the number of cores is 128, where the compute time is  reduced by $100$ seconds when   $lag$ is increased from $3$ to $4$.  The Jacobian lag does not affect the memory usage for all cases because a sparse matrix is always stored regardless of its value updates. The overall algorithm behaves similarly when $lag=2, 3$ and $4$, while the performance with $lag=1$ is much worse. The number of Newton iterations increases when we lag the Jacobian matrix evaluation, but the increase is not significant so that  we  have  the benefit of Jacobian lagging.  In the 1,024-core case, the compute time is decreased by $50\%$ when we lag the Jacobian evaluation from  every  Newton step to every two Newton steps. Continuing lagging the matrix evaluation does not improve the performance a lot.  In all cases, the overall algorithm equipped with the Schwarz preconditioner is able to maintain good scalability. 

\subsection{Submatrix ordering}
The subdomain problems are in the inner most loop of the algorithm, both the convergence and the efficiency of the overall algorithm depend on the performance of the subdomain solver. One critical issue is the ordering of the subdomain matrix.  A few reordering schemes including RCM, ND, 1WD and QMD are considered in this test, and the results are reported in Table~\ref{ordering}. 
\begin{table}
\centering
\caption{Impact of submatrix ordering on the algorithm performance.  A nonlinear system of equations with 12,793,688 unknowns is solved Newton-Krylov-Schwarz  in which  the Schwarz preconditioner is realized with different submatrix reordering schemes.}\label{ordering}
\begin{tabular}{c c c c c c  c c c c}
\cline{1-7}
$np$  &Reordering & NI & LI &  T  &MEM & EFF  \\
\cline{1-7}
128 &  RCM&3.5 &23.6  &  697.6&  616.3 &  100\%\\

128 &  Natural&3.8 &28.1  &  873.2&  699.5 &  100\%\\

128 &  ND&3.6 &28  &  787.2&  696.9 &  100\%\\

128 &  1WD&3.4 &23.6  &  678.9&  616.7 &  100\%\\

128 &  QMD&3.9 &25.6  &  809.5&  660.5 &  100\%\\
\cline{1-7}

256 &  RCM&3.9 &23.7  &  427.2&  307.6 &  82\%\\

256 &  Natural&3.8 &27.8  &  471.6&  348.6 &  93\%\\

256 &  ND&3.9 &26.9  &  465&  343.4 &  85\%\\

256 &  1WD&3.9 &24.2  &  428.2&  306.7 &  79\%\\

256 &  QMD&3.9 &26.7  &  453.8&  326 &  89\%\\
\cline{1-7}

512 &  RCM&3.9 &24.9 &  261&  183 &  67\%\\

512 &  Natural&3.8 &28.2  &  282.1&  210.4 &  77\%\\

512 &  ND&3.9 &27.8  &  279.1&  203 &  71\%\\

512 &  1WD&3.9 &24.5  &  260.2&  181 &  65\%\\

512 &  QMD&3.9 &24.5  &  276&  191.5 &  73\%\\
\cline{1-7}

1024 &  RCM&3.9 &25.3  &  138.8&  84.7 &  63\%\\

1024 &  Natural&3.8 &28.6  &  151.4&  98.9 & 72\% \\

1024 &  ND&3.8 &27.3  &  145.2&  96.5 &  68\%\\

1024 &  1WD&3.8 &25.7  &  134.6&  84.6 &  63\%\\

1024 &  QMD&3.9 &27.1  &  145.3&  91.1 &  70\%\\
\cline{1-7}
\end{tabular}
\end{table}
The number of Newton iterations stays close to a constant, $4$, when different ordering schemes are used since the same linear tolerance is used. Compared with the ``natural" reordering, the number of GMRES iterations is smaller using RCM or 1WD.  All schemes improve the total compute time and the memory usage,  when compared with the ``natural" ordering method. Let us look at  the128-core case,   1WD gives the best   compute time of 678.9 seconds while  the``natural" ordering  is the worst with compute time of $873.2$ that is $20\%$ more. Similarly, for all other processor counts, 1WD performs better than all the other schemes, and 1WD and RCM have a similar performance behavior while 1WD is slightly better. QMD does not help much in this test because it is designed for symmetric matrix while the matrix is highly unsymmetrical in this study.  It is also interesting to see that the memory usage is halved, regardless of the reordering schemes,  when we double the number of processor cores. It indicates that the preconditioner constructed based on the overlapping domain decomposition is scalable in terms of the memory usage. Nevertheless,  for all schemes, the overall algorithm scales well in terms of the compute time  up to 1,024 processor cores.

\subsection{Subdomain overlapping size}
The size of subdomain overlap plays an important role in  the overall algorithm performance since the overlapping size represents  how much information a subdomain receives   from its neighbors.  A larger overlap often improves  the linear solver  convergence in terms of GMRES iterations because it has more information  from its neighbors, but on the other hand it requires  more communication time. In this set of tests, we investigate the algorithm using different overlap, and also with different fill-in level, and  we show the detailed results in Table~\ref{overlap}. Dashed lines are used to separate the ILU(1) and  the ILU(2) results.  
\begin{table}
\centering
\caption{Different overlapping sizes.  A nonlinear system of equations with 12,793,688 unknowns is solved a Newton-Krylov together with the Schwarz preconditioner equipped with different overlapping sizes.}\label{overlap}
\begin{tabular}{c c c c c c  c c c c}
\cline{1-8}
$np$  &$\delta$ &subsolver & NI & LI &  T  &MEM & EFF  \\
\cline{1-8}
128 &  0&ILU(1)&3.9 &33.4  &  806.8&  621 &  100\%\\

128 &  1&ILU(1)&3.5 &23.6  &  691&  616.3 &  100\%\\

128 &  2&ILU(1)&3.2 &23.3  &  643.1&  617.1 &  100\%\\
\hdashline 
128 &  0&ILU(2)&3.9 &27.1  &  902.6&  804.7 & 100\% \\  

128 &  1&ILU(2)&3.6 &17.3  &  797&  800.3 & 100\% \\

128 &  2&ILU(2)&3.3 &16  &  739.7&  793.5 &  100\%\\

\cline{1-8}
256 &  0&ILU(1)&3.9 &36.2  &  439.9&  309.3 &  92\%\\

256 &  1&ILU(1)&3.9 &23.7  &  426.2&  307.6 &  81\%\\

256 &  2&ILU(1)&3.3 &23.2  &  377.8&  308.6 &  85\%\\
\hdashline 
256 &  0&ILU(2)&3.9 &31.3  &  494.7&  398.4 &  91\%\\

256 &  1&ILU(2)&3.9 &18.2 &  466.4&  397 &  85\%\\

256 &  2&ILU(2)&3.5 &15.7  &  429&  394.3 &  86\%\\

\cline{1-8}
512 &  0&ILU(1)&3.9 &39 &  266.2&  168.9 &  76\%\\

512 &  1&ILU(1)&3.9 &24.9  &  260.8&  183 &  66\%\\

512 &  2&ILU(1)&3.7 &23.7  &   269.2&  203.3 &  60\%\\
\hdashline 
512 &  0&ILU(2)&3.9 &33.7  &  297.7&  218.8 &  76\%\\

512 &  1&ILU(2)&3.9 &18.5  &   286.8&  242.1 &  69\%\\

512 &  2&ILU(2)&3.7&16.4  &   289.5&  270.1 &  64\%\\
\cline{1-8}
1024 &  0&ILU(1)&3.9&40.6  &  138.8&  76&  73\%\\

1024 &  1&ILU(1)&3.9 &25.3  &  138.7&  84.7 &  63\%\\

1024 &  2&ILU(1)&3.4 &23.8  &  131&  99.2 &  61\%\\
\hdashline 
1,024 &  0&ILU(2)&3.9 &36.3 &  154.2&  95.7 &  73\%\\

1,024 &  1&ILU(2)&3.9 &20.4  &  154.1&  109.2 &  65\%\\

1,024 &  2&ILU(2)&3.7 &16.8  &  160.7&  130.5 &  58\%\\
\cline{1-8}
\end{tabular}
\end{table}

In  the 128-core case with ILU(2),  when we increase the overlap,   the number of Newton iterations is gently reduced, and the number of GMRES iterations is decreased by $10\%$ with the increase of overlap from $0$ to $1$, then it does not decrease much any more when we further increase the overlap from 1 to 2.  The compute time is decreased with an increase in the overlap for the 128-core and 256-core cases, but  when we use 512 and 1024 processor cores, the compute time  is increased even though the number of GMRES iterations is actually reduced because more processor cores implies more communication cost.  $\delta =2$ is the best choice when we use 128 and 256 cores, while $\delta=1$ is the best parameter for 512 and 1,024 processor cores.  We have a similar observation for ILU(1) as well,  a larger overlap  results in a better convergence in terms of the number of GMRES iterations for all processor counts. The compute time is decreased when we increase the overlap  for small processor  counts, and it becomes similar for different overlap when we increase the number of processor cores.  Sometimes, the compute time increases as we increase the  overlap, for example, the compute time with $\delta=2$ is higher than that of  $\delta=1$ in the 512-core case.   Generally speaking, it is a good idea to use a large overlap  when the number of processor cores is small, and a smaller overlap when the number of processor cores is large.    

\subsection{Scalability with a large number of processor cores}
\begin{figure}
   \centering
   \includegraphics[width=2.8in]{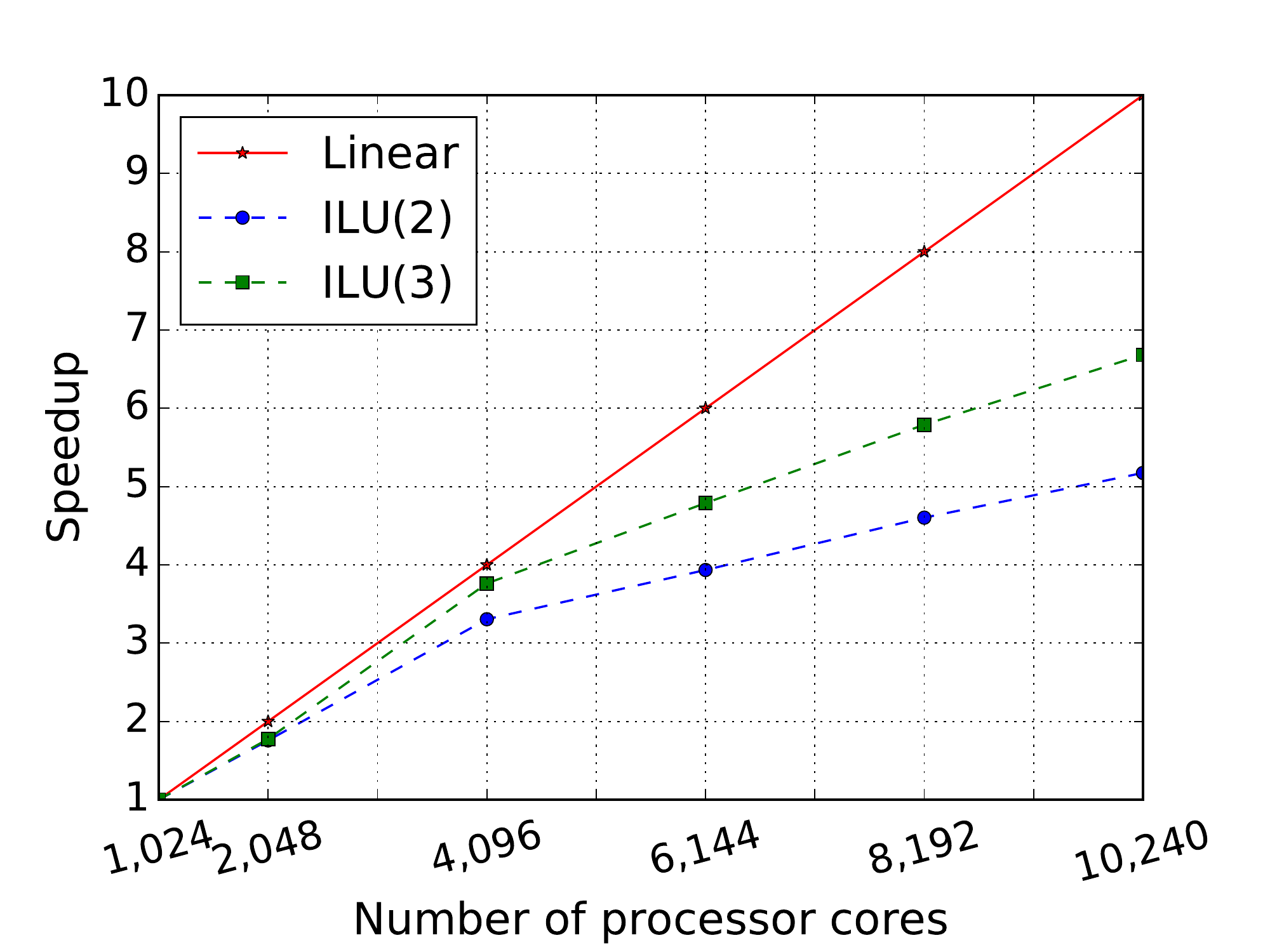} 
   \includegraphics[width=2.8in]{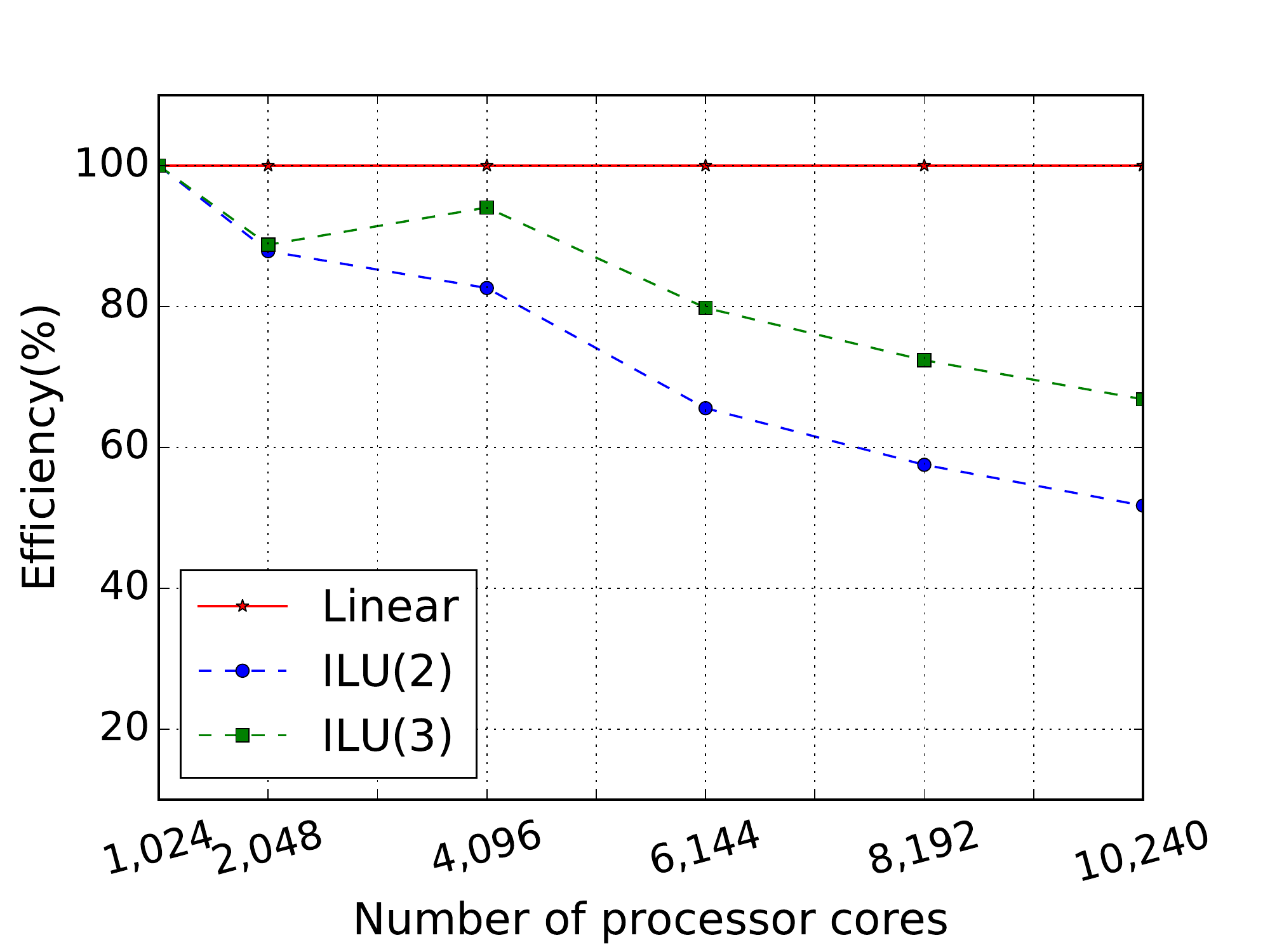} 
   \caption{Speedup and parallel efficiency of the overall algorithm using up to 10,240 processor cores. Right: speedup; left: parallel efficiency. } \label{snessolve_speedup}
\end{figure}

In this subsection, we study the strong scalability of the overall algorithm  on a supercomputer with more than 10,000 processor cores. The mesh used for this test is larger than the ones used previously, and it has 75,717,784 mesh elements,  14,276,963 mesh vertices, and the problem has 90,551,052 unknowns.   We summarize the results in Table~\ref{scalability_large_scale} for the cases of 1,024, 2,048, 4,096, 6,144, 8,192 and 10,240 processor cores.  Two subdomain solvers, ILU(2) and ILU(3), are used.  ILU(0) and  ILU(1) do not work  for these large problems.  
\begin{table}
\centering
\caption{Strong scalability with up to 10,240 processor cores. Newton-Krylov-Schwarz is used to solve a nonlinear system of equations with 90,551,052 unknowns. }\label{scalability_large_scale}
\begin{tabular}{c c c c c c  c c c c}
\cline{1-7}
$np$  &subsolver & NI & LI &  T  &MEM & EFF  \\
\cline{1-7}
1,024 &  ILU(2)&3.1 &27.1  &  1587.5&  1366.9 &  100\%\\

1,024 &  ILU(3)&3.3 &26.1  &  2382.6&  1755.4 &  100\%\\
\cline{1-7}
2,048 &  ILU(2)&3.1 &27.7  &   903.4&  1079.3 &  88\%\\

2,048 &  ILU(3)&3.6 &24.4  &  1341.9&  1304.1 &  89\%\\
\cline{1-7}
4,096 &  ILU(2)&3.1 &26.8  &  480.3&  654.6 &  82\%\\

4,096 &  ILU(2)&3.2 &24.8  &  633.3&  755.3 &  94\%\\
\cline{1-7}
6,144 &  ILU(2)&3 &33  &  403.5&  515.4 &  66\%\\

6,144 &  ILU(3)&3 &28.8  &  497.5&  556 &  80\%\\
\cline{1-7}
8,192 &  ILU(2)&3.1 &33.7  &   344.9&  517.6 &  58\%\\

8,192 &  ILU(3)&3 &31.7  &   411.5&  562.3 &  72\%\\
\cline{1-7}
10,240 &  ILU(2)&3.1 &35.5  &   306.8&  451.2 &  52\%\\

10,240 &  ILU(3)&3.2 &28.8  &   356.5&  480.1 &  67\%\\
\cline{1-7}
\end{tabular}
\end{table}
From Table~\ref{scalability_large_scale}, we see that  the number of Newton iterations stays close to a small constant for all cases except  at 2,048 cores with ILU(2), where the number of Newton iterations per time step is $3.6$. When we increase the number of processor cores, the number of Newton iterations doesn't change much indicating  that the overall algorithm is  scalable in terms of the Newton iteration.  The number of GMRES iterations is    close to 27 for ILU(2) and 24 for ILU(3) when the number of processor cores ranges from 1,024 to 4,096,  and it is increased by $20\%$ for both ILU(2) and ILU(3) when the processor count is equal to or larger than 6,144, which results in  a reduction in the parallel efficiency by about $15\%$. However, we still have a parallel efficiency of $67\%$ for ILU(3) and $52\%$ for ILU(2) even when the number of processor cores is more than 10,000. The memory usage is decreased properly for most cases.  For example, the memory usage is reduced by $40\%$ for both  ILU(2) and ILU(3) when the number of processor cores is increased from 2,048 to 4,096.    The speedup and parallel efficiency of the overall algorithm are shown in Fig.~\ref{snessolve_speedup}. 

\begin{figure}
   \centering
   \includegraphics[width=2.8in]{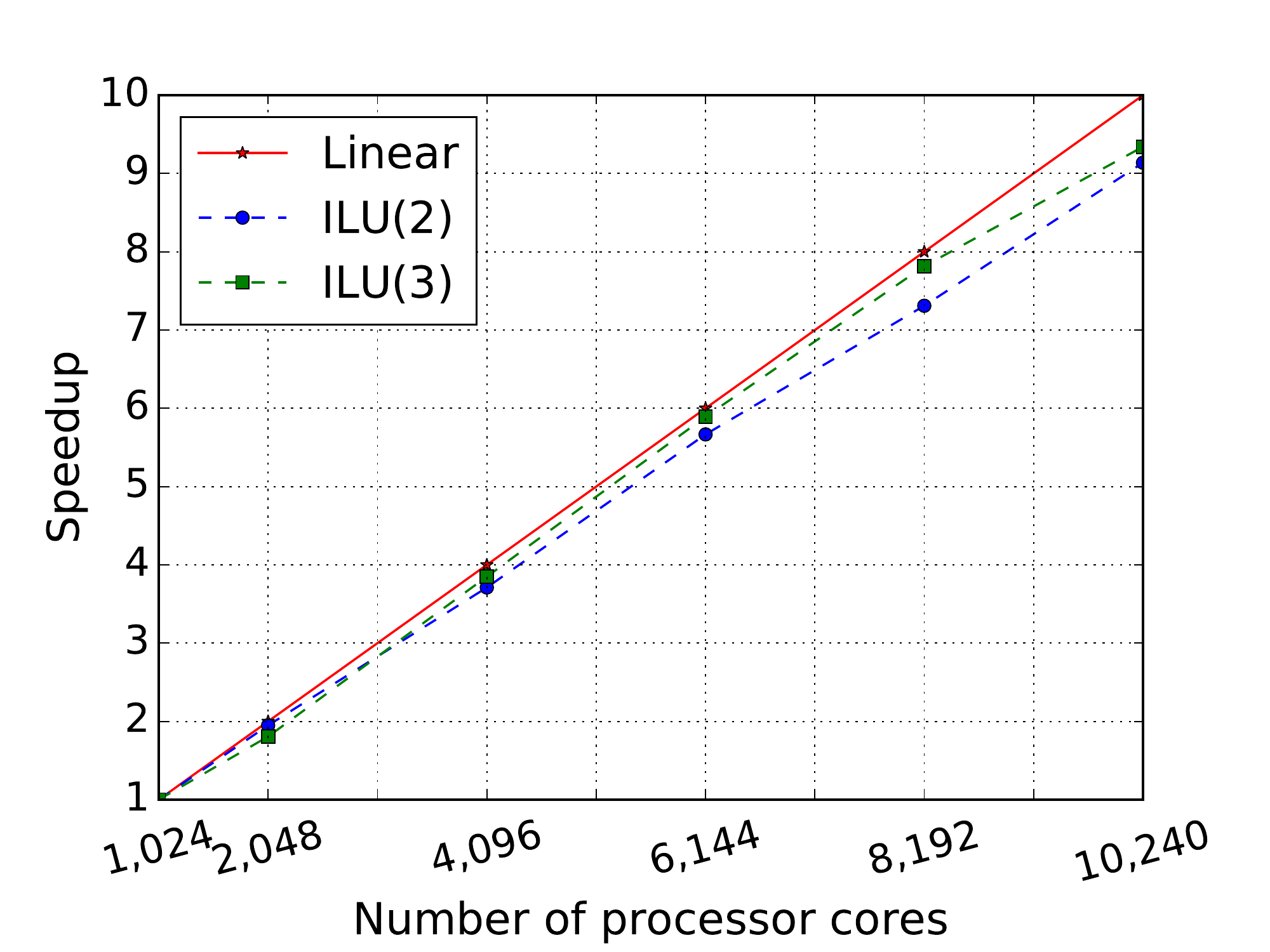} 
   \includegraphics[width=2.8in]{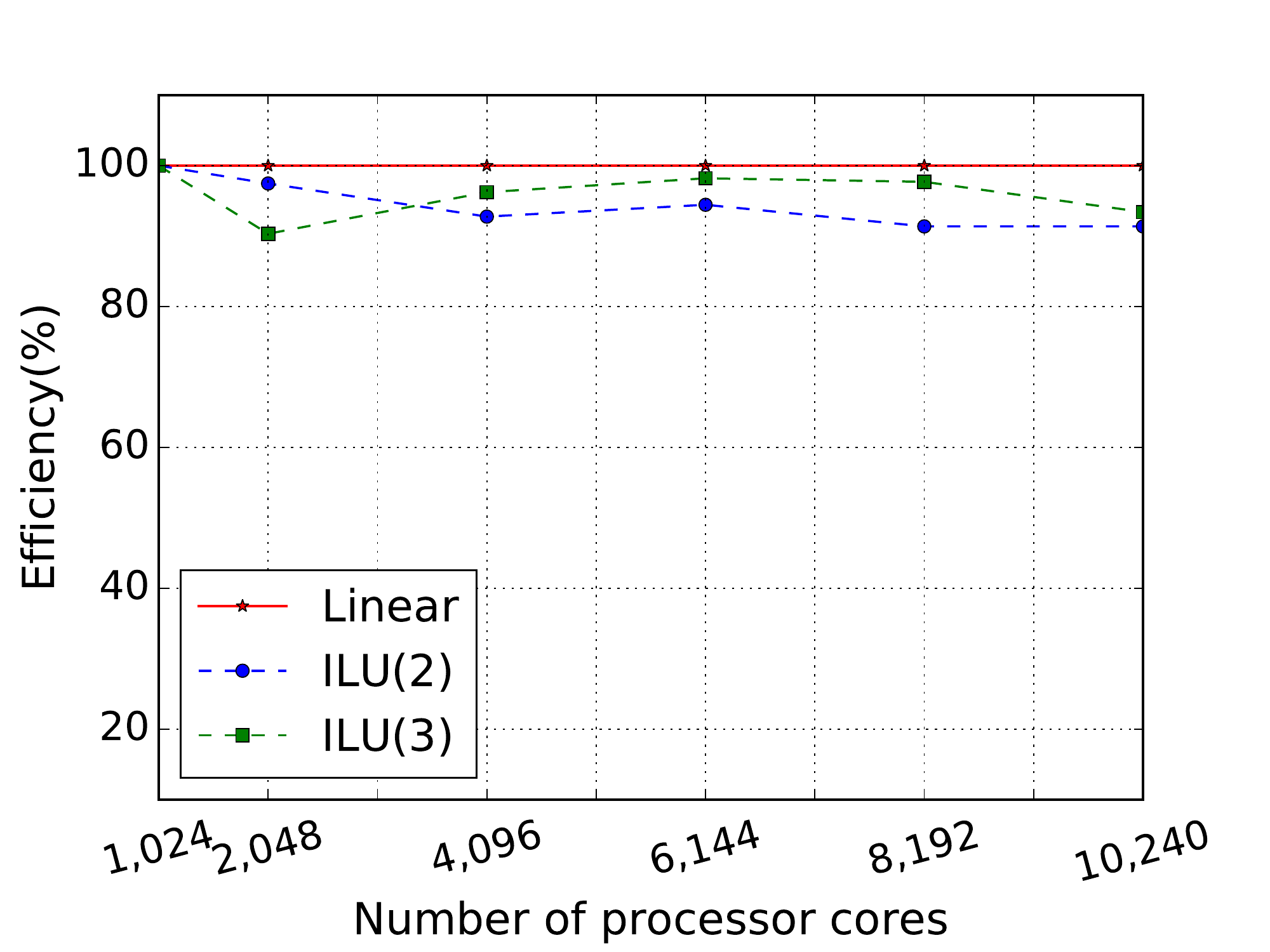} \\
   \caption{Speedup and parallel efficiency of the function evaluation using up to 10,240 processor cores. Right: speedup; left: parallel efficiency. } \label{functioneval_speedup}
\end{figure}

\begin{figure}
   \centering
    \includegraphics[width=2.8in]{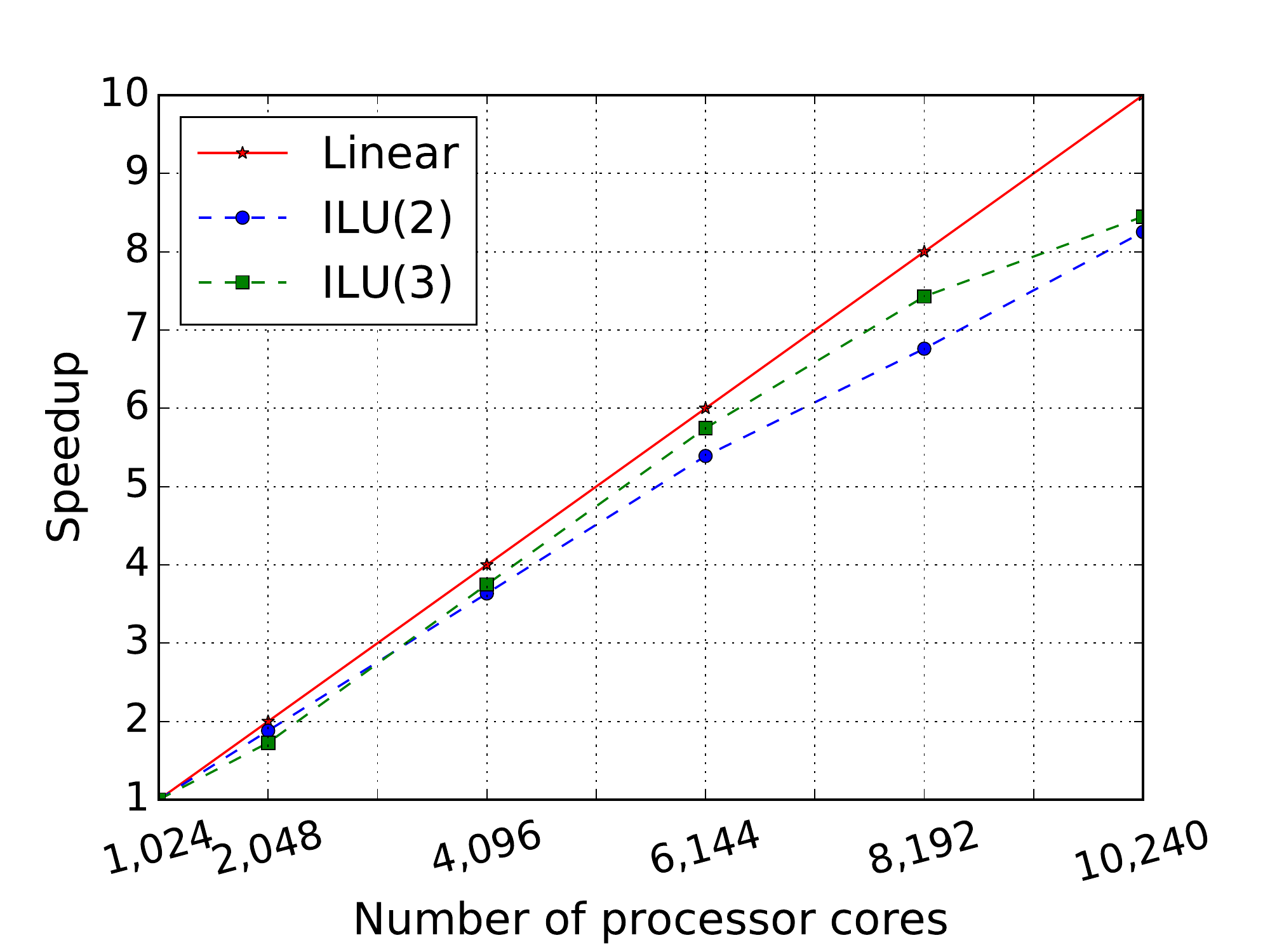} 
   \includegraphics[width=2.8in]{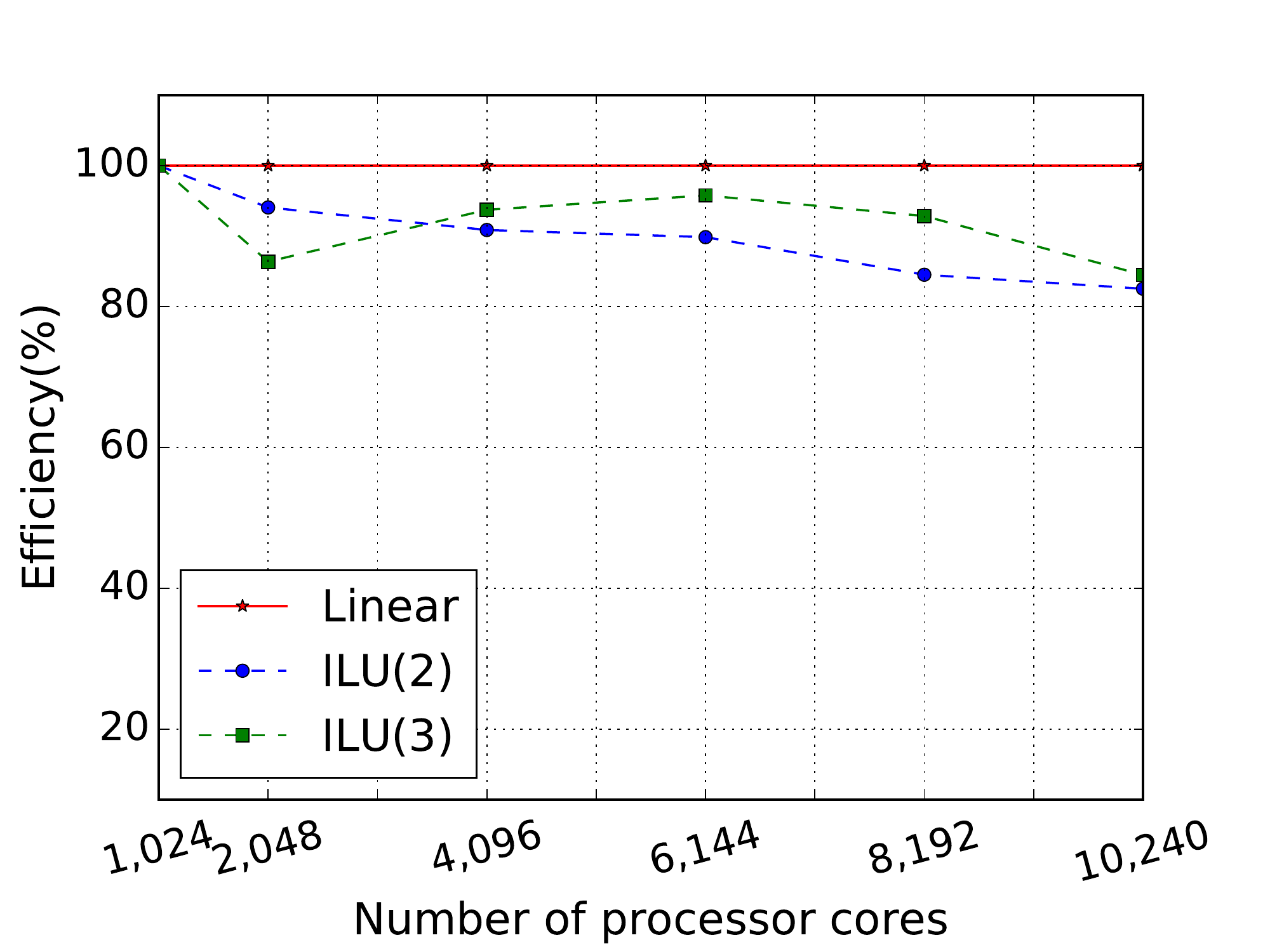} \\
   \caption{Speedup and parallel efficiency of the overall algorithm using up to 10,240 processor cores. Right: speedup; left: parallel efficiency. } \label{jacobianeval_speedup}
\end{figure}

\begin{figure}
   \centering
    \includegraphics[width=2.8in]{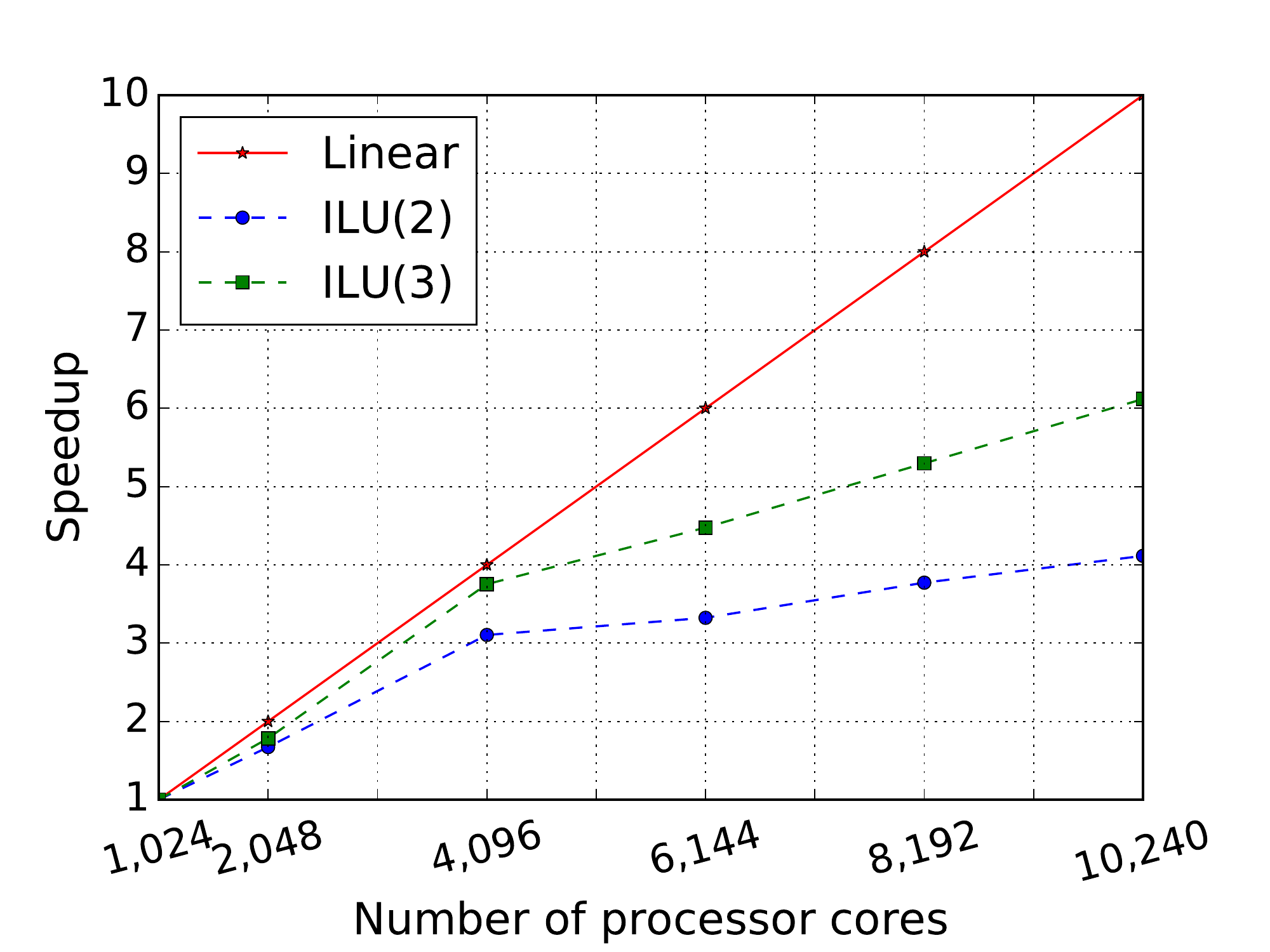} 
   \includegraphics[width=2.8in]{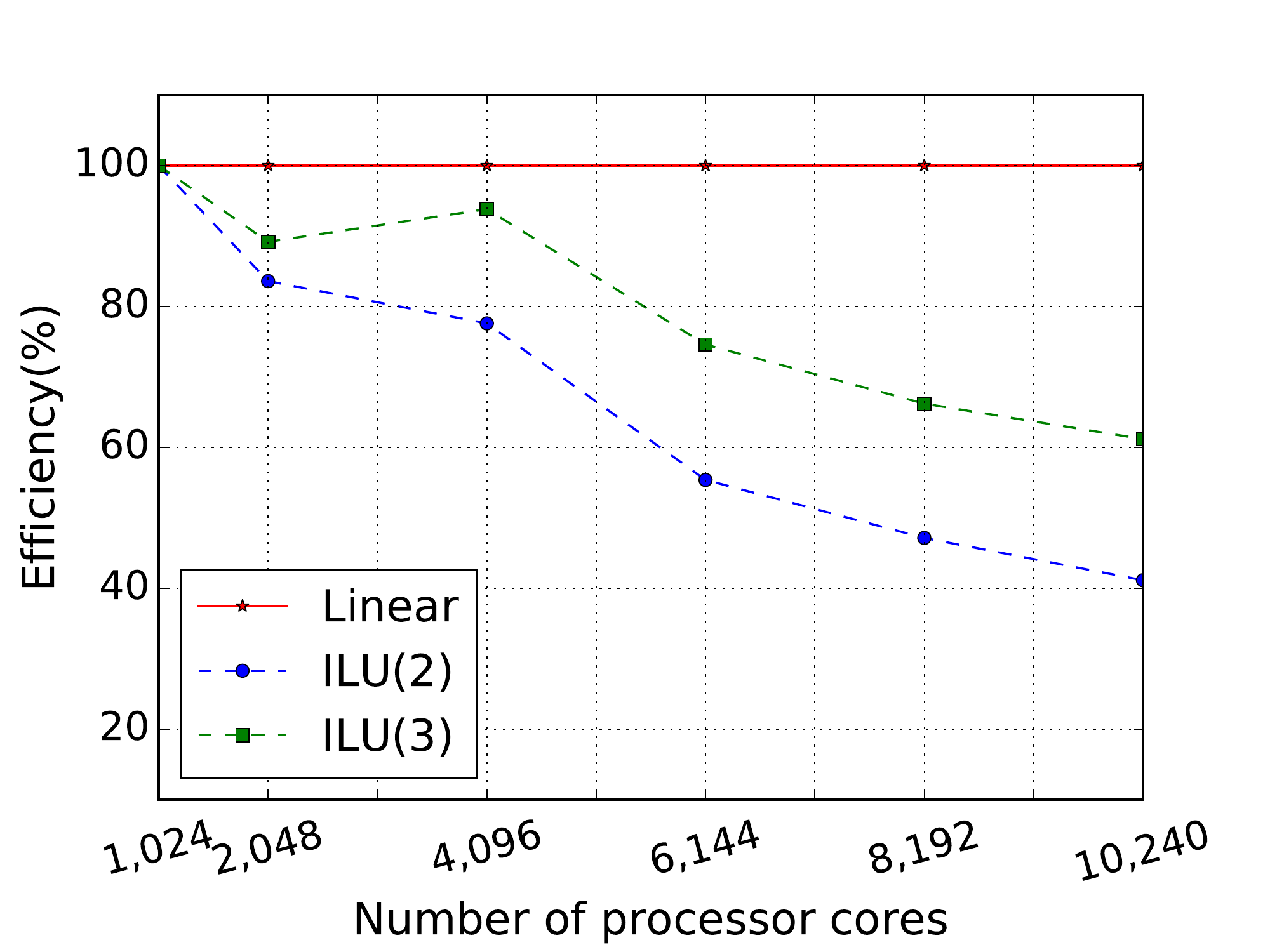} \\
   \caption{Speedup and parallel efficiency of the linear solver using up to 10,240 processor cores. Right: speedup; left: parallel efficiency. } \label{kspsolve_speedup}
\end{figure}

\begin{figure}
   \centering
    \includegraphics[width=2.8in]{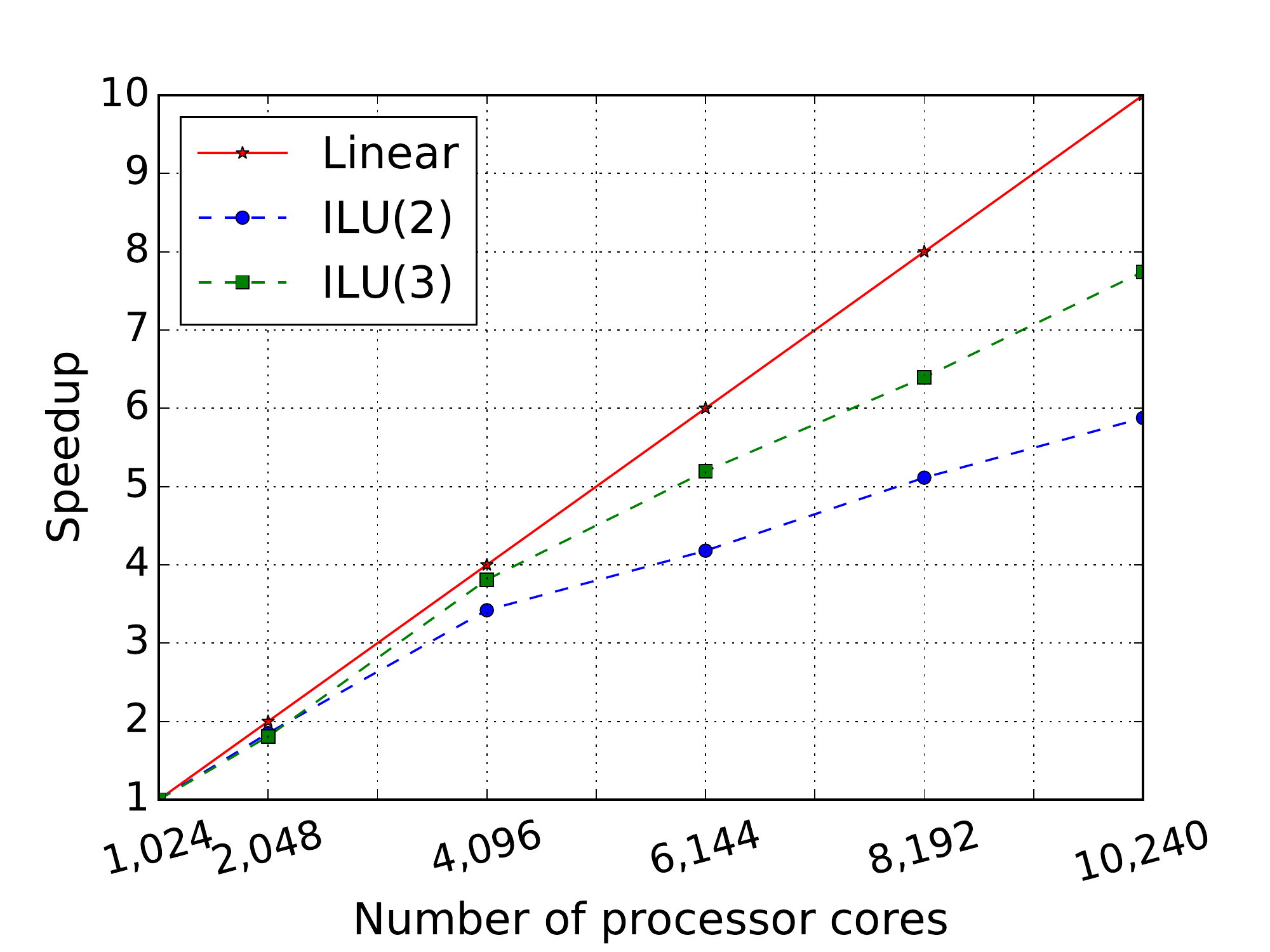} 
   \includegraphics[width=2.8in]{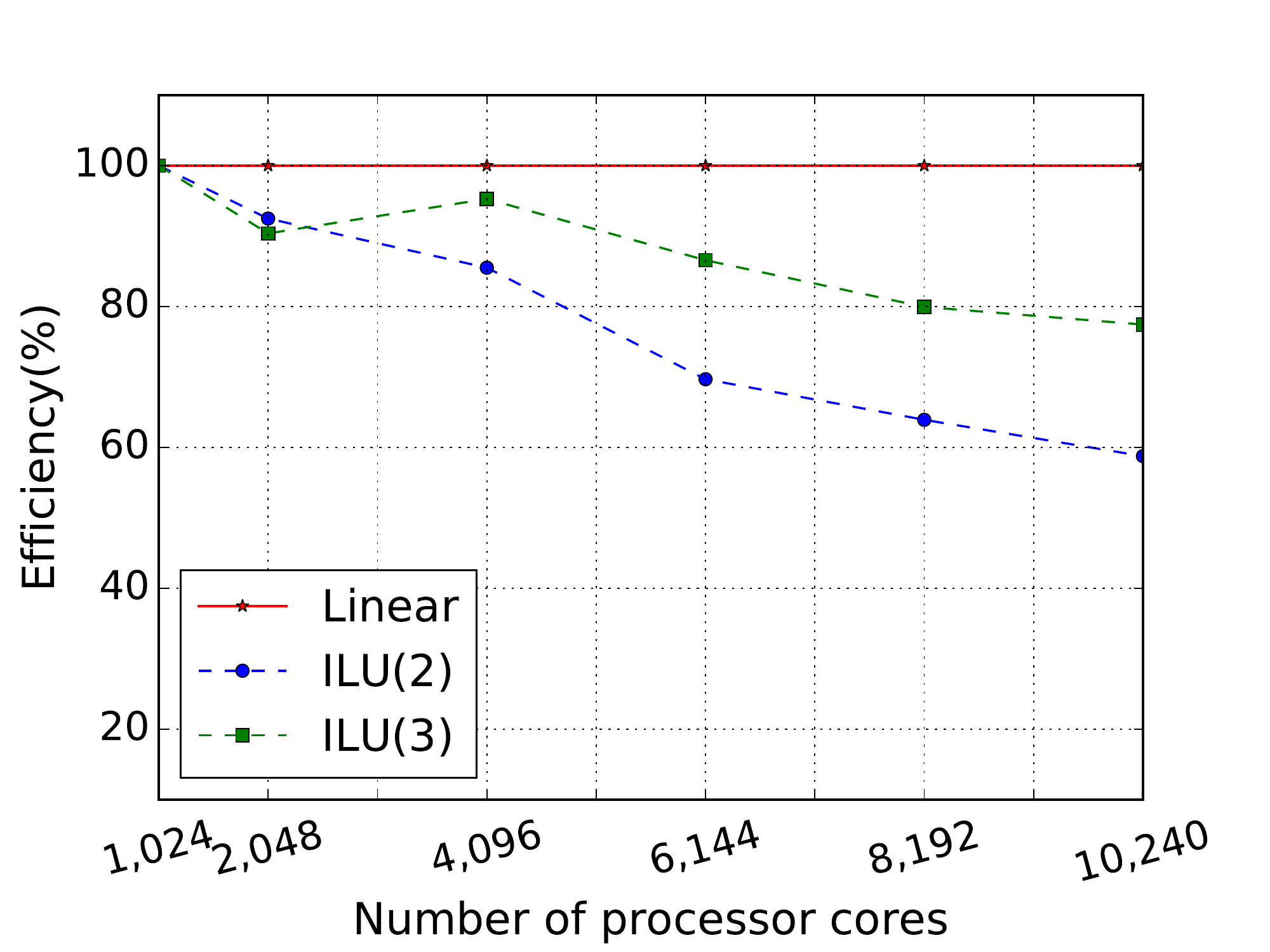} \\
   \caption{Speedup and parallel efficiency of the preconditioner application  using up to 10,240 processor cores. Right: speedup; left: parallel efficiency. } \label{pcapply_speedup}
\end{figure}

To further understand the algorithm performance, we summarize the compute times spent on the individual components of the algorithm in Table~\ref{scalability_details}.  Different components of the algorithm have   different properties  so that the speedup and parallel efficiency  are different. Some  components, such as the Jacobian and function evaluations, are perfectly scalable, and other component, for example preconditioner setup, is difficult to scale.  For convenience,  let us introduce the notations used in Table~\ref{scalability_details}. ``KSPSolve" is the compute time spent on the linear solver,  ``KSPSetUp" denotes the time on the linear solver setup, ``PCSetUp" represents the  time for the preconditioner setup, ``PCApply" is the time used in the application of the preconditioner,  ``FuncEval" denotes the function evaluation  time, and ``JacEval" is the time spent on the  Jacobian evaluation.  Data for ILU(2) and ILU(3) are separated by a dashed line, and the results for different processor counts are separated by a solid line.   The record consists of two rows;  the second row is the actual compute time of  the individual component and the first row  is the  proportion of the total compute time in percentage. The total compute time, ``T", is composed of the time spent on the linear solver,  the function evaluation and the Jacobian evaluation, and the preconditioner setup and application is part of the linear solver.  
\begin{table}
\centering
\caption{Strong scalability of the algorithm components   with up to 1,0240 processor cores. Newton-Krylov-Schwarz is used to solve a nonlinear system of equations with 90,551,052 unknowns. }\label{scalability_details}
\begin{tabular}{c c c c c c  c c c c}
\cline{1-9}
$np$  &subsolver & T & KSPSolve &KSPSetUp& PCSetUP  &PCApply & FuncEval & JacEval  \\
\cline{1-9}
--&--                 & 100\% & 61\%   & 0.1\%& 6\% & 51\% & 5\% &  35\%\\ 
1,024 &  ILU(2)&1587.5 &960.8  &  1.5 &  93 & 827.4  & 73.1 & 561.2\\
\hdashline 
--&--                 & 100\% & 72\%   & 0.1\%& 8\% & 62\% & 3\% &  25\%\\ 
1,024 &  ILU(3)&2382.6 &1716.2  &  2.4 &  199.7 & 1476.8  & 76.6 & 596.5\\
\cline{1-9}
--&--                 & 100\% & 64\%   & 0.2\%& 12\% & 50\% & 4\% &  33\%\\ 
2,048 &  ILU(2)&903.4 &574.6  &  1.5 &  107.6 & 447.3  & 37.5 &  298.3\\
\hdashline 
--&--                 & 100\% & 72\%   & 0.1\%& 9\% & 61\% & 3\% &  26\%\\ 
2,048 &  ILU(3)&1341.9 &962.2  &  1.6 &  126.1 & 817  & 42.4& 345.3\\
\cline{1-9}
--&--                 & 100\% & 64\%   & 0.2\%& 12\% & 50\% & 4\% &  32\%\\ 
4,096 &  ILU(2)&480.3 &309.5 &  0.9 & 58.6  & 241.9 & 19.7 & 154.4 \\
\hdashline 
--&--                 & 100\% & 72\%   & 0.1\%& 10\% & 61\% & 3\% &  25\%\\ 
4,096 &  ILU(3)&633.3 &457.3 &  0.9 & 62.1  & 387.5 & 19.9 & 159.1 \\
\cline{1-9}
--&--                 & 100\% & 72\%   & 0.2\%& 20\% & 49\% & 3\% &  26\%\\ 
6,144  &  ILU(2)&403.5 &289.1  &  0.7 & 81.9  & 197.9  & 12.9 & 104.1 \\
\hdashline 
--&--                 & 100\% & 77\%   & 0.1\%& 19\% & 57\% & 3\% &  21\%\\ 
6,144  &  ILU(3)&497.5&383.4  &  0.7 & 93.6  & 284.3  & 13 &  103.8 \\
\cline{1-9}
--&--                 & 100\% & 74\%   & 0.1\%& 25\% & 47\% & 3\% &  24\%\\ 
8,192 &  ILU(2)&344.9 &254.7 &  0.5 & 86.8  & 161.8  & 10 & 83 \\
\hdashline 
--&--                 & 100\% & 79\%   & 0.1\%& 21\% & 47\% & 2\% &  20\%\\ 
8,192 &  ILU(2)&411.5 &324  &  0.5 & 88 & 230.9  & 9.8 & 80.3 \\
\cline{1-9}
--&--                 & 100\% & 76\%   & 0.1\%& 29\% & 46\% & 3\% &  22\%\\ 
10,240 &  ILU(2)&306.8 &233.5  &  0.5 &  87.6  & 140.8  & 8 & 68 \\
\hdashline 
--&--                 & 100\% & 79\%   & 0.1\%& 24\% & 53\% & 2\% &  20\%\\ 
10,240 &  ILU(3)&356.5 &280.5  &  0.5 &  86.2  & 190.7  & 8.2 & 70.6 \\
\cline{1-9}
\end{tabular}
\end{table}
Theoretically,  the Jacobian and function evaluation should be perfectly scalable as long as the number of linear iterations and Newton iterations does  not increase much when we increase the number of processor cores. From the last and the second to the last columns of Table~\ref{scalability_details},  we observe that the compute time spent on  the Jacobian evaluation and the function evaluation are  almost halved when we double the number of processor cores, which indicates that both the function evaluation and the Jacobian evaluation are ideally scalable.  This phenomenon is also observed from Fig.~\ref{functioneval_speedup} and~\ref{jacobianeval_speedup}. 
Among  the algorithmic components, ``KSPSolve" takes most of the  compute time, that is, it takes $61\%$ of the total compute time with ILU(2) at 1,024 cores and  increases to $76\%$ at 10,240 cores.  Compared with the ILU(2) case, the proportion is a little more in the ILU(3) case, where $72\%$ of the total compute time is used for solving the Jacobian system when the number of processor cores is 1,240, and it is increased a little to  $79\%$ for 10,240 processor cores. The speedup and parallel efficiency of the linear solver,  shown in Fig~\ref{kspsolve_speedup}, is similar to that of the overall algorithm   because it takes most of the total compute time.  
The linear solver consists of the vector orthogonalization, the preconditioner setup, and the preconditioner application, especially the preconditioner setup and application take most of the linear solver time.   Therefore, the design and development of the preconditoner is critical to have the overall algorithm scalable.   The preconditioner setup takes less than or around $10\%$ of the total compute when the number of processor cores is less than or equal to 4,096 processor cores, and it takes up around $20\%$ when the number of processor cores is  more than 4,096   but the overall  algorithm still has  the parallel efficiencies of   $52\%$ with ILU(2) and $67\%$ with ILU(3) at 10,240 processor cores.  The precondiitoner application take $50\%$ of the total compute time for all cases, and the proportion does not  change with the increase of the processor count for both ILU(2) and ILU(3) indicates that preconditioner application has almost the same speedup and parallel efficiency, shown in Fig~\ref{pcapply_speedup}, as the overall algorithm.  In all, all components except the precondiitoner setup are scalable so that the overall algorithm scales reasonably well with up to 10,240 processor cores. 

\subsection{Mesh preparation}
\begin{figure}
   \centering
   \includegraphics[width=2.8in]{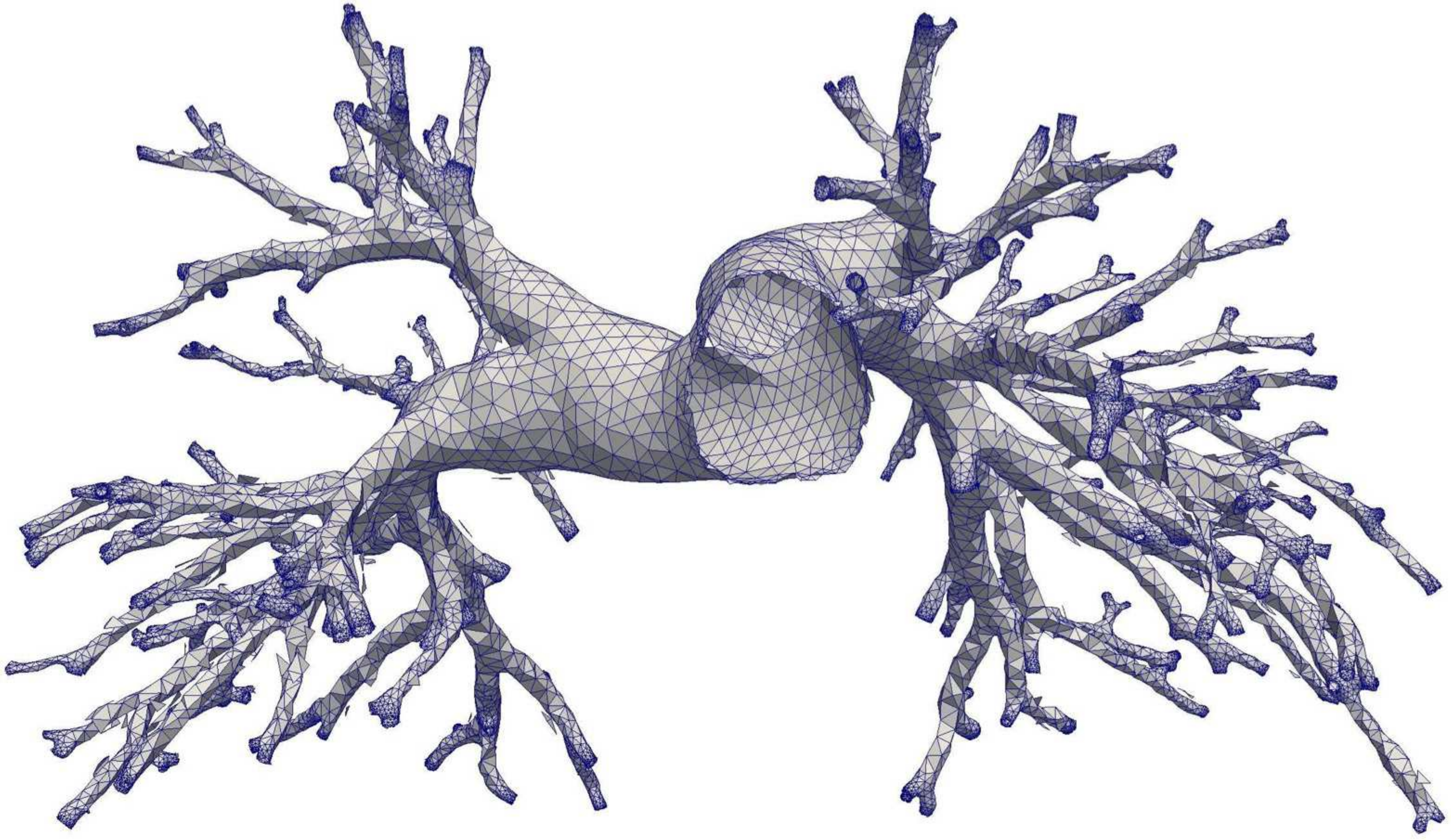} 
   \includegraphics[width=2.8in]{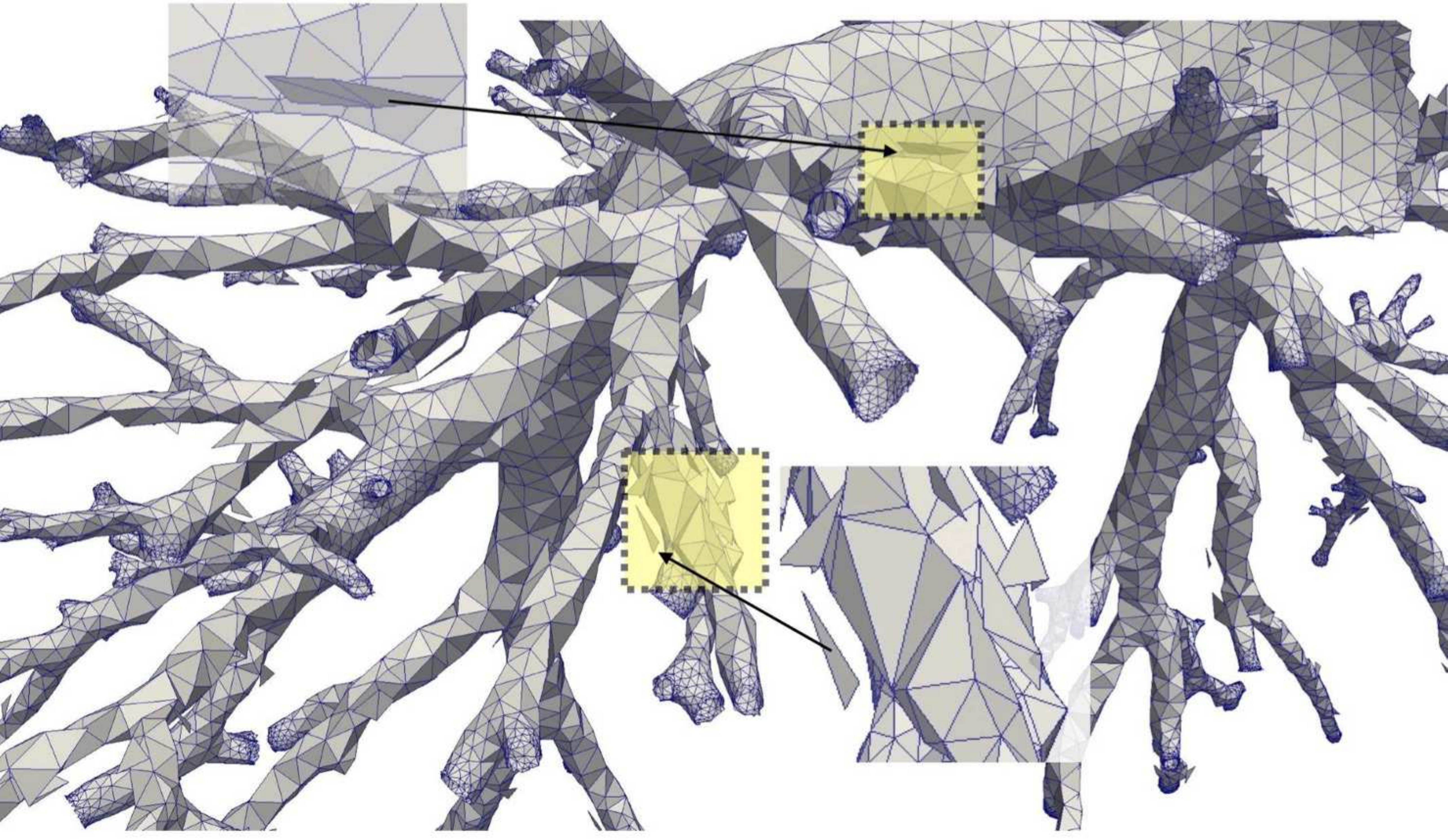} \\
   \includegraphics[width=2.8in]{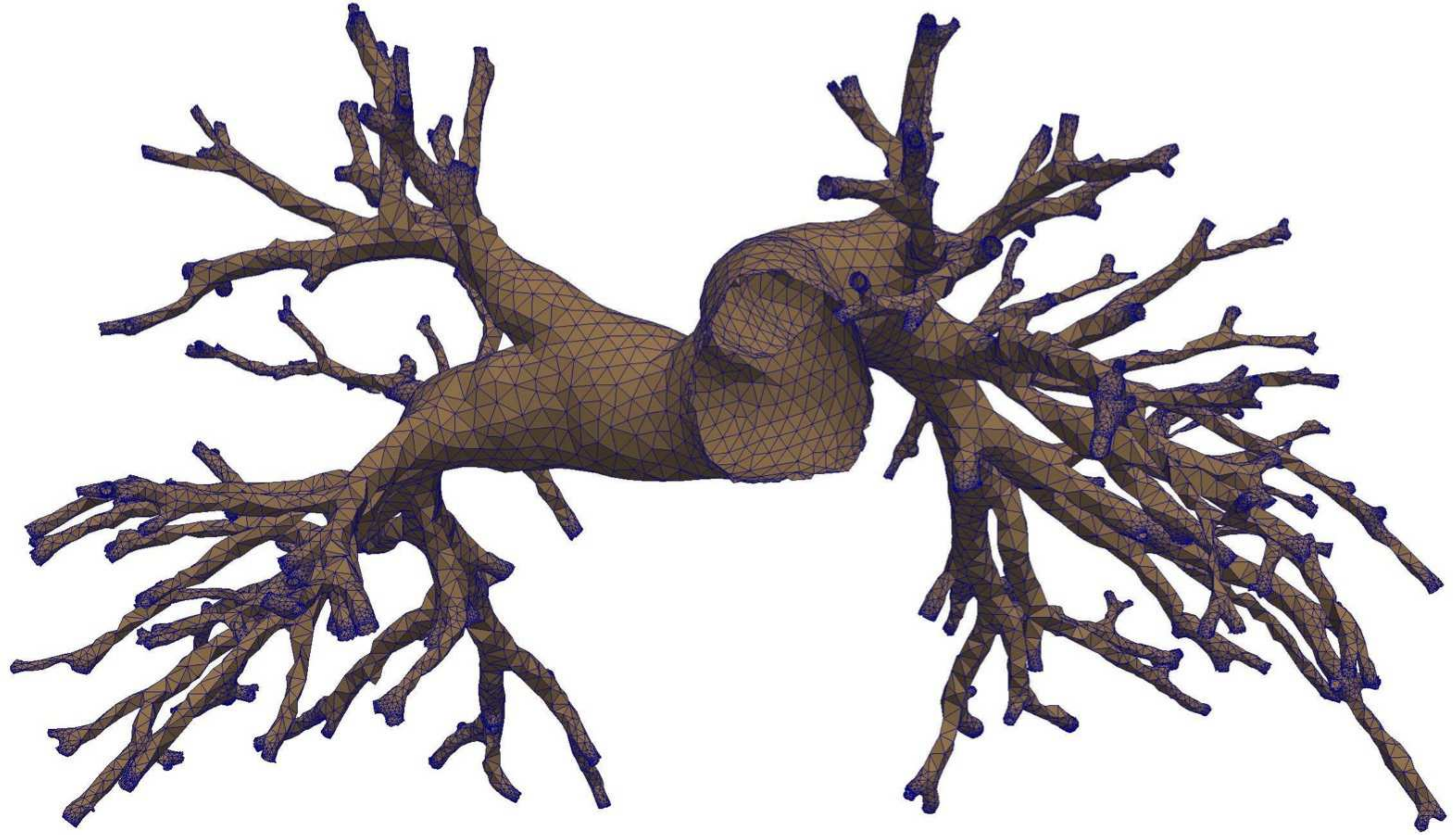} 
   \includegraphics[width=2.8in]{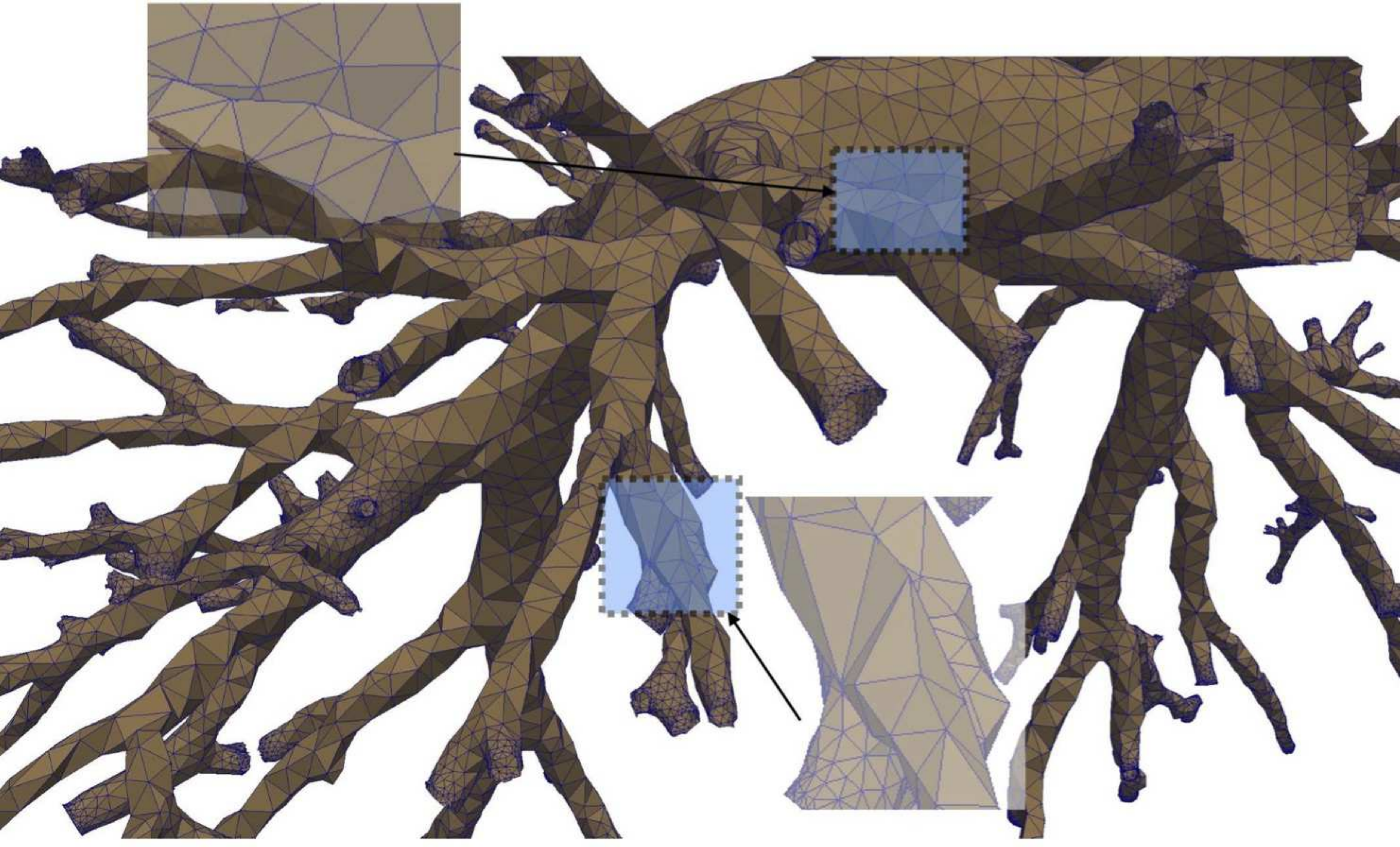} 
   \caption{Interface meshes. Top: invalid interface  containing several interior elements, bottom: the reconstructed interface  mesh where the illegal elements have been removed.  } \label{interface_mesh}
\end{figure}

\begin{figure}
   \centering
   \includegraphics[width=2.5in]{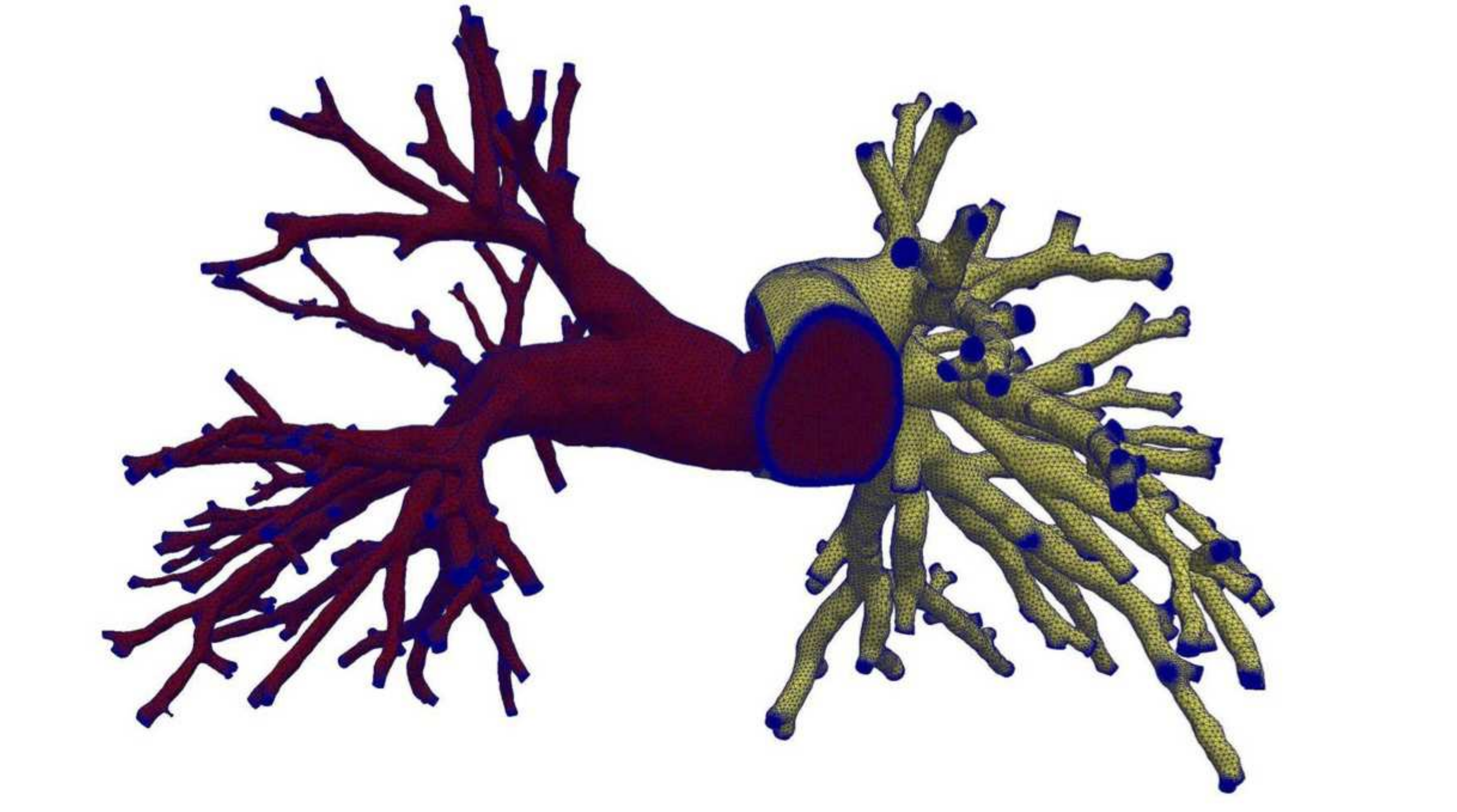} 
   \includegraphics[width=2.5in]{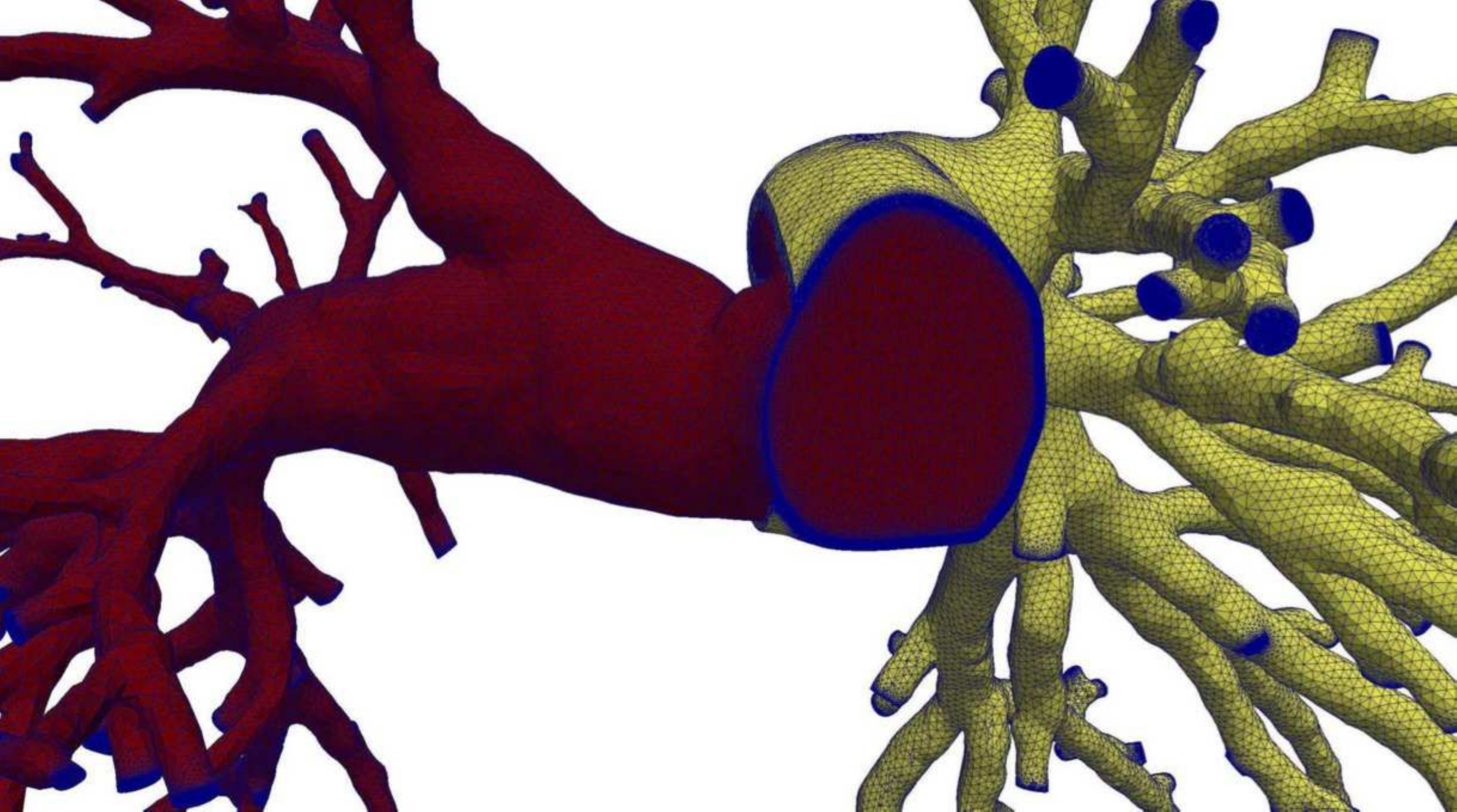} \\
   \includegraphics[width=2.5in]{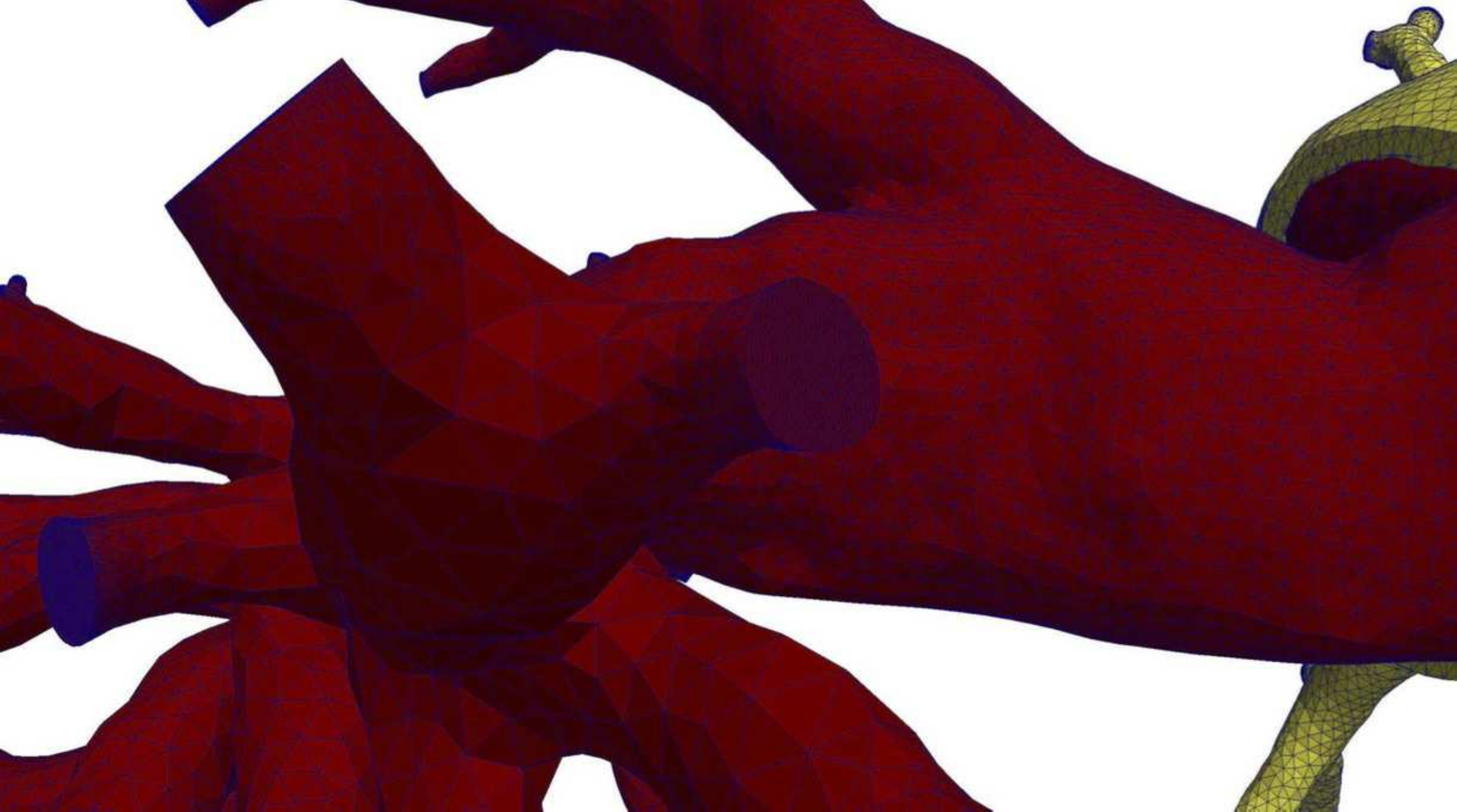} 
   \includegraphics[width=2.5in]{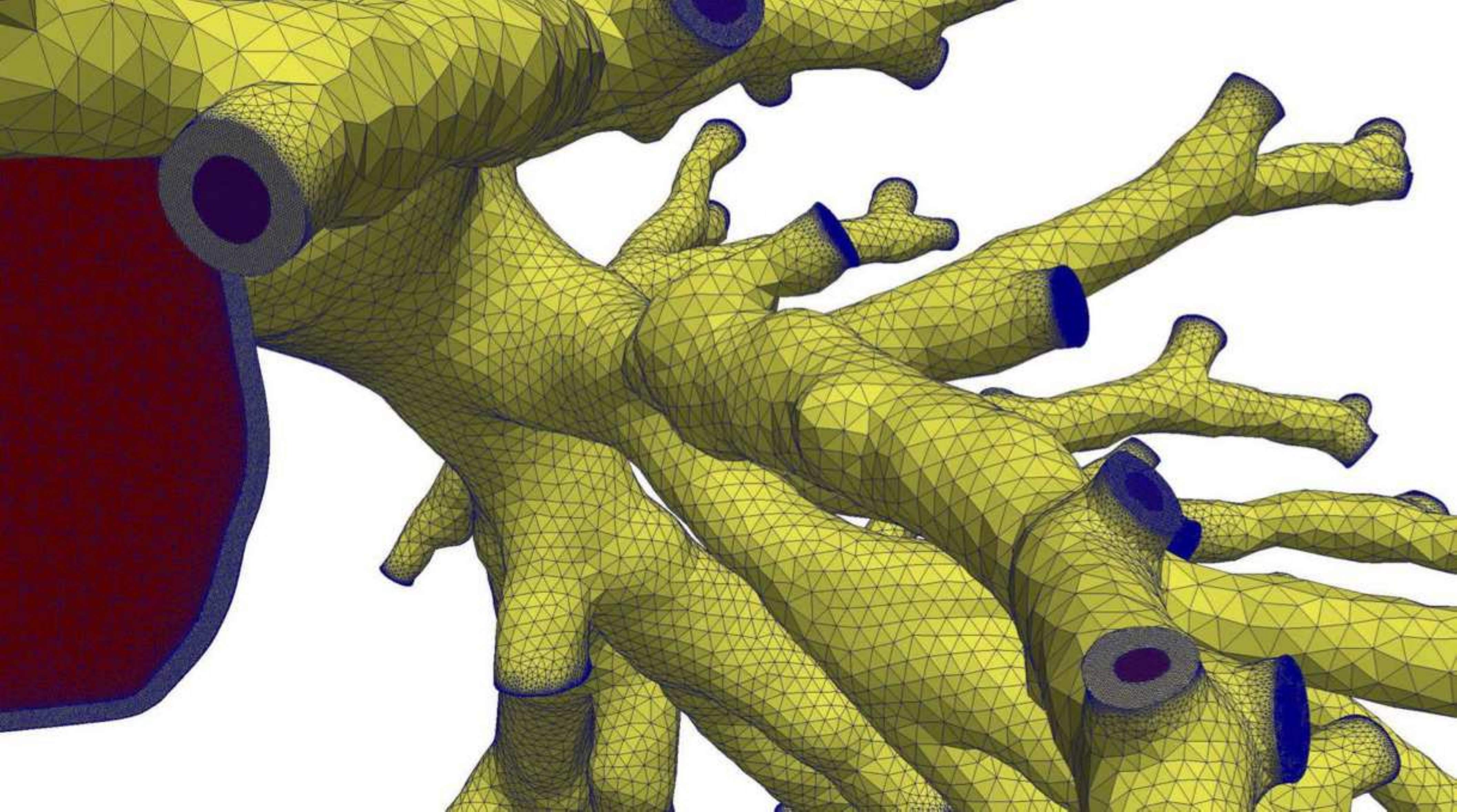} 
   \caption{A sample mesh for fluid-structure interaction simulation.} \label{fsi_mesh}
\end{figure}

 A high quality mesh is very important for the accuracy of the simulation, and also for the rapid convergence of the iterative methods used in the simulation. In addition to high quality fluid and solid meshes, the quality of the interface mesh is also important.  For arteries with a small number of branches, good meshes are quite easy to generate, but when the computational domain is complex, such as the complete pulmonary artery, the FSI mesh generation is nontrivial and often the mesh obtained from the meshing tools such  as ANSYS \cite{Ansys2017Fluent}  doesn't have the required quality.  This is especially the true for the mesh at the fluid-structure  interface.  More precisely speaking, the interface mesh produced by the software  sometimes contains  elements that belong  to  the interior of  the fluid or the solid domain.  To overcome this difficulty, we introduce an interface mesh reconstruction  scheme that removes wrong interface elements   and creates a new interface mesh   by transversing through all fluid mesh elements.  The basic idea of the algorithm  is that we  walk through the fluid mesh, and for each tetrahedron element, we  mark one of its surface triangles  as an interface element if the triangle  is shared by a solid element. Let $\oo_{h,s} = \{K_s\}$ and $\oo_{h,f} = \{K_f \}$ be  the solid  and  fluid meshes consisting of non-overlapping tetrahedrons. Each fluid element is composed of four    surface triangles, $\{T_f\}$. $S_I$ represents the interface mesh constructed from the solid and fluid volume meshes.  The detailed method is summarized in Algorithm \ref{algorithm_IMR}. A sample interface mesh is shown in Fig.~\ref{interface_mesh}.
\begin{algorithm}
  \caption{Interface Mesh Reconstructing}\label{algorithm_IMR}
  \begin{algorithmic}[1]
    \item  $S_I=\{\}$ 
   \For {$K_f \in \oo_{h,f}$}
      \For {$T_f \in K_f$}
       \State Find the neighboring element, $K$, that shares $T_f$ with $K_f$
       \If {$K \in \oo_{h,s}$}
        \State  $S_I = S_I + T_f$
       \EndIf
      \EndFor 
   \EndFor
   \item return $S_I$
  \end{algorithmic}
\end{algorithm}
In Fig.~\ref{interface_mesh}, it is easy to see that there are several illegal interface elements that are floating; i.e., not attached to the interface.  This is observed from the  top right picture, and it is repaired  using Algorithm~\ref{algorithm_IMR},  and the new interface mesh is shown in the bottom left picture where the illegal interface elements  are removed.  A valid   FSI mesh is shown in Fig.~\ref{fsi_mesh}. 

\section{Final remarks}
Numerical simulation of blood flows in a compliant pulmonary artery is challenging because of the highly nonlinear nature of mathematical models for the fluid flow and the arterial wall, the complex configurations of the arterial network, and also the size of the discretized problem. In this paper, we developed a highly parallel algorithm for solving the monolithically coupled fluid-structure  system on a supercomputer with more than 10,000 processors. Using this technology, the simulation of unsteady blood flows in  a full three-dimensional, patient-specific pulmonary artery can be obtained in less than a few hours. In order to apply the techniques for clinical applications, more works are needed, such as more realistic boundary conditions, and materials parameters.

\section*{Acknowledgments}
This manuscript has been authored by Battelle Energy Alliance, LLC under Contract No. DE-AC07-05ID14517 with the U.S. Department of Energy. The United States Government retains and the publisher, by accepting the article for publication, ascknowledges that the United States Government retains a nonexclusive, paid-up, irrevocable, world-wide license to publish or reproduce the published form of this manuscript, or allow others to do so, for United States Government purposes.

%\section*{References}

%\bibliographystyle{wileyj}
%\bibliography{pulmonaryfsi}

\begin{thebibliography}{10}
\providecommand{\url}[1]{\texttt{#1}}
\providecommand{\urlprefix}{URL }
\expandafter\ifx\csname urlstyle\endcsname\relax
  \providecommand{\doi}[1]{doi:\discretionary{}{}{}#1}\else
  \providecommand{\doi}{doi:\discretionary{}{}{}\begingroup
  \urlstyle{rm}\Url}\fi

\bibitem{wang2016global}
Wang H, Naghavi M, Allen C, Barber R, Carter A, Casey D, Charlson F, Chen A,
  Coates M, Coggeshall M, \emph{et~al.}. Global, regional, and national life
  expectancy, all-cause mortality, and cause-specific mortality for 249 causes
  of death, 1980--2015: a systematic analysis for the {Global Burden of Disease
  Study} 2015. \emph{The Lancet}  2016; \textbf{388}(10053):1459--1544.

\bibitem{kheyfets2015patient}
Kheyfets VO, Rios L, Smith T, Schroeder T, Mueller J, Murali S, Lasorda D,
  Zikos A, Spotti J, Reilly~Jr JJ, \emph{et~al.}. Patient-specific
  computational modeling of blood flow in the pulmonary arterial circulation.
  \emph{Comput. Methods Programs Biomed.}  2015; \textbf{120}(2):88--101.

\bibitem{crosetto2011fluid}
Crosetto P, Reymond P, Deparis S, Kontaxakis D, Stergiopulos N, Quarteroni A.
  Fluid--structure interaction simulation of aortic blood flow. \emph{Comput.
  Fluids}  2011; \textbf{43}(1):46--57.

\bibitem{verdugo2017efficient}
Verdugo F, Roth CJ, Yoshihara L, Wall WA. Efficient solvers for coupled models
  in respiratory mechanics. \emph{Int. J. Numer. Methods Biomed. Eng.}  2017;
  \textbf{33}(2).

\bibitem{farhat1998load}
Farhat C, Lesoinne M, Le~Tallec P. Load and motion transfer algorithms for
  fluid/structure interaction problems with non-matching discrete interfaces:
  {M}omentum and energy conservation, optimal discretization and application to
  aeroelasticity. \emph{Comput. Methods Appl. Mech. Eng.}  1998;
  \textbf{157}(1-2):95--114.

\bibitem{kong2016scalability}
Kong F, Cai XC. Scalability study of an implicit solver for coupled
  fluid-structure interaction problems on unstructured meshes in {3D}.
  \emph{Int. J. High Perform. Comput. Appl.}  2018; \textbf{32}(2):207--219.

\bibitem{kong2017scalable}
Kong F, Cai XC. A scalable nonlinear fluid--structure interaction solver based
  on a {Schwarz} preconditioner with isogeometric unstructured coarse spaces in
  {3D}. \emph{J. Comput. Phys.}  2017; \textbf{340}:498--518.

\bibitem{bazilevs2006isogeometric}
Bazilevs Y, Calo VM, Zhang Y, Hughes TJ. Isogeometric fluid--structure
  interaction analysis with applications to arterial blood flow. \emph{Comput.
  Mech.}  2006; \textbf{38}(4-5):310--322.

\bibitem{farhat2006provably}
Farhat C, Van~der Zee KG, Geuzaine P. Provably second-order time-accurate
  loosely-coupled solution algorithms for transient nonlinear computational
  aeroelasticity. \emph{Comput. Methods Appl. Mech. Eng.}  2006;
  \textbf{195}(17-18):1973--2001.

\bibitem{kongfully}
Kong F, Wang Y, Schunert S, Peterson JW, Permann CJ, Andr{\v{s}} D, Martineau
  RC. A fully coupled two-level {Schwarz} preconditioner based on smoothed
  aggregation for the transient multigroup neutron diffusion equations.
  \emph{Numer. Linear Algebra Appl.}  2018;
  \doi{https://doi.org/10.1002/nla.2162}.

\bibitem{causin2005added}
Causin P, Gerbeau JF, Nobile F. Added-mass effect in the design of partitioned
  algorithms for fluid--structure problems. \emph{Comput. Methods Appl. Mech.
  Eng.}  2005; \textbf{194}(42-44):4506--4527.

\bibitem{souli2000ale}
Souli M, Ouahsine A, Lewin L. {ALE} formulation for fluid--structure
  interaction problems. \emph{Comput. Methods Appl. Mech. Eng.}  2000;
  \textbf{190}(5-7):659--675.

\bibitem{spilker2007morphometry}
Spilker RL, Feinstein JA, Parker DW, Reddy VM, Taylor CA. Morphometry-based
  impedance boundary conditions for patient-specific modeling of blood flow in
  pulmonary arteries. \emph{Ann. Biomed. Eng.}  2007; \textbf{35}(4):546--559.

\bibitem{tang2011three}
Tang BT, Fonte TA, Chan FP, Tsao PS, Feinstein JA, Taylor CA. Three-dimensional
  hemodynamics in the human pulmonary arteries under resting and exercise
  conditions. \emph{Ann. Biomed. Eng.}  2011; \textbf{39}(1):347--358.

\bibitem{qureshi2014numerical}
Qureshi MU, Vaughan GD, Sainsbury C, Johnson M, Peskin CS, Olufsen MS, Hill N.
  Numerical simulation of blood flow and pressure drop in the pulmonary
  arterial and venous circulation. \emph{Biomech. Model. Mechanobiol.}  2014;
  \textbf{13}(5):1137--1154.

\bibitem{su2012impact}
Su Z, Hunter KS, Shandas R. Impact of pulmonary vascular stiffness and
  vasodilator treatment in pediatric pulmonary hypertension: 21
  patient-specific fluid--structure interaction studies. \emph{Comput. Methods
  Programs Biomed.}  2012; \textbf{108}(2):617--628.

\bibitem{hunter2006simulations}
Hunter KS, Lanning CJ, Chen SYJ, Zhang Y, Garg R, Ivy DD, Shandas R.
  Simulations of congenital septal defect closure and reactivity testing in
  patient-specific models of the pediatric pulmonary vasculature: a {3D}
  numerical study with fluid-structure interaction. \emph{J. Biomech. Eng.}
  2006; \textbf{128}(4):564--572.

\bibitem{CFD_ACE}
{CFD-ACE+}. \emph{https://www.esi-group.com}  2017; .

\bibitem{yang2007fluid}
Yang X, Liu Y, Yang J. Fluid-structure interaction in a pulmonary arterial
  bifurcation. \emph{J. Biomech.}  2007; \textbf{40}(12):2694--2699.

\bibitem{Ansys2017Fluent}
{ANSYS}. \emph{http://www.ansys.com}  2017; .

\bibitem{wu2014fully}
Wu Y, Cai XC. A fully implicit domain decomposition based {ALE} framework for
  three-dimensional fluid--structure interaction with application in blood flow
  computation. \emph{J. Comput. Phys.}  2014; \textbf{258}:524--537.

\bibitem{kong2016parallel}
Kong F. {A Parallel Implicit Fluid-structure Interaction Solver with
  Isogeometric Coarse Spaces for 3D Unstructured Mesh Problems with Complex
  Geometry}. Ph{D} {T}hesis, University of Colorado Boulder 2016.

\bibitem{kong2017efficient}
Kong F, Kheyfets V, Finol E, Cai XC. An efficient parallel simulation of
  unsteady blood flows in patient-specific pulmonary artery. \emph{Int. J.
  Numer. Methods Biomed. Eng.}  2018; \doi{https://doi.org/10.1002/cnm.2952}.

\bibitem{kong2016highly}
Kong F, Cai XC. A highly scalable multilevel {Schwarz} method with boundary
  geometry preserving coarse spaces for {3D} elasticity problems on domains
  with complex geometry. \emph{SIAM J. Sci. Comput.}  2016;
  \textbf{38}(2):C73--C95.

\bibitem{howell2009applied}
Howell P, Kozyreff G, Ockendon J. \emph{{Applied Solid Mechanics}}, vol.~43.
  Cambridge University Press, 2009.

\bibitem{taylor1998finite}
Taylor CA, Hughes TJ, Zarins CK. Finite element modeling of blood flow in
  arteries. \emph{Comput. Methods Appl. Mech. Eng.}  1998;
  \textbf{158}(1-2):155--196.

\bibitem{whiting2001stabilized}
Whiting CH, Jansen KE. A stabilized finite element method for the
  incompressible {Navier-Stokes} equations using a hierarchical basis.
  \emph{Int. J. Numer. Methods Fluids}  2001; \textbf{35}(1):93--116.

\bibitem{barker2010scalable}
Barker AT, Cai XC. Scalable parallel methods for monolithic coupling in
  fluid-structure interaction with application to blood flow modeling. \emph{J.
  Comput. Phys.}  2010; \textbf{229}(3):642--659.

\bibitem{dembo1982inexact}
Dembo RS, Eisenstat SC, Steihaug T. Inexact {Newton} methods. \emph{SIAM J.
  Numer. Anal.}  1982; \textbf{19}(2):400--408.

\bibitem{saad2003iterative}
Saad Y. \emph{{Iterative Methods for Sparse Linear Systems}}, vol.~82. SIAM,
  2003.

\bibitem{smith2004domain}
Smith B, Bjorstad P, Gropp W. \emph{{Domain Decomposition: Parallel Multilevel
  Methods for Elliptic Partial Differential Equations}}. Cambridge University
  Press, 2004.

\bibitem{dennis1996numerical}
Dennis~Jr JE, Schnabel RB. \emph{{Numerical Methods for Unconstrained
  Optimization and Nonlinear Equations}}, vol.~16. SIAM, 1996.

\bibitem{toselli2006domain}
Toselli A, Widlund O. \emph{{Domain Decomposition Methods: Algorithms and
  Theory}}, vol.~34. Springer Science \& Business Media, 2006.

\bibitem{karypis1997parmetis}
Karypis G, Schloegel K, Kumar V. {Parmetis: Parallel Graph Partitioning and
  Sparse Matrix Ordering Library}. \emph{Technical {R}eport}, Department of
  Computer Science, University of Minnesota 1997.

\bibitem{petsc-user-ref}
Balay S, Abhyankar S, Adams MF, Brown J, Brune P, Buschelman K, Dalcin L,
  Eijkhout V, Gropp WD, Kaushik D, \emph{et~al.}. {{PETS}c Users Manual}.
  \emph{Technical {R}eport ANL-95/11 - Revision 3.9}, Argonne National
  Laboratory 2018. \urlprefix\url{http://www.mcs.anl.gov/petsc}.

\end{thebibliography}

\end{document}